\documentclass[prd,amsmath,amssymb,longbibliography,superscriptaddress,twocolumn,nofootinbib,10pt]{revtex4-1}
\pdfoutput=1
\usepackage{graphicx}
\usepackage{dcolumn}
\usepackage{bm}
\usepackage{amssymb}
\usepackage{latexsym}
\usepackage{booktabs}
\usepackage{amsmath}
\usepackage{multirow}

\usepackage[colorlinks=true, linkcolor=blue, citecolor=blue]{hyperref}

\begin{document}

\title{Simultaneous determination of the Hubble constant and cosmic baryon density: Forecasts for the synergy between FRBs and emerging probes}

\author{Peng-Ju Wu}\thanks{Email}\email{wupengju@nxu.edu.cn}
\affiliation{School of Physics, Ningxia University, Yinchuan 750021, China}

\author{Bo-Yang Zhang}
\affiliation{School of Physics and Electronic Information Engineering, Ningxia Normal University, Guyuan 756000, China}

\author{Ji-Guo Zhang}
\affiliation{Liaoning Key Laboratory of Cosmology and Astrophysics, College of Sciences, Northeastern University, Shenyang 110819, China}

\author{Guo-Hong Du}
\affiliation{Liaoning Key Laboratory of Cosmology and Astrophysics, College of Sciences, Northeastern University, Shenyang 110819, China}

\author{Shang-Jie Jin}
\affiliation{Liaoning Key Laboratory of Cosmology and Astrophysics, College of Sciences, Northeastern University, Shenyang 110819, China}
\affiliation{Department of Physics, University of Western Australia, Perth WA 6009, Australia}
\affiliation{Research Center for the Early Universe, Graduate School of Science, The University of Tokyo, Tokyo 113-0033, Japan}

\author{Xin Zhang}\thanks{Corresponding author}\email{zhangxin@mail.neu.edu.cn}
\affiliation{Liaoning Key Laboratory of Cosmology and Astrophysics, College of Sciences, Northeastern University, Shenyang 110819, China}
\affiliation{Key Laboratory of Data Analytics and Optimization for Smart Industry (Ministry of Education), Northeastern University, Shenyang 110819, China}
\affiliation{National Frontiers Science Center for Industrial Intelligence and Systems Optimization, Northeastern University, Shenyang 110819, China}

\begin{abstract}
Two of the most pressing challenges in cosmology are the persistent discrepancy in measurements of the Hubble constant, referred to as the Hubble tension, and the deficit of baryons in the local Universe, known as the missing baryon problem. Fast radio bursts (FRBs) provide a unique probe of both the Hubble constant $H_0$ and the cosmic baryon density $\Omega_{\rm b}$. However, constraints from FRBs alone suffer from a severe $H_0$--$\Omega_{\rm b}$ degeneracy that prevents them from addressing either problem. We show that this degeneracy can be broken by combining FRBs with other emerging probes whose degeneracy directions differ in the $H_0$--$\Omega_{\rm b}$ plane. Specifically, we quantify three multi-messenger approaches: FRBs paired with gravitational wave (GW) standard sirens, strong gravitational lensing (SGL) time delays, and 21 cm intensity mapping (IM) surveys. The combinations FRB+GW, FRB+SGL, and FRB+21\,cm\,IM each deliver simultaneous constraints on $H_0$ and $\Omega_{\rm b}$ better than (1\%, 1.5\%) in the $\Lambda$CDM model, and when dynamical dark energy is introduced, the constraining precision degrades gracefully as model complexity increases. Furthermore, within a model-independent framework, both FRB+GW and FRB+SGL constrain $H_0$ and $\Omega_{\rm b}$ to precisions better than (1.5\%, 3\%). These precision levels are based on nominal observational expectations and would improve significantly under optimistic observational scenarios.
\end{abstract}

\maketitle
\section{Introduction}\label{sec1}
A profound crisis in modern cosmology arises from a set of persistent observational tensions that challenge the robustness of the standard $\Lambda$ cold dark matter ($\Lambda$CDM) paradigm. Among these, two discrepancies have garnered particular attention for both their statistical significance and their implications for fundamental physics: the Hubble tension and the missing baryon problem \cite{Verde:2019ivm,Riess:2021jrx,Fukugita:1997bi,Cen:1998hc,Shull:2011aa}. The Hubble tension reflects the $>5\sigma$ discrepancy between the Hubble constant value measured from the cosmic microwave background (CMB) data by the Planck satellite $H_0 = 67.4 \pm 0.5 \, \text{km/s/Mpc}$ \cite{Planck:2018vyg}, and the higher value derived from distance-ladder measurements by the SH0ES collaboration $H_0 = 73.04 \pm 1.04 \, \text{km/s/Mpc}$ \cite{Riess:2021jrx}. The missing baryon problem highlights the apparent shortfall in the observed inventory of baryonic matter at low redshifts relative to the cosmic baryon density predicted by the CMB analyses ($\Omega_{\rm b} \sim 0.05$) \cite{Planck:2018vyg}, though with far less statistical significance than the Hubble tension. In both cases, the tension lies between early- and late-Universe determinations based on distinct methodologies and assumptions. Resolving them requires precise, independent measurements from alternative probes to discern whether the discrepancies originate in new physics, unmodeled astrophysics, or hidden systematics.

Fast radio bursts (FRBs)---millisecond-duration radio transients that appear randomly across the sky---emerge as promising probes for this purpose \cite{Hagstotz:2021jzu,Wu:2021jyk,Liu:2022bmn,James:2022dcx,Baptista:2023uqu,Fortunato:2023deh,Gao:2023izj,Wei:2023avr,Zhao:2022yiv,Zhang:2023gye,Wang:2025ugc,Zhang:2024rra,Yang:2024vqq,Kalita:2024xae,Sales:2025shu,Gao:2025fcr,Zhang:2025thh,McQuinn:2013tmc,Deng:2013aga,Macquart:2020lln,Yang:2022ftm,Connor:2024mjg,Zhang:2025yhi,Liu:2025fdf}. As an FRB propagates through the ionized intergalactic medium (IGM), its signal interacts with free electrons, resulting in a frequency-dependent time delay quantified by the dispersion measure (DM). Observed DMs of FRBs significantly exceed the contribution expected from the Milky Way, indicating their extragalactic origins \cite{Lorimer:2007qn,Thornton:2013iua,Petroff:2019tty}, with the exception of one Galactic burst \cite{CHIMEFRB:2020abu,Bochenek:2020zxn}. Since the IGM constitutes the dominant baryonic reservoir along the line of sight, the IGM component of DM accumulates with redshift and thus can be considered as an indicator of the distance to the FRB source. This quantity is proportional to the product \(H_0\Omega_{\rm b}\), so that FRB observations alone can constrain only this degenerate combination, precluding independent determination of either parameter. To unlock the potential of FRBs in addressing cosmological tensions, this degeneracy must be broken. This can be achieved by combining FRBs with complementary probes that either provide an independent measurement of \(H_0\) or exhibit a distinct degeneracy direction in the \(H_0\)--\(\Omega_{\rm b}\) plane.

In this paper, we assess the potential of future FRB observations in combination with three emerging cosmological probes---gravitational wave (GW) standard sirens, strong gravitational lensing (SGL) time delays, and neutral hydrogen 21~cm intensity mapping (IM) surveys---to provide simultaneous measurements of \(H_0\) and \(\Omega_{\rm b}\). GW standard sirens provide a direct measurement of the luminosity distance \(D_{\rm L}\) \cite{Schutz:1986gp,Holz:2005df}, which scales as \(D_{\rm L} \propto 1/H_0\). This offers a distance estimate independent of the electromagnetic cosmic distance ladder, although it depends on accurate modeling of gravitational waveforms for calibration. SGL time delays, on the other hand, yield the so-called time-delay distance \(D_{\Delta t}\), a geometric quantity constructed from angular diameter distances to the lens, to the source, and between them \cite{Treu:2016ljm}. Since all angular diameter distances scale linearly with \(1/H_0\), this composite distance inherits the same \(1/H_0\) dependence. In contrast, 21\,cm\,IM measures the baryon acoustic oscillation (BAO) feature in the large-scale structure, providing simultaneous constraints on both the angular diameter distance \(D_{\rm A}\) and the Hubble parameter \(H\) at the effective redshift of the survey \cite{Bull:2014rha}. While \(D_{\rm L} \propto 1/H_0\) and  \(H(z) \propto H_0\), the physical interpretation of these observables depends critically on the absolute scale of the BAO ruler, i.e., the sound horizon at the baryon drag epoch $r_{\rm d}$. This scale is determined by early-Universe physics and is itself a sensitive function of both \(H_0\) and \(\Omega_{\rm b}\) \cite{Chen:2018dbv}. Consequently, measurements of \(D_{\rm A}/r_{\rm d}\) and \(H(z)r_{\rm d}\) do not directly yield \(H_0\) and \(\Omega_{\rm b}\) in isolation; instead, they yield constraints that exhibit a strong degeneracy between the two parameters. The degeneracy direction between \(H_0\) and \(\Omega_{\rm b}\) in this case likely differs from that in the FRB measurements. By combining FRBs with these complementary probes, the individual parameter degeneracies may be broken, enabling simultaneous determination of \(H_0\) and \(\Omega_{\rm b}\).

Our primary objective is to assess how effectively the proposed synergistic approaches can break the characteristic degeneracy between \( H_0 \) and \( \Omega_{\mathrm{b}} \) inherent in the FRB measurements, thereby deriving constraints on these two fundamental parameters. We will evaluate the performance of these synergies under both nominal and optimistic scenarios. This paper is structured as follows. In Section~\ref{sec2}, we describe the simulation of future observational data for the four emerging probes based on specific experimental configurations. In Section~\ref{sec3}, we present the constraint results and make relevant discussions. Finally, we summarize our conclusions in Section~\ref{sec4}.

\section{Data simulation}\label{sec2}
In this section, we generate mock observations for four emerging probes using representative next-generation facilities. For simulations, we adopt the flat $\Lambda$CDM model with Planck 2018 parameters: $H_0 = 67.26\,{\rm km/s/Mpc}$, $\Omega_{\rm m} = 0.317$, $\Omega_{\rm b} = 0.0495$, and $\Omega_K = 0$ \cite{Planck:2018vyg}.

\subsection{SKA-era FRB observations}\label{sec21}
The DM contribution from IGM relates to cosmology, and its average value is calculated by
\begin{align}
\label{FRB-DM-IGM}
\overline{{\rm DM}}_{\rm IGM}(z)=\displaystyle{\frac{3cH_0\Omega_{\rm b}}{8\pi Gm_{\rm p}}}\int_{0}^{z}\displaystyle{\frac{(1+z')f_{\rm IGM}(z')f_{\rm e} (z'){\rm d}z'}{E(z')}},
\end{align}
where $E(z)\equiv H(z)/H_0$ is the dimensionless Hubble parameter, $c$ is the speed of light, $G$ is Newton's constant, $m_{\rm p}$ is the proton mass, $f_{\rm e}(z)=Y_{\rm H}\,\chi_{\rm e,H}+Y_{\rm He}\,\chi_{\rm e,He}/2$, with $Y_{\rm H}=3/4$ and $Y_{\rm He}=1/4$ being the mass fractions of hydrogen and helium, and $\chi_{\rm e,H}=1$ and $\chi_{\rm e,He}=1$ being the ionization fractions, respectively. It has been suggested that $f_{\rm IGM}$ evolves with redshift, with values of $\simeq0.9$ at $z\geq1.5$ \cite{Meiksin:2007rz} and $\simeq0.82$ at $z\leq0.4$ \cite{Shull:2011aa}. While a constant value of $f_{\rm IGM}\simeq0.83$ has been frequently adopted, such a treatment overlooks the physical complexity and redshift dependence of the IGM baryon fraction. To incorporate the above redshift trends while maintaining a continuous form, we employ:
\begin{equation}
f_{\rm IGM}(z) =
\begin{cases}
0.053\,z + 0.82, & z \le 1.5,\\[4pt]
0.9, & z > 1.5.
\end{cases}
\end{equation}

The uncertainty of ${\rm DM}_{\rm IGM}$ can be written as
\begin{align}
\sigma_{{\rm DM}_{\rm IGM}}=\left[\sigma_{\rm obs}^2 + \sigma_{\rm MW}^2 +\sigma_{\rm IGM}^2 +  \left(\displaystyle{\frac{\sigma_{\rm host}}{1+z}}\right)^2\right]^{1/2}.
\end{align}
Here, we adopt $\sigma_{\rm obs}=0.5\,\rm{pc\,cm^{-3}}$ from published data \citep{Petroff:2019tty,CHIMEFRB:2021srp}, $\sigma_{\rm MW}=10\,\rm{pc\,cm^{-3}}$ for the Milky Way contribution, $\sigma_{\rm IGM} \simeq 173.8z^{0.4}\,\rm{pc\,cm^{-3}}$ accounting for the baryon inhomogeneity in IGM \cite{Jaroszynski:2020kqy}, and $\sigma_{\rm host}=60\,\rm{pc\,cm^{-3}}$ for the host galaxy contribution \cite{Macquart:2020lln,James:2022dcx}. In this work, the host contribution enters the error budget through the term $\sigma_{\rm host}$. In real observations, however, the host DM must be estimated and subtracted from the total DM. If this subtraction is systematically biased, the resulting ${\rm DM}_{\rm IGM}$ will be offset, thereby biasing the inferred cosmological parameters. Future high-resolution observations, such as those expected from SKA, together with multi-wavelength follow-up of localized FRBs, can provide host galaxy properties (e.g., stellar mass, star formation rate, inclination) that can be used to construct improved priors for the host DM or to model it as a function of host properties. In addition, hierarchical Bayesian methods that simultaneously fit the cosmological parameters and the parameters describing the host DM distribution can mitigate such bias \cite{Macquart:2020lln}. When applying real FRB data to constrain $H_0$ and $\Omega_{\rm b}$, it is prudent to test the sensitivity to different host DM priors, account for selection effects, and validate the analysis with simulations.

In order to mock detectable FRBs, we need to choose a redshift distribution for FRBs. The common assumption is that FRBs' population tracks the star formation history (SFH). However, the CHIME/FRB catalog has shown that the FRB population does not strictly track the cosmic star formation history. In this paper, we adopt the power-law model of redshift distribution to simulate the FRB data, and the FRB event rate is
\begin{align}
N_{\rm SFH}(z)=(1+z)^{\gamma}\mathcal{N}_{\rm SFH}\displaystyle{\frac{\dot{\rho}_*(z)D_{\rm C}^2(z)}{H(z)(1+z)}e^{{-D_{\rm L}^2(z)}/{[2D_{\rm L}^2(z_{\rm cut})]}}},
\end{align}
where $(1+z)^{\gamma}$ refers to the delay with respect to SFH with $\gamma=-1.1$, $\mathcal{N}_{\rm SFH}$ is a normalization factor, $D_{\rm C}$ is the comoving distance, and $z_{\rm cut} = 1$ is a cutoff, which reflects the decline in detected FRBs beyond it due to the instrumental signal-to-noise threshold. The density evolution form of SFH is parameterized as \citep{Madau:2016jbv,Qiang:2021ljr}
\begin{align}
\dot{\rho}_*(z)=\displaystyle{\frac{(1+z)^{2.6}}{1+((1+z)/3.2)^{6.2}}}.
\end{align}
For cosmological constraints, we adopt the FRB detection event rate by the upcoming Square Kilometer Array (SKA) in Ref.~\cite{Zhang:2023gye}. The 10-year observation of the SKA would detect $\mathcal{O}(10^5-10^6)$ FRBs. Assuming that 10\% of the detected FRBs can be well localized to confirm their host galaxies, the 10-year operation of SKA is expected to yield $\mathcal{O}(10^4-10^5)$ FRBs with redshifts available for parameter inference. In this work, we consider two representative cases: a nominal scenario with $N_{\rm FRB} = 10^4$ and an optimistic scenario with $N_{\rm FRB} = 10^5$.

\subsection{ET-era GW observations}\label{sec22}
GWs can serve as standard sirens, since the GW waveform carries the information of the luminosity distance to the source. We simulate the GW standard sirens based on the third-generation GW detector, Einstein Telescope (ET), and assume that the GWs are from the binary neutron star (BNS) mergers. For the redshift distribution of BNSs, we adopt the form \cite{Zhao:2010sz}
\begin{align}
P(z)\propto \displaystyle{\frac{4\pi D_{\rm C}^2(z)R(z)}{H(z)(1+z)}},
\end{align}
where $R(z)$ is the time evolution of the burst rate with the form \cite{Schneider:2000sg,Cutler:2009qv}
\begin{eqnarray}
R(z) =
\begin{cases}
1+2z,                                 &  z\leq 1, \\
\displaystyle{\frac{3}{4}}(5-z),      &  1 < z < 5,\\
0,                                    &  z \geq 5.
\end{cases}
\end{eqnarray}
We calculate the central value of luminosity distance in the $\Lambda$CDM model. The total errors of $D_{\rm L}$ consist of the instrumental error, weak lensing error, and peculiar velocity error,
\begin{align}
\sigma_{D_{\rm L}}=\sqrt{(\sigma_{D_{\rm L}}^{\rm inst})^2 + (\sigma_{D_{\rm L}}^{\rm lens})^2 + (\sigma_{D_{\rm L}}^{\rm pv})^2}.
\end{align}
For the $\sigma_{D_{\rm L}}^{\rm inst}$ calculation, we refer the readers to Refs.~\cite{Zhao:2010sz,Zhang:2019loq,Jin:2020hmc,Wu:2022dgy,Jin:2023sfc,Wang:2021srv,Xiao:2025mcg,Du:2025odq}. The errors caused by the weak lensing and peculiar velocity of the GW source can be found in Refs.~\cite{Hirata:2010ba,Gordon:2007zw}. The $D_{\rm L}$-redshift relation can be established, once the electromagnetic counterpart of the GW source is detected by optical telescopes. ET is expected to detect $1000$ GW events from BNSs within $z\lesssim5$, as the expectation of its 10-year operation \cite{Zhao:2010sz,Cai:2016sby}. We simulate a sample of $1000$ GW standard sirens.

\subsection{LSST-era SGL observations}\label{sec23}
SGL is a rare astronomical phenomenon in which photons from a distant source are deflected by the gravity of intervening mass overdensities (e.g., galaxies, groups, clusters) as they propagate to Earth's detectors. SGL time delays provide a geometric probe of the universe. When the background source exhibits intrinsic flux variability, monitoring the lensed images allows a measurement of the arrival time differences between multiple images \cite{Treu:2016ljm}. For a pair of images \(i\) and \(j\), the time delay is given by
\begin{align}
\Delta t_{ij} = \frac{D_{\Delta t}}{c} \left[ \frac{(\vec{\theta}_i - \vec{\beta})^2}{2} - \psi(\vec{\theta}_i) - \frac{(\vec{\theta}_j - \vec{\beta})^2}{2} + \psi(\vec{\theta}_j) \right],
\end{align}
where \(\vec{\beta}\) and \(\vec{\theta}_i\) are the angular positions of the source and the image $i$, respectively, and \(\psi\) is the lensing potential. The cosmological information is encapsulated in the time-delay distance, defined as
\begin{align}
D_{\Delta t} \equiv (1 + z_{\rm l}) \frac{D_{\rm l} D_{\rm s}}{D_{\rm ls}},
\end{align}
where \(z_{\rm l}\) is the lens redshift, \(D_{\rm l}\), \(D_{\rm s}\), and \(D_{\rm ls}\) are the angular diameter distances to the lens, to the source, and between the lens and source, respectively. A measurement of the time delay, combined with models for the lens potential and the source position, yields a constraint on the distance ratio \(D_{\rm l} D_{\rm s} / D_{\rm ls}\). Since this ratio scales inversely with the Hubble constant, time-delay observations provide a direct avenue to constrain \(H_0\). We simulate a sample of 55 time-delay events based on the Large Synoptic Survey Telescope (LSST) \cite{LSST:2008ijt}. We generate the event redshifts using the code from Ref.~\cite{Collett:2015roa} and calculate the time-delay distances within the fiducial model, assuming a \(5\%\) relative uncertainty for each measurement \cite{Qi:2018atg}.

\subsection{HIRAX-era 21\,cm\,IM  observations}\label{sec24}
The 21\,cm\,IM technique enables the measurement of the BAO signal by mapping the collective 21 cm emission from the neutral hydrogen (H\,{\sc i}) in galaxies. We mock 21\,cm\,IM observations based on the Hydrogen Intensity and Real-time Analysis eXperiment (HIRAX) \cite{Newburgh:2016mwi}. The modeling begins with the mean 21~cm brightness temperature, given by
\begin{equation}
\overline{T}_{\mathrm{b}}(z) = 188 \, h \, \Omega_{\mathrm{HI}}(z) \frac{(1+z)^2}{E(z)} \; \mathrm{mK},
\end{equation}
where $\Omega_{\mathrm{HI}}(z)$ is the neutral hydrogen density fraction. In observational coordinates $(\boldsymbol{q}={\boldsymbol{k}}_{\perp}D_{\rm C}, y=k_{\parallel}D_{\nu})$, the signal covariance can be written as~\cite{Bull:2014rha}
\begin{equation}
\small
C^{\rm S}(\boldsymbol{q},y) = \frac{\overline{T}_{\rm b}^2(z_i)\alpha_{\perp}^2\alpha_{\parallel}}{D_{\rm C}^2D_\nu} \left(b_{\mathrm{HI}} + f\mu^2\right)^2 \exp\left(-k^2\mu^2\sigma_{\mathrm{NL}}^2\right) \times P(k),
\end{equation}
where $z_i$ is the central redshift of the $i$-th redshift bin, $k$ is the wave vector, $\mu$ is the cosine between \(\vec{k}\)  and the line of sight direction, $D_{\nu}(z) = c(1+z)^2 / H(z)$, and the scaling factors are defined as $\alpha_{\perp} \equiv D_{\rm A}^{\mathrm{fid}}(z)/D_{\rm A}(z)$ and $\alpha_{\parallel} \equiv H(z)/H^{\mathrm{fid}}(z)$. Here, $b_{\mathrm{HI}}$ is the H\,{\sc i} bias, $f(z)$ is the linear growth rate, $\sigma_{\mathrm{NL}}$ parametrizes the non-linear dispersion scale, and $P(k)$ is the matter power spectrum, which can be generated by {\tt CAMB} \cite{Lewis:1999bs}. The total covariance is modeled as the sum of three components: $C^{\mathrm{T}} = C^{\mathrm{S}} + C^{\mathrm{N}} + C^{\mathrm{F}}$.

The instrumental noise covariance is given by \cite{Bull:2014rha}
\begin{equation}
C^{\mathrm{N}}(\boldsymbol{q},y) = \frac{\sigma_{\mathrm{pix}}^2 V_{\mathrm{pix}}}{D_{\mathrm{C}}^2 D_{\nu}} B_{\parallel}^{-1} B_{\perp}^{-2},
\label{noise_cov}
\end{equation}
where $\sigma_{\mathrm{pix}}$ is the pixel noise, $V_{\mathrm{pix}} = D_{\mathrm{C}}^2\,\mathrm{FoV} \times D_{\nu}\,\delta\nu/\nu_{21}$ is the pixel volume with $\nu_{21}=1420.4\,\mathrm{MHz}$ the rest frequency of the 21 cm line, and \(B_{\parallel}\) and \(B_{\perp}\) describe the frequency and angular responses of the instrument, respectively. For HIRAX, the pixel noise is calculated by
\begin{equation}
\sigma_{\mathrm{pix}} = \frac{T_{\mathrm{sys}}}{\sqrt{n_{\mathrm{pol}}\,t_{\mathrm{tot}}\,\delta\nu\,(\mathrm{FoV}/S_{\mathrm{area}})}} \frac{\lambda^2}{A_{\mathrm{e}}\sqrt{\mathrm{FoV}}} \frac{1}{\sqrt{n(\boldsymbol{u})\,N_{\mathrm{b}}}},
\label{pixel_noise}
\end{equation}
where $T_{\mathrm{sys}}$ is the system temperature, $n_{\mathrm{pol}}=2$ is the number of polarization channels, $\delta\nu$ is the channel bandwidth, $\lambda$ is the observed wavelength, $t_{\mathrm{tot}}=10,000\,\mathrm{hr}$ is the observing time, $S_{\mathrm{area}}=15,000\,\mathrm{deg}^2$ is the survey area, $A_{\mathrm{e}}=\eta\pi(D_{\mathrm{d}}/2)^2$ is the effective collecting area of each receiver, with $D_{\mathrm{d}}=6\,\mathrm{m}$ the dish diameter and $\eta=0.7$ the efficiency factor, $\mathrm{FoV}\approx(\lambda/D_{\mathrm{d}})^2$, $N_{\mathrm{b}}=1$ is the number of beams, and $n(\boldsymbol{u})$ is the baseline density, with $\boldsymbol{u}$ the baseline vector. The system temperature is
\begin{equation}
T_{\mathrm{sys}} = T_{\mathrm{rec}} + T_{\mathrm{gal}} + T_{\mathrm{CMB}},
\end{equation}
where $T_{\mathrm{rec}}=50\,\mathrm{K}$ is the receiver noise temperature, $T_{\mathrm{gal}}\approx25\,\mathrm{K}\times(408\,\mathrm{MHz}/\nu)^{2.75}$ is the Galactic contribution, and $T_{\mathrm{CMB}}\approx2.73\,\mathrm{K}$ is the CMB temperature. Assuming a Gaussian channel bandpass, the effective beam in the parallel direction is \cite{Bull:2014rha}
\begin{equation}
B_{\parallel}(y) = \exp\left(-\frac{(y\,\delta\nu/\nu_{21})^2}{16\ln 2}\right),
\end{equation}
while the transverse effective beam is accounted for through the baseline density.

The residual foreground covariance has the form \cite{Bull:2014rha}
\begin{equation}
C^{\mathrm{F}}(\boldsymbol{q},y) = \varepsilon_{\mathrm{FG}}^2 \sum_{X} A_{X} \left(\frac{\ell_p}{2\pi q}\right)^{n_{X}} \left(\frac{\nu_p}{\nu_i}\right)^{m_{X}},
\label{fg_cov}
\end{equation}
where \(X\) denotes a foreground component, with \(A_X\), \(n_X\), and \(m_X\) being its amplitude, angular power-law index, and frequency power-law index, respectively. Here \(\ell_p=1000\) and \(\nu_p=130\,\mathrm{MHz}\) are the pivot multipole and pivot frequency at which these parameters are evaluated \cite{Santos:2004ju}, and \(\nu_i = \nu_{21}/(1+z_i)\) is the observing frequency corresponding to the \(i\)-th bin. The parameter \(\varepsilon_{\mathrm{FG}}\) is the foreground removal efficiency, with \(\varepsilon_{\mathrm{FG}}=1\) corresponding to no removal and \(\varepsilon_{\mathrm{FG}}=0\) to perfect removal. We adopt \(\varepsilon_{\mathrm{FG}}=10^{-5}\) and \(10^{-6}\) as the nominal and optimistic scenarios, respectively. Note that in this parameterization, the residual foreground is suppressed by a constant factor \(\varepsilon_{\rm FG}\) across all \(k\) modes and frequencies, assuming uniform suppression. In reality, however, foreground subtraction is expected to be highly non-uniform: different \(k_\parallel\) and \(k_\perp\) modes are affected differently, and the cleaning process may also remove part of the 21 cm signal itself, leading to signal loss that modifies the signal covariance. These effects can degrade the BAO measurement and weaken the cosmological constraints. Consequently, our forecasts for 21\,cm\,IM should be regarded as benchmarks based on an idealized foreground model. A full assessment would require realistic foreground simulations and end-to-end analyses that account for mode-dependent suppression and signal loss.

Measurement uncertainties on the cosmological observables are forecast using the Fisher matrix formalism. We focus on the parameter set $\boldsymbol{p} = \{D_{\rm A}(z), H(z), f\sigma_8(z)\}$ within independent redshift bins of width $\Delta z = 0.1$ over $0.8 < z < 2.5$. For each redshift bin, the Fisher matrix is computed as:
\begin{equation}
F_{ij} = \frac{1}{8\pi^2} V_{\mathrm{bin}} \int_{-1}^{1} d\mu \int_{k_{\mathrm{min}}}^{k_{\mathrm{max}}} k^2 {\rm d}k \; \frac{\partial \ln C^{\mathrm{T}}}{\partial p_i} \frac{\partial \ln C^{\mathrm{T}}}{\partial p_j},
\end{equation}
where $V_{\mathrm{bin}}$ is the comoving survey volume of the bin. The resulting Fisher matrices serve as inverse covariances for $\{D_{\rm A}, H, f\sigma_8\}$. The central values of the observables are computed within the $\Lambda$CDM framework. Note that for cosmological parameter inference, we restore the sound horizon scale. That is, we employ the dimensionless observables $\{D_{\rm A}/r_{\rm d}, c/(H\times r_{\rm d}), f\sigma_8\}$ to constrain cosmological parameters. Here, the comoving sound horizon at the baryon drag epoch is calculated as
\begin{equation}
r_{\rm d} = \int_{z_{\rm d}}^{\infty} \frac{c_{\rm s}(z)}{H(z)} \, {\rm d}z,
\label{sound}
\end{equation}
where the sound speed is \(c_{\rm s}(z) = c / \sqrt{3[1 + R_{\rm b} \, a(z)]}\) with \(R_{\rm b} = 3\Omega_{\rm b} / (4\Omega_\gamma)\), and the drag redshift \(z_{\rm d}\) is computed following \cite{Chen:2018dbv}:
\begin{equation}
\begin{aligned}
z_{\rm d} &= 1048 \left[1 + 0.00124 (\Omega_{\rm b} h^2)^{-0.738}\right] \left[1 + g_1(\Omega_{\rm m} h^2)^{g_2}\right], \\
g_1 &= \frac{0.0783}{(\Omega_{\rm b} h^2)^{0.238} \left[1 + 39.5 (\Omega_{\rm b} h^2)^{0.763}\right]}, \\
g_2 &= \frac{0.56}{1 + 21.1 (\Omega_{\rm b} h^2)^{1.81}}.
\end{aligned}
\end{equation}

\begin{figure*}
\includegraphics[scale=0.4]{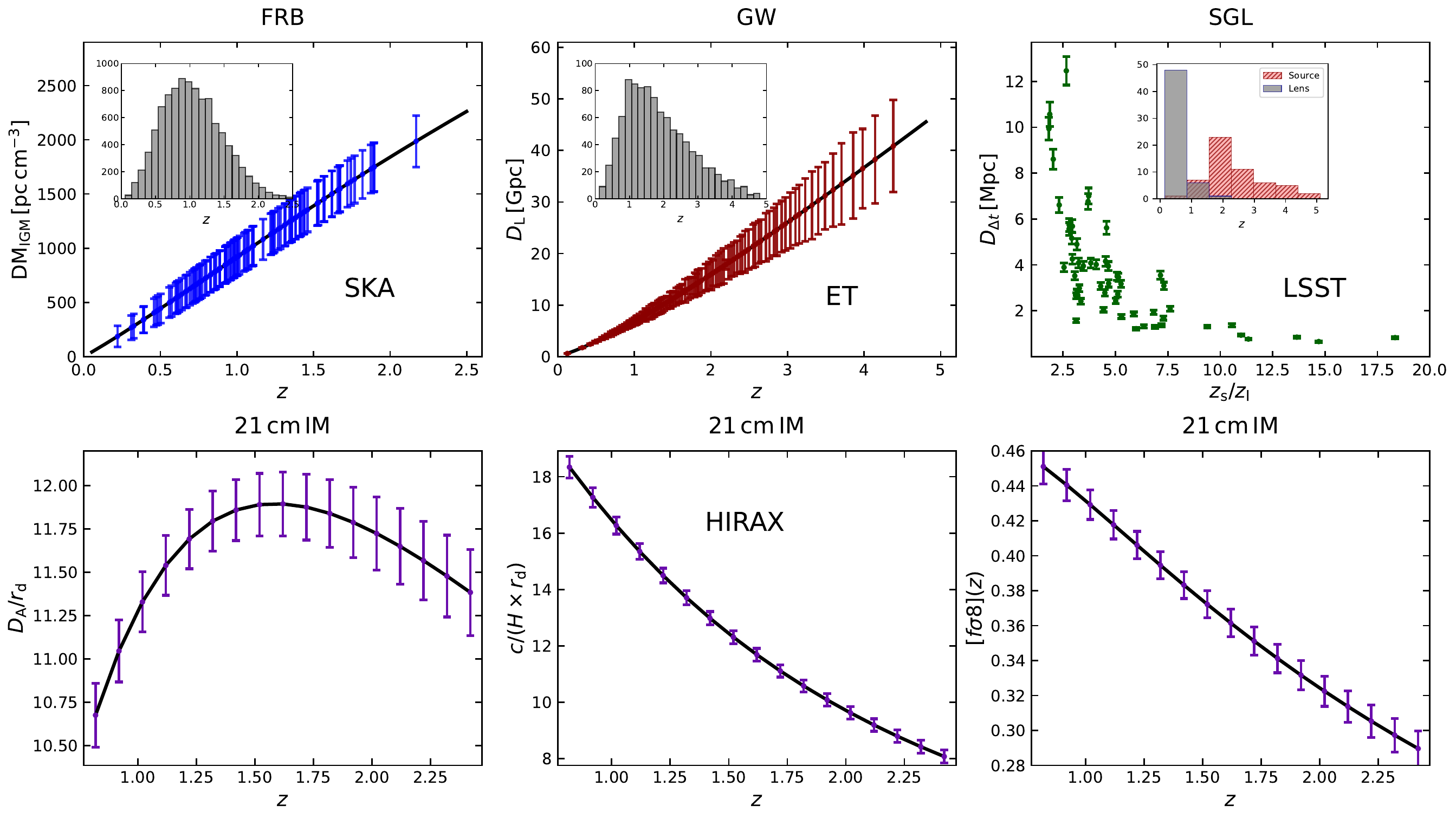}
\centering
\caption{Simulated observational data for four emerging cosmological probes: FRBs from SKA (top left); GW standard sirens from ET (top middle); SGL time delays from LSST (top right); and 21\,cm\,IM from HIRAX (bottom three panels). For the FRB, GW, and SGL data, we show both measurement uncertainties and redshift distributions of the simulated samples. Note that the central values of the mock data points are fixed to the fiducial cosmology (i.e., they assume the fiducial model as the true underlying cosmology). Measurement errors are included, but no statistical realizations that would shift the central values away from the fiducial model are included.}
\label{mock-data}
\end{figure*}

\renewcommand{\arraystretch}{1.3}
\begin{table*}
\centering
\caption{Summary of forecast assumptions for each probe.}
\label{tab:forecast_assumptions}
\begin{tabular}{|l|l|c|m{11.5cm}|}
\hline
\textbf{Probe} & \textbf{Instrument} & \textbf{Survey duration} & \textbf{Key assumptions} \\
\hline
FRB            & SKA                 & 10 years                   & (i)~10\% of detected FRBs can be localized to their host galaxies; \newline (ii) The operation is expected to yield $\mathcal{O}(10^4-10^5)$ FRBs with redshifts.\\
\hline
GW             & ET                  & 10 years                    & (i)~GWs can be localized to their host galaxies from electromagnetic counterparts; \newline (ii) The operation is expected to yield 1,000 GW events with redshifts.\\
\hline
SGL            & LSST                & 10 years                    & (i)~Galaxy-galaxy SGLs with the singular isothermal ellipsoid lens model; \newline (ii) The operation is expected to yield 55 time-delay events with redshifts. \\
\hline
21\,cm\,IM     & HIRAX               & $10,000$ hours       & (i)~Foregrounds can be subtracted sufficiently to extract the 21 cm signal. \newline (ii) The ratio of residual to original foreground amplitude reaches $10^{-5}$--$10^{-6}$. \\
\hline
\end{tabular}
\end{table*}

In Fig.~\ref{mock-data}, we present the simulated observational data from FRBs, GW standard sirens, SGL time delays, and 21\,cm\,IM surveys, based on specific experiments. In generating the mock data, we fix the fiducial (central) values and do not include their statistical fluctuations. This is justified for three reasons: (i) forecast analyses focus on parameter uncertainties rather than central values; (ii) fixing a common fiducial cosmology avoids artificial tensions when combining the four probes; and (iii) it ensures that the contours clearly illustrate how parameter degeneracies are broken through probe synergy. The key forecast assumptions for each probe are summarized in Table~\ref{tab:forecast_assumptions}.

In this work, we break the $H_0$--$\Omega_{\rm b}$ degeneracy by directly combining the likelihoods of different probes. This is justified because FRB, GW, SGL, and 21\,cm\,IM trace distinct aspects of the Universe, so the statistical errors of their observables are largely uncorrelated. Nevertheless, these probes can be linked via underlying physical mechanisms. FRBs and 21\,cm\,IM are connected through their common dependence on the large-scale matter distribution: FRBs have been proposed as cosmological distance probes that enable clustering studies of large-scale structure \cite{Masui:2015ola,Maity:2024zwa}, establishing a statistical relationship with the density field traced by 21\,cm\,IM. The association between FRBs and GWs has been explored extensively, especially in the context of compact binary mergers \cite{Mingarelli:2015bpo,Wang:2016dgs,Margalit:2019hke,Most:2023unc}; see Ref.~\cite{Zhang:2024rqu} for a review. FRBs and SGL systems can be linked through strongly lensed FRBs, where the time delay provides an independent geometric distance complementing the DM-derived distance \cite{Li:2017mek,Liu:2019jka,Wucknitz:2020spz,Zhao:2021jeb,Gao:2022ifq}. Beyond FRB-centered connections, GW and SGL are related through strongly lensed GW events \cite{Wang:1996as,Nakamura:1997sw,Liao:2017ioi,Diego:2021fyd,Wei:2017emo,Huang:2023prq,Sereno:2011ty}. The cross-correlation between 21\,cm\,IM and GWs has also been explored \cite{Zazzera:2025ord,Scelfo:2022lsx,Scelfo:2021fqe}, along with that between 21\,cm\,IM and gravitational lensing (mainly weak lensing, not strong lensing as studied here) \cite{Sangka:2024vfg,Euclid:2026ulk}. These physical connections are of interest for future work and are not pursued further here.

\section{Results and discussion}\label{sec3}
\begin{table}
    \centering
    \caption{Cosmological constraints on the $\Lambda$CDM, $w$CDM, $w_0w_a$CDM, and cosmography models from multi-messenger probes: FRB, GW, SGL, 21\,cm\,IM, and their combinations. Here, $H_0$ is in units of km/s/Mpc. All uncertainties denote 68.3\% marginalized confidence intervals (1$\sigma$). The notation ``N/A'' indicates cases where the data provide insufficient constraining power to yield a meaningful limit on the parameter.}
    \label{tab:errors}
    \begin{tabular}{lcccccc}
        \hline
        Probe & $\sigma(H_0)$ & $\sigma(\Omega_{\rm b})$  & $\sigma(\Omega_{\rm m})$ & $\sigma(w/w_0)$ & $\sigma(w_a)$ \\
        \hline
        \multicolumn{6}{c}{\textbf{$\Lambda$CDM model}} \\
        \hline
        FRB only                 & N/A               &  N/A                 & 0.012           & --        & -- \\
        GW only                  & 0.53              & N/A                   & 0.014              & --        & -- \\
        SGL only                 & 0.76                  & N/A                   & N/A               & --        & -- \\
        21\,cm\,IM only        & N/A                 & 0.011            & 0.0052            & --         & -- \\
        FRB+GW                & 0.38                 & 0.00057            & 0.0089        & --       & -- \\
        FRB+SGL               & 0.47                 & 0.00058           & 0.012            & --    & -- \\
        FRB+21\,cm\,IM      & 0.59               & 0.00035           & 0.0047            & --   & -- \\
        \hline
        \multicolumn{6}{c}{\textbf{$w$CDM model}} \\
        \hline
        FRB only                & N/A                   &   N/A                  & 0.014            & 0.16                & -- \\
        GW only                 & 1.2                    & N/A                     & 0.020                & 0.15                & -- \\
        SGL only                & 2.2                    & N/A                      &  0.20                  & 1.2                   & -- \\
        21\,cm\,IM only       & N/A                    & 0.011                & 0.0053                & 0.034             & -- \\
        FRB+GW                & 0.94                    & 0.0015                & 0.011            & 0.096            & -- \\
        FRB+SGL               & 1.1                    & 0.0017               & 0.013            & 0.11             & -- \\
        FRB+21\,cm\,IM     & 1.0                     & 0.00099                 & 0.0046            & 0.033 & -- \\
        \hline
        \multicolumn{6}{c}{\textbf{$w_0w_a$CDM model}} \\
        \hline
        FRB only            & N/A                       &   N/A                  & 0.074                & 0.46             & 2.1 \\
        GW only             & 1.8                       & N/A                 & 0.059                   & 0.28               & 1.6 \\
        SGL only            & 2.8                       & N/A                 & 0.13                    & 0.84             & N/A \\
        21\,cm\,IM only    & N/A                    & 0.014                  & 0.029              & 0.23             & 0.69 \\
        FRB+GW           & 1.6                      & 0.0024                 & 0.059                & 0.19             & 1.2 \\
        FRB+SGL          & 2.3                      & 0.0036                 & 0.067            & 0.29                 & 1.5 \\
        FRB+21\,cm\,IM     & 1.7                & 0.0026                & 0.024                & 0.18                & 0.56 \\

        \hline
        \multicolumn{6}{c}{\textbf{Cosmography (model-independent)}} \\
        \hline
         & $\sigma(H_0)$ & $\sigma(\Omega_{\rm b})$ & $\sigma(q_0)$ & $\sigma(j_0)$ & \\
        \hline
        FRB only                & N/A                       &  N/A                  & 0.14              & 0.78                    \\
        GW only                 & 0.93                      & N/A                   & 0.11              & 0.66                  \\
        SGL only                & 1.1                      & N/A                   & 0.47              & N/A                   \\
        FRB+GW               & 0.83                     & 0.0013               & 0.095              & 0.59                   \\
        FRB+SGL             & 0.93                     & 0.0014                & 0.11              & 0.62                   \\
                \hline
    \end{tabular}
\end{table}

\begin{figure*}
\includegraphics[scale=0.45]{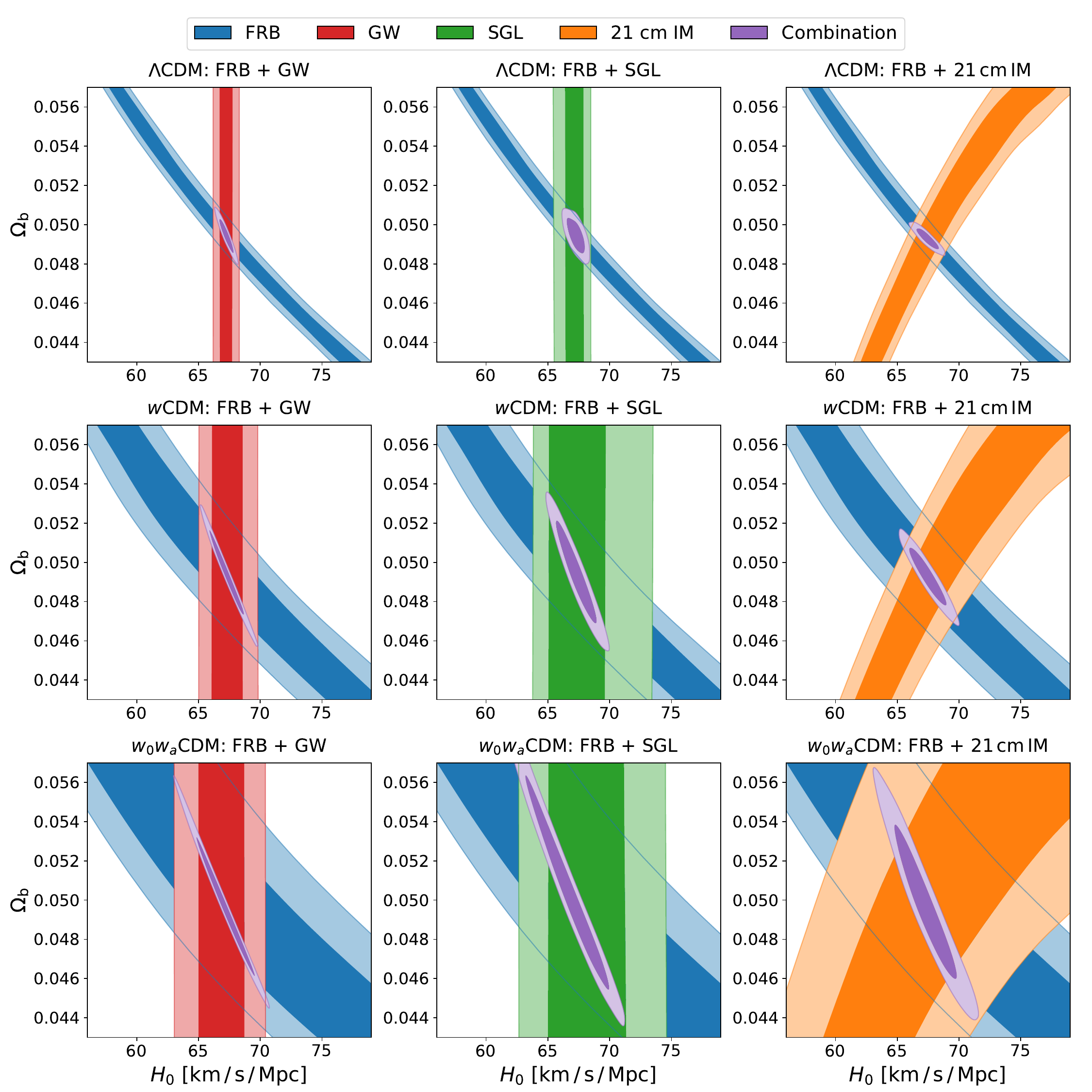}
\centering
\caption{Two-dimensional posterior distribution contours in the $H_0-\Omega_{\rm b}$ plane for the $\Lambda$CDM, $w$CDM, and $w_0w_a$CDM models: constraints from the simulated FRB, GW, SGL, 21\,cm\,IM, FRB+GW, FRB+SGL, and FRB+21\,cm\,IM data.}
\label{Contours-3x3}
\end{figure*}

\begin{figure*}
\includegraphics[scale=0.72]{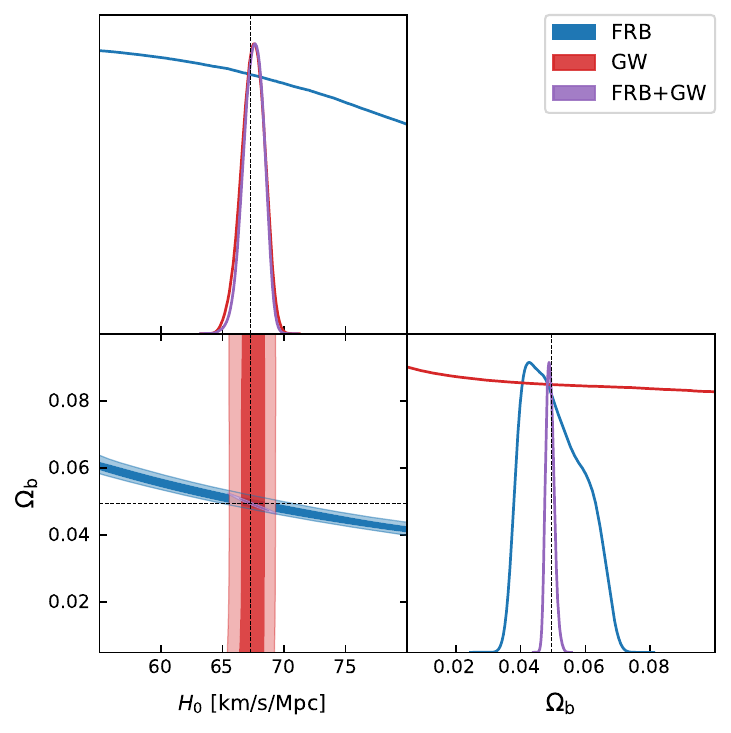}
\includegraphics[scale=0.72]{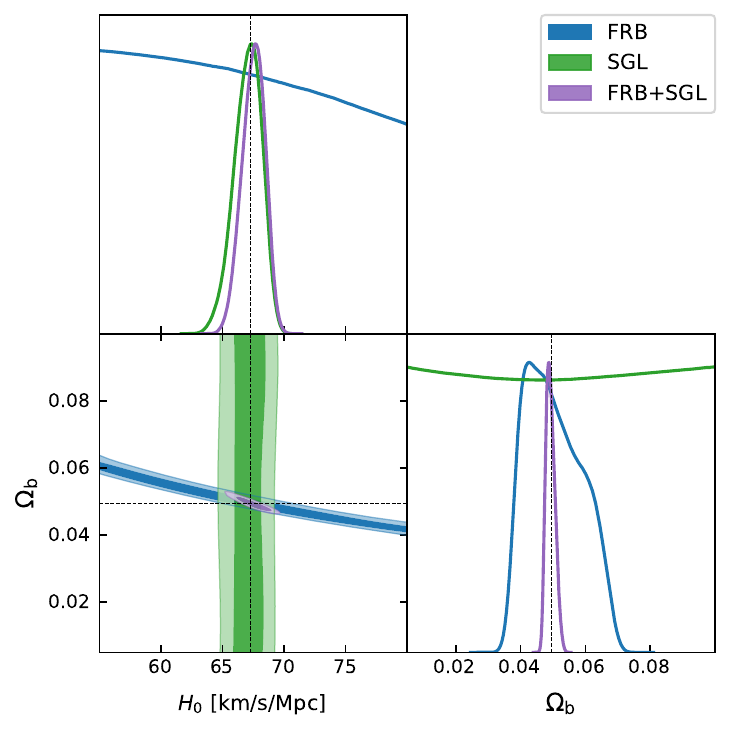}
\centering
\caption{Constraints on cosmographic parameters derived from FRB, GW, and FRB+GW in the left panel, and from FRB, SGL, and the joint FRB+SGL probe in the right panel.}
\label{cosmography}
\end{figure*}

In this section, we report the constraint results from four emerging probes and their synergies. We consider three cosmological models: (i) the $\Lambda$CDM model with a constant dark energy equation of state (EoS) $w(z) = -1$; (ii) the $w$CDM model---the simplest dynamical dark energy model with a constant but free EoS $w(z) = w$; and (iii) the $w_0w_a$CDM model---a parameterized dynamical dark energy model with a time-varying EoS given by $w(z) = w_0 + w_a z/(1+z)$ \cite{Chevallier:2000qy,Linder:2002et}. We adopt the Markov chain Monte Carlo method to infer the probability distributions of cosmological parameters by maximizing the likelihood $\mathcal{L} \propto \exp (-\chi^2/2)$, with $\chi^2$ being the chi-squared statistic, which quantifies the discrepancy between the theoretical predictions and the observational data. We measure the convergence of chains by checking that all parameters have $R-1<0.01$, where $R$ is the potential scale reduction factor of the Gelman-Rubin diagnostics. The $1\sigma$ (68.3\% confidence level) and $2\sigma$ (95.4\% confidence level) posterior distribution contours for $H_0$ and $\Omega_{\rm b}$ are shown in Figs.~\ref{Contours-3x3} and \ref{cosmography}, and the $1\sigma$ errors for the marginalized parameter constraints are summarized in Table~\ref{tab:errors}.

In the $\Lambda$CDM framework, combining FRBs with complementary probes yields significant improvements in constraining both $H_0$ and $\Omega_{\rm b}$, successfully breaking the parameter degeneracy inherent in FRB dispersion measures. The FRB+GW synergy delivers the most powerful constraints on $H_0$, achieving $\sigma(H_0)=0.38$~km/s/Mpc and $\sigma(\Omega_{\rm b})=0.00057$, corresponding to precisions of $0.56\%$ and $1.2\%$, respectively. The $H_0$ constraint tightens by $28\%$ compared to the GW-only result $\sigma(H_0)=0.53$~km/s/Mpc. FRB+SGL provides $\sigma(H_0)=0.47$~km/s/Mpc and $\sigma(\Omega_{\rm b})=0.00058$, corresponding to relative errors of $0.70\%$ and $1.2\%$, respectively. The $H_0$ constraint tightens by $38\%$ compared to the SGL-only result $\sigma(H_0)=0.76$~km/s/Mpc. Meanwhile, the FRB+21\,cm\,IM combination, which provides the tightest $\Omega_{\rm b}$ constraint, yields $\sigma(H_0)=0.59$~km/s/Mpc and $\sigma(\Omega_{\rm b})=0.00035$ ($0.87\%$ and $0.71\%$ relative errors). All three combinations can constrain $H_0$ to sub-percent precision and $\Omega_{\rm b}$ to around 1\% precision.

In the $w$CDM model, introducing dark energy equation of state variability degrades constraints as the additional parameter covaries with cosmic expansion history. The FRB+GW combination still provides the strongest constraints on $H_0$, achieving $\sigma(H_0)=0.94$~km/s/Mpc and $\sigma(\Omega_{\rm b})=0.0015$, corresponding to precisions of $1.4\%$ and $3.0\%$, respectively. The $H_0$ constraint tightens by $22\%$ compared to the GW-only measurement $\sigma(H_0)=1.2$ km/s/Mpc. The FRB+SGL synergy yields $\sigma(H_0)=1.1$~km/s/Mpc and $\sigma(\Omega_{\rm b})=0.0017$ ($1.6\%$ and $3.4\%$ relative errors), with the $H_0$ constraint improving by $50\%$ over the SGL-only measurement $\sigma(H_0)=2.2$~km/s/Mpc. In addition, FRB+21\,cm\,IM, which provides the tightest $\Omega_{\rm b}$ constraint, produces $\sigma(H_0)=1.0$~km/s/Mpc and $\sigma(\Omega_{\rm b})=0.00099$ ($1.5\%$ and $2.0\%$ relative errors). In this extended model, all three combinations constrain $H_0$ to $\sim1.5\%$ precision, and $\Omega_{\rm b}$ to between $2\%$ and $3.5\%$ precision.

In the $w_0w_a$CDM model, allowing for a time-varying dark energy EoS parameterized by $w_0$ and $w_a$ further expands parameter degeneracies. Nevertheless, FRB synergies continue to yield competitive constraints. FRB+GW delivers $\sigma(H_0)=1.6$~km/s/Mpc and $\sigma(\Omega_{\rm b})=0.0024$, corresponding to precisions of $2.4\%$ and $4.8\%$, respectively. The $H_0$ constraint tightens by $11\%$ compared to the GW-only result $\sigma(H_0)=1.8$~km/s/Mpc. FRB+SGL provides $\sigma(H_0)=2.3$~km/s/Mpc and $\sigma(\Omega_{\rm b})=0.0036$ ($3.4\%$ and $7.3\%$ relative errors), with the $H_0$ constraint improving by $18\%$ over the SGL-only result $\sigma(H_0)=2.8$~km/s/Mpc. FRB+21\,cm\,IM achieves $\sigma(H_0)=1.7$~km/s/Mpc and $\sigma(\Omega_{\rm b})=0.0026$ ($2.5\%$ and $5.2\%$ relative errors). Even in this extended parameter space, all three combinations constrain $H_0$ to $\sim2.5\%$--$3.5\%$ and $\Omega_{\rm b}$ to $\sim5\%$--$7\%$ precision.

The constraining power on parameters beyond $H_0$ and $\Omega_{\mathrm{b}}$ varies significantly across different synergies. Among them, the FRB+21\,cm\,IM synergy consistently yields the tightest constraints across all models. It provides the strongest limits on $\Omega_{\mathrm{m}}$ across all cosmological models, on the constant dark energy EoS parameter $w$ in the $w$CDM model, and on the EoS parameters $w_0$ and $w_a$ in the $w_0 w_a$CDM model, significantly outperforming both FRB+GW and FRB+SGL. The FRB+GW combination delivers the second-best constraints, with FRB+SGL offering the weakest constraints among the three. It is worth noting that SGL admits two distinct cosmological applications: the time delay method, which is employed in this work, and the velocity dispersion method, which holds considerable promise for constraining $\Omega_{\mathrm{m}}$ and the dark energy EoS \cite{Qi:2022kfg}. Future FRB+SGL analyses could exploit both approaches simultaneously, thereby enhancing the constraining power of this synergy on cosmological parameters. Overall, for observational strategies aiming to simultaneously sharpen the measurements of $H_0$ and $\Omega_{\mathrm{b}}$ while also achieving substantial gains in constraining other cosmological parameters, FRB+21\,cm\,IM stands out as a particularly promising approach.

It should be stressed that the constraining power of 21\,cm\,IM is highly dependent on the performance of foreground subtraction. In this work, we model the residual foreground contribution through the term $C^{\mathrm{F}}$, parameterized by $\varepsilon_{\rm FG}$, which quantifies the cleaning efficiency as the ratio of residual to original foreground amplitude \cite{Bull:2014rha}. Smaller values of $\varepsilon_{\rm FG}$ thus correspond to more effective foreground removal. For the baseline analysis, we adopt $\varepsilon_{\rm FG}=10^{-5}$, representing the nominal expectation. An optimistic foreground suppression scenario with $\varepsilon_{\rm FG}=10^{-6}$ can lead to notably improved constraints. Specifically, in the $\Lambda$CDM model, the FRB+21\,cm\,IM combination yields $\sigma(H_0)=0.53$~km/s/Mpc, $\sigma(\Omega_{\rm b})=0.00032$ and $\sigma(\Omega_{\rm m})=0.0038$; in the $w$CDM model, it yields $\sigma(w)=0.027$; and in the $w_0w_a$CDM model, it yields $\sigma(w_0)=0.15$ and $\sigma(w_a)=0.47$. Overall, these results highlight that the future potential of the FRB+21\,cm\,IM synergy is primarily limited by foreground removal capabilities. As discussed in Sec.~\ref{sec24}, our foreground model assumes uniform suppression and neglects potential signal loss, so these constraints should be viewed as optimistic benchmarks.

All the models discussed above rely on specific dark energy and dark matter EoS parameterizations. To circumvent the limitations of model-dependent assumptions, the cosmographic approach provides a valuable alternative. The cosmographic parameters are defined as derivatives of the scale factor with respect to cosmic time:
\begin{align}
q \equiv -\frac{1}{a H^2} \frac{{\rm d}^2 a}{{\rm d} t^2},\quad
j \equiv \frac{1}{a H^3} \frac{{\rm d}^3 a}{{\rm d} t^3},\quad
s \equiv \frac{1}{a H^4} \frac{{\rm d}^4 a}{{\rm d} t^4},
\end{align}
The traditional Taylor expansion of the Hubble parameter around $z=0$,
\begin{align}
H(z) & \simeq H_0 + H_0(1+q_0)z + \frac{1}{2}H_0(j_0 - q_0^2)z^2 \nonumber\\
 &+ \frac{1}{6}H_0\left(3q_0^3 + 3q_0^2 - 4j_0q_0 - 3j_0 - s_0\right)z^3 ,
\end{align}
diverges for $z \gtrsim 1$, making it unsuitable for high-redshift data. To utilize high-redshift data, one can consider the rational Pad\'e approximation \cite{Gruber:2013wua,Wei:2013jya,Capozziello:2020ctn}. The error of an $(n,m)$-order Pad\'e approximant scales as $\mathcal{O}(z^{m+n+1})$, the same as that of the $(m+n)$-order Taylor expansion. It is found that the polynomials $P(2,1)$, $P(2,2)$ and $P(3,2)$ best approximate the cosmological functions. To avoid overfitting, we adopt the $P(2,1)$ Pad\'e approximant, as higher-order expansions would introduce additional parameters that significantly weaken the constraining power on $H_0$ and $\Omega_{\rm b}$. The $(2,1)$ Pad\'e approximant takes the form \cite{DAgostino:2023cgx}
\begin{align}\label{eq:Pade21}
H(z) \simeq \frac{2(1+z)^2 \left[3(1-q_0) + \left(1 + j_0 - q_0 - 3q_0^2\right)z \right]^2}{A_0 + A_1 z + A_2 z^2},
\end{align}
with coefficients given by
\begin{align}
\begin{aligned}
A_0 &= 18(q_0-1)^2,\\
A_1 &= 6(q_0-1)\left(-5 - 2j_0 + 8q_0 + 3q_0^2\right),\\
A_2 &= 14 + 7j_0 + 2j_0^2 - 10(4+j_0)q_0 + (17-9j_0)q_0^2 \\
&\ \ \ + 18q_0^3 + 9q_0^4.
\end{aligned}
\end{align}
The approximant maintains the same accuracy as the third-order Taylor expansion while providing a significantly larger radius of convergence, making it suitable for analyses that include high-redshift observations.

In this framework, FRB+GW achieves $\sigma(H_0)=0.83$~km/s/Mpc ($1.2\%$) and $\sigma(\Omega_{\rm b})=0.0013$ ($2.6\%$), while FRB+SGL yields $\sigma(H_0)=0.93$ km/s/Mpc ($1.4\%$) and $\sigma(\Omega_{\rm b})=0.0014$ ($2.8\%$). Importantly, as shown in Fig.~\ref{cosmography}, the derived parameters are centered on the fiducial values, as expected for forecasts in which the mock observables are generated at the fiducial cosmology without statistical fluctuations. The constraints on $H_0$ and $\Omega_{\mathrm{b}}$ are comparable to those obtained in the $w$CDM model: both FRB+GW and FRB+SGL constrain $H_0$ and $\Omega_{\rm b}$ to better than (1.5\%, 3\%) precision. In cosmography, the FRB+21\,cm\,IM synergy cannot be used, because the expansion cannot be reliably extrapolated to the early-Universe regime required for computing the sound horizon at the baryon drag epoch, as shown in Eq.~\ref{sound}. In this scenario, we can only treat the sound horizon as a free parameter. However, in that case, 21\,cm\,IM cannot provide meaningful constraints on \(H_0\) and \(\Omega_{\rm b}\), and there is no degeneracy between these two parameters. Consequently, combining 21\,cm\,IM with FRBs does not significantly improve the constraints on these two parameters. If introducing a prior on the sound horizon from the early universe, we can constrain \(H_0\) and thus break the degeneracy between \(H_0\) and \(\Omega_{\rm b}\) inherent in FRBs. However, this approach contradicts our original goal of addressing the Hubble tension or missing baryon problem without relying on early-universe observations. To achieve this goal, any connection with CMB or BBN observations should be avoided.

The above results are obtained using the Pad\'e approximant. We note that a variety of alternative model-independent approaches have been developed. For instance, the Gaussian process method can provide a non-parametric framework for reconstructing the distance-redshift relation and its derivative directly from data \cite{Seikel:2012uu}; the artificial neural network methods can learn the underlying expansion history in a model-independent manner \cite{Wang:2019vxv}, and genetic algorithms can identify optimal functional forms for cosmological quantities without assuming a specific parameterization \cite{Nesseris:2012tt}. These methods have been widely applied to cosmological studies. In this work, we employ the Pad\'e approximant as a first step toward model-independent constraints. The application of these non-parametric techniques to the multi-messenger synergies explored here would be a natural and valuable extension for future studies.

Our forecasts for FRB observations are based on the upcoming SKA. The 10-year operation of SKA is expected to yield $\mathcal{O}(10^5-10^6)$ FRBs~\cite{Zhang:2023gye}. In this work, we assume that 10\% of the detected FRBs can be localized to their host galaxies, and adopt a nominal scenario with $N_{\rm FRB}=10^4$ FRBs with redshifts available for parameter inference. Given that future surveys may detect a larger number of FRBs (for instance, an optimistic scenario with $N_{\rm FRB}=10^5$), it is natural to explore how such an increase would enhance the constraining power. To address this, we evaluate the performance of the three synergies under the optimistic FRB catalog size. In the $\Lambda$CDM model, with $N_{\rm FRB}=10^5$, FRB+GW yields constraints of $\sigma(H_0)=0.24$~km/s/Mpc and $\sigma(\Omega_{\rm b})=0.00027$, FRB+SGL provides $\sigma(H_0)=0.46$~km/s/Mpc and $\sigma(\Omega_{\rm b})=0.00037$, and FRB+21\,cm\,IM gives $\sigma(H_0)=0.52$~km/s/Mpc and $\sigma(\Omega_{\rm b})=0.00034$. As can be seen, all three combinations constrain $H_0$ and $\Omega_{\rm b}$ to better than $1\%$ precision. In the cosmographic approach, the FRB+GW yields $\sigma(H_0)=0.54$~km/s/Mpc and $\sigma(\Omega_{\rm b})=0.00080$, and FRB+SGL yields $\sigma(H_0)=0.64$~km/s/Mpc and $\sigma(\Omega_{\rm b})=0.00083$. Both combinations constrain $H_0$ and $\Omega_{\rm b}$ to better than $1\%$ and 2\% precision, respectively. These improvements demonstrate that a larger FRB sample would significantly enhance the constraining power, albeit with the stringent requirement of coordinated optical and radio observations to reliably determine the host galaxies of FRBs.

Our forecasts assume relatively idealized conditions. In reality, several systematic effects may broaden the constraints or introduce biases. The host-galaxy DM contribution constitutes a major source of uncertainty. In our simulation, we adopted $\sigma_{\rm host}=60~\text{pc}\,\text{cm}^{-3}$, but if the typical scatter proves to be as large as $\sim100~\text{pc}\,\text{cm}^{-3}$ \cite{Zhu:2020prt}, it would translate into an additional distance uncertainty at low redshifts and could broaden the $H_0$ constraints relative to our forecasts. Localization precision is also critical: our analysis assumes that 10\% of detected FRBs can be precisely localized to their host galaxies, but if a significant fraction only have arcminute-level rather than arcsecond-level localizations, the number of events with reliable redshift identifications would be substantially reduced, degrading the overall constraining power. For the Milky Way, we have included a statistical uncertainty of $\sigma_{\rm MW}=10~\text{pc}\,\text{cm}^{-3}$ in our error budget. However, the subtraction of the Galactic DM contribution introduces an additional calibration floor, as systematic differences between popular Galactic electron density models are known to be non-negligible \cite{Cordes:2002wz,Yao:2017kcp}. This model-to-model variation is not explicitly modeled in our forecasts and would need to be assessed in real data analysis, for instance by comparing or marginalizing over multiple Galactic electron models. These considerations indicate that our Fisher-matrix forecasts should be viewed as optimistic benchmarks representing the achievable precision under relatively idealized conditions, and that a full quantitative treatment of all systematics is needed when applying this framework to actual observations. We have also tested the sensitivity of our results to the choice of FRB redshift distribution. Our fiducial power-law model with $\gamma=-1.1$ is motivated by the CHIME/FRB catalog, which has shown that the FRB population does not strictly track the cosmic star formation history. To assess the impact of this assumption, we repeated our analysis using an SFH-tracking distribution, which assumes that FRBs trace star formation without a time delay. We find that the $1\sigma$ constraints on $H_0$ and $\Omega_{\rm b}$ change by less than 5\% between the two models. Switching to this alternative model does not significantly change the constraint precisions, and the qualitative conclusion that all three synergies break the parameter degeneracy remains robust.

In this work, we generate mock observations for four emerging probes based on specific experimental setups (each probe being associated with a particular telescope). However, each probe actually has multiple planned observational facilities, and the observational constraints for a single probe can be enhanced through multi-experiment joint observations. Improving the measurement capabilities of individual probes is an important and worthwhile direction for further breaking parameter degeneracies. For FRB observations, future facilities such as FASTA \cite{Xue:2023hix}, DSA-2000 \cite{Hallinan:2019qyo}, CHORD \cite{Vanderlinde:2019tjt} and BURSTT \cite{Lin:2022wgp} will all be powerful tools for FRB detection and localization. For GW observations, ground-based detectors like CE \cite{Reitze:2019iox}, the space-based observatory LISA \cite{LISA:2017pwj,Robson:2018ifk,LISACosmologyWorkingGroup:2022jok}, Taiji \cite{Ruan:2018tsw}, TianQin \cite{TianQin:2015yph,Liu:2020eko,Wang:2019ryf,TianQin:2020hid}, DECIGO \cite{Kawamura:2011zz} and pulsar timing array experiments (such as SKA) will play crucial roles. Furthermore, the multi-band joint GW observations could improve the constraints \cite{Jin:2023zhi}. For a brief review of GW standard siren cosmology, we refer the reader to Ref.~\cite{Jin:2025dvf}. For SGL observations, CSST is also expected to discover a large number of SGL systems \cite{Cao:2023bnl}. For 21\,cm\,IM observations, facilities like SKA \cite{Santos:2015gra,SKA:2018ckk,Wu:2021vfz}, CHIME \cite{CHIME:2022dwe}, and FASTA \cite{Pan:2024xoj} will also be valuable, and joint survey strategies are worth pursuing \cite{Wu:2022jkf,Pan:2024xoj}. We also note that a single facility can carry out multiple types of observations. For instance, SKA is designed to conduct both FRB and 21\,cm\,IM surveys. In this work, we choose one representative facility for each probe to keep the forecast transparent and to focus on the cosmological synergy itself, rather than on the instrumental implementation. We do not consider the case of a single facility performing multiple observations simultaneously. This is an interesting direction, and we leave the cosmological potential of such combined observations with a single experimental design for future work.

In addition to the emerging probes outlined above, conventional probes also have the potential to break the $H_0$--$\Omega_{\rm b}$ degeneracy when jointly analyzed with FRBs. The 21\,cm\,IM technique enables measurements of the BAO signal, and optical galaxy surveys such as DESI \cite{DESI:2025zgx} and Euclid \cite{Amendola:2016saw} offer alternative BAO measurement approaches. GW standard sirens measure luminosity distances, and Type Ia supernovae (SNe Ia) could also offer luminosity distance constraints. Furthermore, gamma-ray bursts (GRBs) have been employed to measure luminosity distance at redshifts far beyond the reach of SNe Ia \cite{Cao:2022ugh,Cao:2021irf,Cao:2022wlg,Dainotti:2023bwq,Dainotti:2022ked,Dainotti:2022wli,Bargiacchi:2023rfd}. The association of GRB\,170817A with GW\,170817 \cite{Troja:2017nqp} has highlighted their multi-messenger potential, and cosmographic techniques have been applied to GRB data \cite{Bargiacchi:2023rfd}. Nevertheless, SNe Ia and GRB measurements require empirical calibration: SNe Ia cosmology needs accurate calibration of the absolute magnitude, while GRB distances rely on luminosity correlations anchored by low-redshift sources. These calibration requirements introduce additional systematic uncertainties in the cosmological analysis. In contrast, the emerging probes investigated in this work offer several advantages. First, 21\,cm\,IM surveys can map cosmic volumes at redshifts beyond the detection range of existing galaxy surveys, yielding complementary BAO measurements whose systematic errors differ from those of optical surveys. Second, GW standard sirens yield absolute distance measurements that avoid the distance-ladder calibration required by SNe Ia and GRB observations. Third, SGL time-delay observations provide geometric distance estimates that are also independent of the distance ladder. Joint combinations of these emerging probes with FRB data not only break the parameter degeneracy but also deliver independent constraints with distinct systematics, making them complementary to, rather than redundant with, future SNe Ia and galaxy surveys. Overall, the multi-probe synergies explored in this study provide a promising pathway to simultaneous percent-level precision measurements of $H_0$ and $\Omega_{\rm b}$, thereby helping to address the Hubble tension and the missing baryon problem.

\section{Conclusions}\label{sec4}
We propose that FRBs, in synergy with other emerging cosmological probes, can serve as a powerful tool for simultaneously measuring two key parameters: the Hubble constant $H_0$ and the cosmic baryon density $\Omega_{\rm b}$. Precisely determining these two quantities is essential for addressing the Hubble tension and the missing baryon problem. By leveraging synergies with GW standard sirens, SGL time delays, and 21\,cm\,IM surveys, we effectively break the degeneracies inherent in FRB dispersion measures, thereby yielding high-precision constraints on $H_0$ and $\Omega_{\rm b}$.

Quantitatively, within the $\Lambda$CDM framework, the three multi-probe synergies (FRB+GW, FRB+SGL, and FRB+21\,cm\,IM) each constrain the parameters ($H_0$, $\Omega_{\rm b}$) to percent-level precision better than (1\%, 1.5\%). As one allows for dynamical dark energy, the constraining power degrades gracefully with increasing model complexity. Notably, even in the cosmographic framework that makes no assumptions about the nature of dark energy and dark matter, the FRB+GW and FRB+SGL combinations maintain high robustness, delivering constraints on $H_0$ and $\Omega_{\rm b}$ at precisions better than (1.5\%, 3\%). These benchmarks are based on nominal observational expectations and would improve substantially under optimistic scenarios. Furthermore, among the observational strategies examined, FRB+21\,cm\,IM emerges as a particularly promising approach for jointly sharpening constraints on $H_0$ and $\Omega_{\rm b}$ while also offering significant gains in the determination of other cosmological parameters. This potential, however, depends on effective foreground removal.

Our analysis is based on specific experimental designs and observational programs. In practice, a single probe can be implemented through multiple experiments, and its measurement capability can be strengthened by combining data from different experiments or observational bands. Such combinations within a single probe, together with the cross-probe synergies studied here, will significantly reduce statistical uncertainties. This therefore provides an approach to simultaneously constrain $H_0$ and $\Omega_{\rm b}$, and may open a promising avenue toward addressing both the Hubble tension and the missing baryon problem.

\begin{acknowledgments}
We thank the anonymous referee for constructive comments on this work. This work was supported by the National SKA Program of China (Grants Nos. 2022SKA0110200 and 2022SKA0110203), the National Natural Science Foundation of China (Grants Nos. 12533001, 12575049, and 12473001), the 111 Project (Grant No. B16009), and the Natural Science Foundation of Ningxia Province, China (Grant No.~2026AAC030079).
\end{acknowledgments}

\section*{Data availability}
There are no publicly available research data or software supporting this manuscript. Requests for further information or data should be sent to the authors.

\bibliography{refs}

\begin{thebibliography}{134}%
\makeatletter
\providecommand \@ifxundefined [1]{%
 \@ifx{#1\undefined}
}%
\providecommand \@ifnum [1]{%
 \ifnum #1\expandafter \@firstoftwo
 \else \expandafter \@secondoftwo
 \fi
}%
\providecommand \@ifx [1]{%
 \ifx #1\expandafter \@firstoftwo
 \else \expandafter \@secondoftwo
 \fi
}%
\providecommand \natexlab [1]{#1}%
\providecommand \enquote  [1]{``#1''}%
\providecommand \bibnamefont  [1]{#1}%
\providecommand \bibfnamefont [1]{#1}%
\providecommand \citenamefont [1]{#1}%
\providecommand \href@noop [0]{\@secondoftwo}%
\providecommand \href [0]{\begingroup \@sanitize@url \@href}%
\providecommand \@href[1]{\@@startlink{#1}\@@href}%
\providecommand \@@href[1]{\endgroup#1\@@endlink}%
\providecommand \@sanitize@url [0]{\catcode `\\12\catcode `\$12\catcode
  `\&12\catcode `\#12\catcode `\^12\catcode `\_12\catcode `\%12\relax}%
\providecommand \@@startlink[1]{}%
\providecommand \@@endlink[0]{}%
\providecommand \url  [0]{\begingroup\@sanitize@url \@url }%
\providecommand \@url [1]{\endgroup\@href {#1}{\urlprefix }}%
\providecommand \urlprefix  [0]{URL }%
\providecommand \Eprint [0]{\href }%
\providecommand \doibase [0]{http://dx.doi.org/}%
\providecommand \selectlanguage [0]{\@gobble}%
\providecommand \bibinfo  [0]{\@secondoftwo}%
\providecommand \bibfield  [0]{\@secondoftwo}%
\providecommand \translation [1]{[#1]}%
\providecommand \BibitemOpen [0]{}%
\providecommand \bibitemStop [0]{}%
\providecommand \bibitemNoStop [0]{.\EOS\space}%
\providecommand \EOS [0]{\spacefactor3000\relax}%
\providecommand \BibitemShut  [1]{\csname bibitem#1\endcsname}%
\let\auto@bib@innerbib\@empty
\bibitem [{\citenamefont {Verde}\ \emph {et~al.}(2019)\citenamefont {Verde},
  \citenamefont {Treu},\ and\ \citenamefont {Riess}}]{Verde:2019ivm}%
  \BibitemOpen
  \bibfield  {author} {\bibinfo {author} {\bibfnamefont {L.}~\bibnamefont
  {Verde}}, \bibinfo {author} {\bibfnamefont {T.}~\bibnamefont {Treu}}, \ and\
  \bibinfo {author} {\bibfnamefont {A.~G.}\ \bibnamefont {Riess}},\ }\bibfield
  {title} {\enquote {\bibinfo {title} {{Tensions between the Early and the Late
  Universe}},}\ }\href {\doibase 10.1038/s41550-019-0902-0} {\bibfield
  {journal} {\bibinfo  {journal} {Nature Astron.}\ }\textbf {\bibinfo {volume}
  {3}},\ \bibinfo {pages} {891} (\bibinfo {year} {2019})},\ \Eprint
  {http://arxiv.org/abs/1907.10625} {arXiv:1907.10625 [astro-ph.CO]}
  \BibitemShut {NoStop}%
\bibitem [{\citenamefont {Riess}\ \emph {et~al.}(2022)\citenamefont {Riess}
  \emph {et~al.}}]{Riess:2021jrx}%
  \BibitemOpen
  \bibfield  {author} {\bibinfo {author} {\bibfnamefont {Adam~G.}\ \bibnamefont
  {Riess}} \emph {et~al.},\ }\bibfield  {title} {\enquote {\bibinfo {title} {{A
  Comprehensive Measurement of the Local Value of the Hubble Constant with 1 km
  s$^{-1}$ Mpc$^{-1}$ Uncertainty from the Hubble Space Telescope and the SH0ES
  Team}},}\ }\href {\doibase 10.3847/2041-8213/ac5c5b} {\bibfield  {journal}
  {\bibinfo  {journal} {Astrophys. J. Lett.}\ }\textbf {\bibinfo {volume}
  {934}},\ \bibinfo {pages} {L7} (\bibinfo {year} {2022})},\ \Eprint
  {http://arxiv.org/abs/2112.04510} {arXiv:2112.04510 [astro-ph.CO]}
  \BibitemShut {NoStop}%
\bibitem [{\citenamefont {Fukugita}\ \emph {et~al.}(1998)\citenamefont
  {Fukugita}, \citenamefont {Hogan},\ and\ \citenamefont
  {Peebles}}]{Fukugita:1997bi}%
  \BibitemOpen
  \bibfield  {author} {\bibinfo {author} {\bibfnamefont {M.}~\bibnamefont
  {Fukugita}}, \bibinfo {author} {\bibfnamefont {C.~J.}\ \bibnamefont {Hogan}},
  \ and\ \bibinfo {author} {\bibfnamefont {P.~J.~E.}\ \bibnamefont {Peebles}},\
  }\bibfield  {title} {\enquote {\bibinfo {title} {{The Cosmic baryon
  budget}},}\ }\href {\doibase 10.1086/306025} {\bibfield  {journal} {\bibinfo
  {journal} {Astrophys. J.}\ }\textbf {\bibinfo {volume} {503}},\ \bibinfo
  {pages} {518} (\bibinfo {year} {1998})},\ \Eprint
  {http://arxiv.org/abs/astro-ph/9712020} {arXiv:astro-ph/9712020} \BibitemShut
  {NoStop}%
\bibitem [{\citenamefont {Cen}\ and\ \citenamefont
  {Ostriker}(1999)}]{Cen:1998hc}%
  \BibitemOpen
  \bibfield  {author} {\bibinfo {author} {\bibfnamefont {Renyue}\ \bibnamefont
  {Cen}}\ and\ \bibinfo {author} {\bibfnamefont {Jeremiah~P.}\ \bibnamefont
  {Ostriker}},\ }\bibfield  {title} {\enquote {\bibinfo {title} {{Where are the
  baryons?}}}\ }\href {\doibase 10.1086/306949} {\bibfield  {journal} {\bibinfo
   {journal} {Astrophys. J.}\ }\textbf {\bibinfo {volume} {514}},\ \bibinfo
  {pages} {1--6} (\bibinfo {year} {1999})},\ \Eprint
  {http://arxiv.org/abs/astro-ph/9806281} {arXiv:astro-ph/9806281} \BibitemShut
  {NoStop}%
\bibitem [{\citenamefont {Shull}\ \emph {et~al.}(2012)\citenamefont {Shull},
  \citenamefont {Smith},\ and\ \citenamefont {Danforth}}]{Shull:2011aa}%
  \BibitemOpen
  \bibfield  {author} {\bibinfo {author} {\bibfnamefont {J.~Michael}\
  \bibnamefont {Shull}}, \bibinfo {author} {\bibfnamefont {Britton~D.}\
  \bibnamefont {Smith}}, \ and\ \bibinfo {author} {\bibfnamefont {Charles~W.}\
  \bibnamefont {Danforth}},\ }\bibfield  {title} {\enquote {\bibinfo {title}
  {{The Baryon Census in a Multiphase Intergalactic Medium: 30{\%} of the
  Baryons May Still Be Missing}},}\ }\href {\doibase
  10.1088/0004-637X/759/1/23} {\bibfield  {journal} {\bibinfo  {journal}
  {Astrophys. J.}\ }\textbf {\bibinfo {volume} {759}},\ \bibinfo {pages} {23}
  (\bibinfo {year} {2012})},\ \Eprint {http://arxiv.org/abs/1112.2706}
  {arXiv:1112.2706 [astro-ph.CO]} \BibitemShut {NoStop}%
\bibitem [{\citenamefont {Aghanim}\ \emph {et~al.}(2020)\citenamefont {Aghanim}
  \emph {et~al.}}]{Planck:2018vyg}%
  \BibitemOpen
  \bibfield  {author} {\bibinfo {author} {\bibfnamefont {N.}~\bibnamefont
  {Aghanim}} \emph {et~al.} (\bibinfo {collaboration} {Planck}),\ }\bibfield
  {title} {\enquote {\bibinfo {title} {{Planck 2018 results. VI. Cosmological
  parameters}},}\ }\href {\doibase 10.1051/0004-6361/201833910} {\bibfield
  {journal} {\bibinfo  {journal} {Astron. Astrophys.}\ }\textbf {\bibinfo
  {volume} {641}},\ \bibinfo {pages} {A6} (\bibinfo {year} {2020})},\ \bibinfo
  {note} {[Erratum: Astron.Astrophys. 652, C4 (2021)]},\ \Eprint
  {http://arxiv.org/abs/1807.06209} {arXiv:1807.06209 [astro-ph.CO]}
  \BibitemShut {NoStop}%
\bibitem [{\citenamefont {Hagstotz}\ \emph {et~al.}(2022)\citenamefont
  {Hagstotz}, \citenamefont {Reischke},\ and\ \citenamefont
  {Lilow}}]{Hagstotz:2021jzu}%
  \BibitemOpen
  \bibfield  {author} {\bibinfo {author} {\bibfnamefont {Steffen}\ \bibnamefont
  {Hagstotz}}, \bibinfo {author} {\bibfnamefont {Robert}\ \bibnamefont
  {Reischke}}, \ and\ \bibinfo {author} {\bibfnamefont {Robert}\ \bibnamefont
  {Lilow}},\ }\bibfield  {title} {\enquote {\bibinfo {title} {{A new
  measurement of the Hubble constant using fast radio bursts}},}\ }\href
  {\doibase 10.1093/mnras/stac077} {\bibfield  {journal} {\bibinfo  {journal}
  {Mon. Not. Roy. Astron. Soc.}\ }\textbf {\bibinfo {volume} {511}},\ \bibinfo
  {pages} {662--667} (\bibinfo {year} {2022})},\ \Eprint
  {http://arxiv.org/abs/2104.04538} {arXiv:2104.04538 [astro-ph.CO]}
  \BibitemShut {NoStop}%
\bibitem [{\citenamefont {Wu}\ \emph {et~al.}(2022)\citenamefont {Wu},
  \citenamefont {Zhang},\ and\ \citenamefont {Wang}}]{Wu:2021jyk}%
  \BibitemOpen
  \bibfield  {author} {\bibinfo {author} {\bibfnamefont {Qin}\ \bibnamefont
  {Wu}}, \bibinfo {author} {\bibfnamefont {Guo-Qiang}\ \bibnamefont {Zhang}}, \
  and\ \bibinfo {author} {\bibfnamefont {Fa-Yin}\ \bibnamefont {Wang}},\
  }\bibfield  {title} {\enquote {\bibinfo {title} {{An 8~per{\,}cent
  determination of the Hubble constant from localized fast radio bursts}},}\
  }\href {\doibase 10.1093/mnrasl/slac022} {\bibfield  {journal} {\bibinfo
  {journal} {Mon. Not. Roy. Astron. Soc.}\ }\textbf {\bibinfo {volume} {515}},\
  \bibinfo {pages} {L1--L5} (\bibinfo {year} {2022})},\ \bibinfo {note}
  {[Erratum: Mon.Not.Roy.Astron.Soc. 531, L8 (2024)]},\ \Eprint
  {http://arxiv.org/abs/2108.00581} {arXiv:2108.00581 [astro-ph.CO]}
  \BibitemShut {NoStop}%
\bibitem [{\citenamefont {Liu}\ \emph {et~al.}(2023)\citenamefont {Liu},
  \citenamefont {Yu},\ and\ \citenamefont {Wu}}]{Liu:2022bmn}%
  \BibitemOpen
  \bibfield  {author} {\bibinfo {author} {\bibfnamefont {Yang}\ \bibnamefont
  {Liu}}, \bibinfo {author} {\bibfnamefont {Hongwei}\ \bibnamefont {Yu}}, \
  and\ \bibinfo {author} {\bibfnamefont {Puxun}\ \bibnamefont {Wu}},\
  }\bibfield  {title} {\enquote {\bibinfo {title}
  {{Cosmological-model-independent Determination of Hubble Constant from Fast
  Radio Bursts and Hubble Parameter Measurements}},}\ }\href {\doibase
  10.3847/2041-8213/acc650} {\bibfield  {journal} {\bibinfo  {journal}
  {Astrophys. J. Lett.}\ }\textbf {\bibinfo {volume} {946}},\ \bibinfo {pages}
  {L49} (\bibinfo {year} {2023})},\ \Eprint {http://arxiv.org/abs/2210.05202}
  {arXiv:2210.05202 [astro-ph.CO]} \BibitemShut {NoStop}%
\bibitem [{\citenamefont {James}\ \emph {et~al.}(2022)\citenamefont {James}
  \emph {et~al.}}]{James:2022dcx}%
  \BibitemOpen
  \bibfield  {author} {\bibinfo {author} {\bibfnamefont {C.~W.}\ \bibnamefont
  {James}} \emph {et~al.},\ }\bibfield  {title} {\enquote {\bibinfo {title} {{A
  measurement of Hubble{\textquoteright}s Constant using Fast Radio Bursts}},}\
  }\href {\doibase 10.1093/mnras/stac2524} {\bibfield  {journal} {\bibinfo
  {journal} {Mon. Not. Roy. Astron. Soc.}\ }\textbf {\bibinfo {volume} {516}},\
  \bibinfo {pages} {4862--4881} (\bibinfo {year} {2022})},\ \Eprint
  {http://arxiv.org/abs/2208.00819} {arXiv:2208.00819 [astro-ph.CO]}
  \BibitemShut {NoStop}%
\bibitem [{\citenamefont {Baptista}\ \emph {et~al.}(2024)\citenamefont
  {Baptista}, \citenamefont {Prochaska}, \citenamefont {Mannings},
  \citenamefont {James}, \citenamefont {Shannon}, \citenamefont {Ryder},
  \citenamefont {Deller}, \citenamefont {Scott}, \citenamefont {Glowacki},\
  and\ \citenamefont {Tejos}}]{Baptista:2023uqu}%
  \BibitemOpen
  \bibfield  {author} {\bibinfo {author} {\bibfnamefont {Jay}\ \bibnamefont
  {Baptista}}, \bibinfo {author} {\bibfnamefont {J.~Xavier}\ \bibnamefont
  {Prochaska}}, \bibinfo {author} {\bibfnamefont {Alexandra~G.}\ \bibnamefont
  {Mannings}}, \bibinfo {author} {\bibfnamefont {C.~W.}\ \bibnamefont {James}},
  \bibinfo {author} {\bibfnamefont {R.~M.}\ \bibnamefont {Shannon}}, \bibinfo
  {author} {\bibfnamefont {Stuart~D.}\ \bibnamefont {Ryder}}, \bibinfo {author}
  {\bibfnamefont {A.~T.}\ \bibnamefont {Deller}}, \bibinfo {author}
  {\bibfnamefont {Danica~R.}\ \bibnamefont {Scott}}, \bibinfo {author}
  {\bibfnamefont {Marcin}\ \bibnamefont {Glowacki}}, \ and\ \bibinfo {author}
  {\bibfnamefont {Nicolas}\ \bibnamefont {Tejos}},\ }\bibfield  {title}
  {\enquote {\bibinfo {title} {{Measuring the Variance of the Macquart Relation
  in Redshift{\textendash}Extragalactic Dispersion Measure Modeling}},}\ }\href
  {\doibase 10.3847/1538-4357/ad2705} {\bibfield  {journal} {\bibinfo
  {journal} {Astrophys. J.}\ }\textbf {\bibinfo {volume} {965}},\ \bibinfo
  {pages} {57} (\bibinfo {year} {2024})},\ \Eprint
  {http://arxiv.org/abs/2305.07022} {arXiv:2305.07022 [astro-ph.CO]}
  \BibitemShut {NoStop}%
\bibitem [{\citenamefont {Fortunato}\ \emph {et~al.}(2023)\citenamefont
  {Fortunato}, \citenamefont {Hip{\'o}lito-Ricaldi},\ and\ \citenamefont {dos
  Santos}}]{Fortunato:2023deh}%
  \BibitemOpen
  \bibfield  {author} {\bibinfo {author} {\bibfnamefont {J{\'e}ferson A.~S.}\
  \bibnamefont {Fortunato}}, \bibinfo {author} {\bibfnamefont {Wiliam~S.}\
  \bibnamefont {Hip{\'o}lito-Ricaldi}}, \ and\ \bibinfo {author} {\bibfnamefont
  {Marcelo~V.}\ \bibnamefont {dos Santos}},\ }\bibfield  {title} {\enquote
  {\bibinfo {title} {{Cosmography from well-localized fast radio bursts}},}\
  }\href {\doibase 10.1093/mnras/stad2856} {\bibfield  {journal} {\bibinfo
  {journal} {Mon. Not. Roy. Astron. Soc.}\ }\textbf {\bibinfo {volume} {526}},\
  \bibinfo {pages} {1773--1782} (\bibinfo {year} {2023})},\ \Eprint
  {http://arxiv.org/abs/2307.04711} {arXiv:2307.04711 [astro-ph.CO]}
  \BibitemShut {NoStop}%
\bibitem [{\citenamefont {Gao}\ \emph {et~al.}(2023)\citenamefont {Gao},
  \citenamefont {Zhou}, \citenamefont {Du}, \citenamefont {Zou}, \citenamefont
  {Hu},\ and\ \citenamefont {Xu}}]{Gao:2023izj}%
  \BibitemOpen
  \bibfield  {author} {\bibinfo {author} {\bibfnamefont {Jiaze}\ \bibnamefont
  {Gao}}, \bibinfo {author} {\bibfnamefont {Zhihuan}\ \bibnamefont {Zhou}},
  \bibinfo {author} {\bibfnamefont {Minghui}\ \bibnamefont {Du}}, \bibinfo
  {author} {\bibfnamefont {Rui}\ \bibnamefont {Zou}}, \bibinfo {author}
  {\bibfnamefont {Jianping}\ \bibnamefont {Hu}}, \ and\ \bibinfo {author}
  {\bibfnamefont {Lixin}\ \bibnamefont {Xu}},\ }\bibfield  {title} {\enquote
  {\bibinfo {title} {{A Measurement of Hubble Constant Using Cosmographic
  Approach from Fast Radio Bursts and SNe Ia}},}\ }\href@noop {} {\  (\bibinfo
  {year} {2023})},\ \Eprint {http://arxiv.org/abs/2307.08285} {arXiv:2307.08285
  [astro-ph.CO]} \BibitemShut {NoStop}%
\bibitem [{\citenamefont {Wei}\ and\ \citenamefont
  {Melia}(2023)}]{Wei:2023avr}%
  \BibitemOpen
  \bibfield  {author} {\bibinfo {author} {\bibfnamefont {Jun-Jie}\ \bibnamefont
  {Wei}}\ and\ \bibinfo {author} {\bibfnamefont {Fulvio}\ \bibnamefont
  {Melia}},\ }\bibfield  {title} {\enquote {\bibinfo {title} {{Investigating
  Cosmological Models and the Hubble Tension Using Localized Fast Radio
  Bursts}},}\ }\href {\doibase 10.3847/1538-4357/acefb8} {\bibfield  {journal}
  {\bibinfo  {journal} {Astrophys. J.}\ }\textbf {\bibinfo {volume} {955}},\
  \bibinfo {pages} {101} (\bibinfo {year} {2023})},\ \Eprint
  {http://arxiv.org/abs/2308.05918} {arXiv:2308.05918 [astro-ph.CO]}
  \BibitemShut {NoStop}%
\bibitem [{\citenamefont {Zhao}\ \emph {et~al.}(2022)\citenamefont {Zhao},
  \citenamefont {Zhang}, \citenamefont {Li}, \citenamefont {Zhang},\ and\
  \citenamefont {Zhang}}]{Zhao:2022yiv}%
  \BibitemOpen
  \bibfield  {author} {\bibinfo {author} {\bibfnamefont {Ze-Wei}\ \bibnamefont
  {Zhao}}, \bibinfo {author} {\bibfnamefont {Ji-Guo}\ \bibnamefont {Zhang}},
  \bibinfo {author} {\bibfnamefont {Yichao}\ \bibnamefont {Li}}, \bibinfo
  {author} {\bibfnamefont {Jing-Fei}\ \bibnamefont {Zhang}}, \ and\ \bibinfo
  {author} {\bibfnamefont {Xin}\ \bibnamefont {Zhang}},\ }\bibfield  {title}
  {\enquote {\bibinfo {title} {{FRB dark sirens: Measuring the Hubble constant
  with unlocalized fast radio bursts}},}\ }\href@noop {} {\  (\bibinfo {year}
  {2022})},\ \Eprint {http://arxiv.org/abs/2212.13433} {arXiv:2212.13433
  [astro-ph.CO]} \BibitemShut {NoStop}%
\bibitem [{\citenamefont {Zhang}\ \emph {et~al.}(2023)\citenamefont {Zhang},
  \citenamefont {Zhao}, \citenamefont {Li}, \citenamefont {Zhang},
  \citenamefont {Li},\ and\ \citenamefont {Zhang}}]{Zhang:2023gye}%
  \BibitemOpen
  \bibfield  {author} {\bibinfo {author} {\bibfnamefont {Ji-Guo}\ \bibnamefont
  {Zhang}}, \bibinfo {author} {\bibfnamefont {Ze-Wei}\ \bibnamefont {Zhao}},
  \bibinfo {author} {\bibfnamefont {Yichao}\ \bibnamefont {Li}}, \bibinfo
  {author} {\bibfnamefont {Jing-Fei}\ \bibnamefont {Zhang}}, \bibinfo {author}
  {\bibfnamefont {Di}~\bibnamefont {Li}}, \ and\ \bibinfo {author}
  {\bibfnamefont {Xin}\ \bibnamefont {Zhang}},\ }\bibfield  {title} {\enquote
  {\bibinfo {title} {{Cosmology with fast radio bursts in the era of SKA}},}\
  }\href {\doibase 10.1007/s11433-023-2212-9} {\bibfield  {journal} {\bibinfo
  {journal} {Sci. China Phys. Mech. Astron.}\ }\textbf {\bibinfo {volume}
  {66}},\ \bibinfo {pages} {120412} (\bibinfo {year} {2023})},\ \Eprint
  {http://arxiv.org/abs/2307.01605} {arXiv:2307.01605 [astro-ph.CO]}
  \BibitemShut {NoStop}%
\bibitem [{\citenamefont {Wang}\ \emph {et~al.}(2025)\citenamefont {Wang},
  \citenamefont {Gao},\ and\ \citenamefont {Fan}}]{Wang:2025ugc}%
  \BibitemOpen
  \bibfield  {author} {\bibinfo {author} {\bibfnamefont {Yi-Ying}\ \bibnamefont
  {Wang}}, \bibinfo {author} {\bibfnamefont {Shi-Jie}\ \bibnamefont {Gao}}, \
  and\ \bibinfo {author} {\bibfnamefont {Yi-Zhong}\ \bibnamefont {Fan}},\
  }\bibfield  {title} {\enquote {\bibinfo {title} {{Probing Cosmology with 92
  Localized Fast Radio Bursts and DESI BAO}},}\ }\href {\doibase
  10.3847/1538-4357/adade8} {\bibfield  {journal} {\bibinfo  {journal}
  {Astrophys. J.}\ }\textbf {\bibinfo {volume} {981}},\ \bibinfo {pages} {9}
  (\bibinfo {year} {2025})},\ \Eprint {http://arxiv.org/abs/2501.09260}
  {arXiv:2501.09260 [astro-ph.CO]} \BibitemShut {NoStop}%
\bibitem [{\citenamefont {Zhang}\ \emph
  {et~al.}(2025{\natexlab{a}})\citenamefont {Zhang}, \citenamefont {Jiang},
  \citenamefont {Zhao}, \citenamefont {Qi}, \citenamefont {Zhang},\ and\
  \citenamefont {Zhang}}]{Zhang:2024rra}%
  \BibitemOpen
  \bibfield  {author} {\bibinfo {author} {\bibfnamefont {Ji-Guo}\ \bibnamefont
  {Zhang}}, \bibinfo {author} {\bibfnamefont {Yi-Fan}\ \bibnamefont {Jiang}},
  \bibinfo {author} {\bibfnamefont {Ze-Wei}\ \bibnamefont {Zhao}}, \bibinfo
  {author} {\bibfnamefont {Jing-Zhao}\ \bibnamefont {Qi}}, \bibinfo {author}
  {\bibfnamefont {Jing-Fei}\ \bibnamefont {Zhang}}, \ and\ \bibinfo {author}
  {\bibfnamefont {Xin}\ \bibnamefont {Zhang}},\ }\bibfield  {title} {\enquote
  {\bibinfo {title} {{Combining strongly lensed and unlensed fast radio bursts:
  To be a more precise late-universe probe}},}\ }\href {\doibase
  10.1007/s11433-024-2647-2} {\bibfield  {journal} {\bibinfo  {journal} {Sci.
  China Phys. Mech. Astron.}\ }\textbf {\bibinfo {volume} {68}},\ \bibinfo
  {pages} {280406} (\bibinfo {year} {2025}{\natexlab{a}})},\ \Eprint
  {http://arxiv.org/abs/2411.03126} {arXiv:2411.03126 [astro-ph.CO]}
  \BibitemShut {NoStop}%
\bibitem [{\citenamefont {Yang}\ \emph {et~al.}(2025)\citenamefont {Yang},
  \citenamefont {Hashimoto}, \citenamefont {Hsu}, \citenamefont {Goto},
  \citenamefont {Ling}, \citenamefont {Ho}, \citenamefont {Chen},\ and\
  \citenamefont {Kilerci}}]{Yang:2024vqq}%
  \BibitemOpen
  \bibfield  {author} {\bibinfo {author} {\bibfnamefont {Tsung-Ching}\
  \bibnamefont {Yang}}, \bibinfo {author} {\bibfnamefont {Tetsuya}\
  \bibnamefont {Hashimoto}}, \bibinfo {author} {\bibfnamefont {Tzu-Yin}\
  \bibnamefont {Hsu}}, \bibinfo {author} {\bibfnamefont {Tomotsugu}\
  \bibnamefont {Goto}}, \bibinfo {author} {\bibfnamefont {Chih-Teng}\
  \bibnamefont {Ling}}, \bibinfo {author} {\bibfnamefont {Simon C.~C.}\
  \bibnamefont {Ho}}, \bibinfo {author} {\bibfnamefont {Amos Y.~A.}\
  \bibnamefont {Chen}}, \ and\ \bibinfo {author} {\bibfnamefont {Ece}\
  \bibnamefont {Kilerci}},\ }\bibfield  {title} {\enquote {\bibinfo {title}
  {{Constraining the Hubble constant with scattering in host galaxies of fast
  radio bursts}},}\ }\href {\doibase 10.1051/0004-6361/202450823} {\bibfield
  {journal} {\bibinfo  {journal} {Astron. Astrophys.}\ }\textbf {\bibinfo
  {volume} {693}},\ \bibinfo {pages} {A85} (\bibinfo {year} {2025})},\ \Eprint
  {http://arxiv.org/abs/2411.02249} {arXiv:2411.02249 [astro-ph.CO]}
  \BibitemShut {NoStop}%
\bibitem [{\citenamefont {Kalita}\ \emph {et~al.}(2025)\citenamefont {Kalita},
  \citenamefont {Bhatporia},\ and\ \citenamefont {Weltman}}]{Kalita:2024xae}%
  \BibitemOpen
  \bibfield  {author} {\bibinfo {author} {\bibfnamefont {Surajit}\ \bibnamefont
  {Kalita}}, \bibinfo {author} {\bibfnamefont {Shruti}\ \bibnamefont
  {Bhatporia}}, \ and\ \bibinfo {author} {\bibfnamefont {Amanda}\ \bibnamefont
  {Weltman}},\ }\bibfield  {title} {\enquote {\bibinfo {title} {{Fast Radio
  Bursts as probes of the late-time universe: A new insight on the Hubble
  tension}},}\ }\href {\doibase 10.1016/j.dark.2025.101926} {\bibfield
  {journal} {\bibinfo  {journal} {Phys. Dark Univ.}\ }\textbf {\bibinfo
  {volume} {48}},\ \bibinfo {pages} {101926} (\bibinfo {year} {2025})},\
  \Eprint {http://arxiv.org/abs/2410.01974} {arXiv:2410.01974 [astro-ph.CO]}
  \BibitemShut {NoStop}%
\bibitem [{\citenamefont {Sales}\ \emph {et~al.}(2025)\citenamefont {Sales},
  \citenamefont {de~Farias}, \citenamefont {Queiroz}, \citenamefont {Santos},
  \citenamefont {Batista}, \citenamefont {Oliveira}, \citenamefont {Santana},
  \citenamefont {Wuensche}, \citenamefont {Villela},\ and\ \citenamefont
  {Vieira}}]{Sales:2025shu}%
  \BibitemOpen
  \bibfield  {author} {\bibinfo {author} {\bibfnamefont {L{\'a}zaro~L.}\
  \bibnamefont {Sales}}, \bibinfo {author} {\bibfnamefont {Klecio E.~L.}\
  \bibnamefont {de~Farias}}, \bibinfo {author} {\bibfnamefont {Amilcar~R.}\
  \bibnamefont {Queiroz}}, \bibinfo {author} {\bibfnamefont {Jo{\~a}o R.~L.}\
  \bibnamefont {Santos}}, \bibinfo {author} {\bibfnamefont {Rafael~A.}\
  \bibnamefont {Batista}}, \bibinfo {author} {\bibfnamefont {Ana R.~M.}\
  \bibnamefont {Oliveira}}, \bibinfo {author} {\bibfnamefont {Lucas~F.}\
  \bibnamefont {Santana}}, \bibinfo {author} {\bibfnamefont {Carlos~A.}\
  \bibnamefont {Wuensche}}, \bibinfo {author} {\bibfnamefont {Thyrso}\
  \bibnamefont {Villela}}, \ and\ \bibinfo {author} {\bibfnamefont {Jordany}\
  \bibnamefont {Vieira}},\ }\bibfield  {title} {\enquote {\bibinfo {title}
  {{Model-independent observational constraints with fast radio bursts}},}\
  }\href@noop {} {\  (\bibinfo {year} {2025})},\ \Eprint
  {http://arxiv.org/abs/2507.06975} {arXiv:2507.06975 [astro-ph.CO]}
  \BibitemShut {NoStop}%
\bibitem [{\citenamefont {Gao}\ \emph {et~al.}(2025)\citenamefont {Gao},
  \citenamefont {Wu}, \citenamefont {Hu}, \citenamefont {Yi}, \citenamefont
  {Zhou}, \citenamefont {Wang},\ and\ \citenamefont {Dai}}]{Gao:2025fcr}%
  \BibitemOpen
  \bibfield  {author} {\bibinfo {author} {\bibfnamefont {D.~H.}\ \bibnamefont
  {Gao}}, \bibinfo {author} {\bibfnamefont {Q.}~\bibnamefont {Wu}}, \bibinfo
  {author} {\bibfnamefont {J.~P.}\ \bibnamefont {Hu}}, \bibinfo {author}
  {\bibfnamefont {S.~X.}\ \bibnamefont {Yi}}, \bibinfo {author} {\bibfnamefont
  {X.}~\bibnamefont {Zhou}}, \bibinfo {author} {\bibfnamefont {F.~Y.}\
  \bibnamefont {Wang}}, \ and\ \bibinfo {author} {\bibfnamefont {Z.~G.}\
  \bibnamefont {Dai}},\ }\bibfield  {title} {\enquote {\bibinfo {title}
  {{Measuring the Hubble constant using localized and nonlocalized fast radio
  bursts}},}\ }\href {\doibase 10.1051/0004-6361/202453006} {\bibfield
  {journal} {\bibinfo  {journal} {Astron. Astrophys.}\ }\textbf {\bibinfo
  {volume} {698}},\ \bibinfo {pages} {A215} (\bibinfo {year} {2025})},\ \Eprint
  {http://arxiv.org/abs/2410.03994} {arXiv:2410.03994 [astro-ph.CO]}
  \BibitemShut {NoStop}%
\bibitem [{\citenamefont {Zhang}\ and\ \citenamefont
  {Zhang}(2025)}]{Zhang:2025thh}%
  \BibitemOpen
  \bibfield  {author} {\bibinfo {author} {\bibfnamefont {Zi-Liang}\
  \bibnamefont {Zhang}}\ and\ \bibinfo {author} {\bibfnamefont {Bing}\
  \bibnamefont {Zhang}},\ }\bibfield  {title} {\enquote {\bibinfo {title}
  {{Cosmological Parameter Estimate from Persistent Radio Sources of Fast Radio
  Bursts}},}\ }\href {\doibase 10.3847/2041-8213/adcc30} {\bibfield  {journal}
  {\bibinfo  {journal} {Astrophys. J. Lett.}\ }\textbf {\bibinfo {volume}
  {984}},\ \bibinfo {pages} {L40} (\bibinfo {year} {2025})},\ \Eprint
  {http://arxiv.org/abs/2504.13132} {arXiv:2504.13132 [astro-ph.CO]}
  \BibitemShut {NoStop}%
\bibitem [{\citenamefont {McQuinn}(2014)}]{McQuinn:2013tmc}%
  \BibitemOpen
  \bibfield  {author} {\bibinfo {author} {\bibfnamefont {Matthew}\ \bibnamefont
  {McQuinn}},\ }\bibfield  {title} {\enquote {\bibinfo {title} {{Locating the
  ''missing'' baryons with extragalactic dispersion measure estimates}},}\
  }\href {\doibase 10.1088/2041-8205/780/2/L33} {\bibfield  {journal} {\bibinfo
   {journal} {Astrophys. J. Lett.}\ }\textbf {\bibinfo {volume} {780}},\
  \bibinfo {pages} {L33} (\bibinfo {year} {2014})},\ \Eprint
  {http://arxiv.org/abs/1309.4451} {arXiv:1309.4451 [astro-ph.CO]} \BibitemShut
  {NoStop}%
\bibitem [{\citenamefont {Deng}\ and\ \citenamefont
  {Zhang}(2014)}]{Deng:2013aga}%
  \BibitemOpen
  \bibfield  {author} {\bibinfo {author} {\bibfnamefont {Wei}\ \bibnamefont
  {Deng}}\ and\ \bibinfo {author} {\bibfnamefont {Bing}\ \bibnamefont
  {Zhang}},\ }\bibfield  {title} {\enquote {\bibinfo {title} {{Cosmological
  Implications of Fast Radio Burst/Gamma-Ray Burst Associations}},}\ }\href
  {\doibase 10.1088/2041-8205/783/2/L35} {\bibfield  {journal} {\bibinfo
  {journal} {Astrophys. J. Lett.}\ }\textbf {\bibinfo {volume} {783}},\
  \bibinfo {pages} {L35} (\bibinfo {year} {2014})},\ \Eprint
  {http://arxiv.org/abs/1401.0059} {arXiv:1401.0059 [astro-ph.HE]} \BibitemShut
  {NoStop}%
\bibitem [{\citenamefont {Macquart}\ \emph {et~al.}(2020)\citenamefont
  {Macquart} \emph {et~al.}}]{Macquart:2020lln}%
  \BibitemOpen
  \bibfield  {author} {\bibinfo {author} {\bibfnamefont {J.~P.}\ \bibnamefont
  {Macquart}} \emph {et~al.},\ }\bibfield  {title} {\enquote {\bibinfo {title}
  {{A census of baryons in the Universe from localized fast radio bursts}},}\
  }\href {\doibase 10.1038/s41586-020-2300-2} {\bibfield  {journal} {\bibinfo
  {journal} {Nature}\ }\textbf {\bibinfo {volume} {581}},\ \bibinfo {pages}
  {391--395} (\bibinfo {year} {2020})},\ \Eprint
  {http://arxiv.org/abs/2005.13161} {arXiv:2005.13161 [astro-ph.CO]}
  \BibitemShut {NoStop}%
\bibitem [{\citenamefont {Yang}\ \emph {et~al.}(2022)\citenamefont {Yang},
  \citenamefont {Wu},\ and\ \citenamefont {Wang}}]{Yang:2022ftm}%
  \BibitemOpen
  \bibfield  {author} {\bibinfo {author} {\bibfnamefont {K.~B.}\ \bibnamefont
  {Yang}}, \bibinfo {author} {\bibfnamefont {Q.}~\bibnamefont {Wu}}, \ and\
  \bibinfo {author} {\bibfnamefont {F.~Y.}\ \bibnamefont {Wang}},\ }\bibfield
  {title} {\enquote {\bibinfo {title} {{Finding the Missing Baryons in the
  Intergalactic Medium with Localized Fast Radio Bursts}},}\ }\href {\doibase
  10.3847/2041-8213/aca145} {\bibfield  {journal} {\bibinfo  {journal}
  {Astrophys. J. Lett.}\ }\textbf {\bibinfo {volume} {940}},\ \bibinfo {pages}
  {L29} (\bibinfo {year} {2022})},\ \Eprint {http://arxiv.org/abs/2211.04058}
  {arXiv:2211.04058 [astro-ph.HE]} \BibitemShut {NoStop}%
\bibitem [{\citenamefont {Connor}\ \emph {et~al.}(2025)\citenamefont {Connor}
  \emph {et~al.}}]{Connor:2024mjg}%
  \BibitemOpen
  \bibfield  {author} {\bibinfo {author} {\bibfnamefont {Liam}\ \bibnamefont
  {Connor}} \emph {et~al.},\ }\bibfield  {title} {\enquote {\bibinfo {title}
  {{A gas-rich cosmic web revealed by the partitioning of the missing
  baryons}},}\ }\href {\doibase 10.1038/s41550-025-02566-y} {\bibfield
  {journal} {\bibinfo  {journal} {Nature Astron.}\ }\textbf {\bibinfo {volume}
  {9}},\ \bibinfo {pages} {1226--1239} (\bibinfo {year} {2025})},\ \Eprint
  {http://arxiv.org/abs/2409.16952} {arXiv:2409.16952 [astro-ph.CO]}
  \BibitemShut {NoStop}%
\bibitem [{\citenamefont {Zhang}\ \emph
  {et~al.}(2025{\natexlab{b}})\citenamefont {Zhang}, \citenamefont {Song},
  \citenamefont {Zhao}, \citenamefont {Sun}, \citenamefont {Zhang},\ and\
  \citenamefont {Zhang}}]{Zhang:2025yhi}%
  \BibitemOpen
  \bibfield  {author} {\bibinfo {author} {\bibfnamefont {Ji-Guo}\ \bibnamefont
  {Zhang}}, \bibinfo {author} {\bibfnamefont {Ji-Yu}\ \bibnamefont {Song}},
  \bibinfo {author} {\bibfnamefont {Ze-Wei}\ \bibnamefont {Zhao}}, \bibinfo
  {author} {\bibfnamefont {Wan-Peng}\ \bibnamefont {Sun}}, \bibinfo {author}
  {\bibfnamefont {Jing-Fei}\ \bibnamefont {Zhang}}, \ and\ \bibinfo {author}
  {\bibfnamefont {Xin}\ \bibnamefont {Zhang}},\ }\bibfield  {title} {\enquote
  {\bibinfo {title} {{Cosmic baryon census with fast radio bursts and
  gravitational waves}},}\ }\href@noop {} {\  (\bibinfo {year}
  {2025}{\natexlab{b}})},\ \Eprint {http://arxiv.org/abs/2507.06841}
  {arXiv:2507.06841 [astro-ph.CO]} \BibitemShut {NoStop}%
\bibitem [{\citenamefont {Liu}\ \emph {et~al.}(2025)\citenamefont {Liu},
  \citenamefont {Zhang}, \citenamefont {Yu},\ and\ \citenamefont
  {Wu}}]{Liu:2025fdf}%
  \BibitemOpen
  \bibfield  {author} {\bibinfo {author} {\bibfnamefont {Yang}\ \bibnamefont
  {Liu}}, \bibinfo {author} {\bibfnamefont {Yuchen}\ \bibnamefont {Zhang}},
  \bibinfo {author} {\bibfnamefont {Hongwei}\ \bibnamefont {Yu}}, \ and\
  \bibinfo {author} {\bibfnamefont {Puxun}\ \bibnamefont {Wu}},\ }\bibfield
  {title} {\enquote {\bibinfo {title} {{Constraining the Baryon Fraction in the
  Intergalactic Medium with 92 localized Fast Radio Bursts}},}\ }\href@noop {}
  {\  (\bibinfo {year} {2025})},\ \Eprint {http://arxiv.org/abs/2506.03536}
  {arXiv:2506.03536 [astro-ph.CO]} \BibitemShut {NoStop}%
\bibitem [{\citenamefont {Lorimer}\ \emph {et~al.}(2007)\citenamefont
  {Lorimer}, \citenamefont {Bailes}, \citenamefont {McLaughlin}, \citenamefont
  {Narkevic},\ and\ \citenamefont {Crawford}}]{Lorimer:2007qn}%
  \BibitemOpen
  \bibfield  {author} {\bibinfo {author} {\bibfnamefont {D.~R.}\ \bibnamefont
  {Lorimer}}, \bibinfo {author} {\bibfnamefont {M.}~\bibnamefont {Bailes}},
  \bibinfo {author} {\bibfnamefont {M.~A.}\ \bibnamefont {McLaughlin}},
  \bibinfo {author} {\bibfnamefont {D.~J.}\ \bibnamefont {Narkevic}}, \ and\
  \bibinfo {author} {\bibfnamefont {F.}~\bibnamefont {Crawford}},\ }\bibfield
  {title} {\enquote {\bibinfo {title} {{A bright millisecond radio burst of
  extragalactic origin}},}\ }\href {\doibase 10.1126/science.1147532}
  {\bibfield  {journal} {\bibinfo  {journal} {Science}\ }\textbf {\bibinfo
  {volume} {318}},\ \bibinfo {pages} {777} (\bibinfo {year} {2007})},\ \Eprint
  {http://arxiv.org/abs/0709.4301} {arXiv:0709.4301 [astro-ph]} \BibitemShut
  {NoStop}%
\bibitem [{\citenamefont {Thornton}\ \emph {et~al.}(2013)\citenamefont
  {Thornton} \emph {et~al.}}]{Thornton:2013iua}%
  \BibitemOpen
  \bibfield  {author} {\bibinfo {author} {\bibfnamefont {D.}~\bibnamefont
  {Thornton}} \emph {et~al.},\ }\bibfield  {title} {\enquote {\bibinfo {title}
  {{A Population of Fast Radio Bursts at Cosmological Distances}},}\ }\href
  {\doibase 10.1126/science.1236789} {\bibfield  {journal} {\bibinfo  {journal}
  {Science}\ }\textbf {\bibinfo {volume} {341}},\ \bibinfo {pages} {53--56}
  (\bibinfo {year} {2013})},\ \Eprint {http://arxiv.org/abs/1307.1628}
  {arXiv:1307.1628 [astro-ph.HE]} \BibitemShut {NoStop}%
\bibitem [{\citenamefont {Petroff}\ \emph {et~al.}(2019)\citenamefont
  {Petroff}, \citenamefont {Hessels},\ and\ \citenamefont
  {Lorimer}}]{Petroff:2019tty}%
  \BibitemOpen
  \bibfield  {author} {\bibinfo {author} {\bibfnamefont {E.}~\bibnamefont
  {Petroff}}, \bibinfo {author} {\bibfnamefont {J.~W.~T.}\ \bibnamefont
  {Hessels}}, \ and\ \bibinfo {author} {\bibfnamefont {D.~R.}\ \bibnamefont
  {Lorimer}},\ }\bibfield  {title} {\enquote {\bibinfo {title} {{Fast Radio
  Bursts}},}\ }\href {\doibase 10.1007/s00159-019-0116-6} {\bibfield  {journal}
  {\bibinfo  {journal} {Astron. Astrophys. Rev.}\ }\textbf {\bibinfo {volume}
  {27}},\ \bibinfo {pages} {4} (\bibinfo {year} {2019})},\ \Eprint
  {http://arxiv.org/abs/1904.07947} {arXiv:1904.07947 [astro-ph.HE]}
  \BibitemShut {NoStop}%
\bibitem [{\citenamefont {Andersen}\ \emph {et~al.}(2020)\citenamefont
  {Andersen} \emph {et~al.}}]{CHIMEFRB:2020abu}%
  \BibitemOpen
  \bibfield  {author} {\bibinfo {author} {\bibfnamefont {B.~C.}\ \bibnamefont
  {Andersen}} \emph {et~al.} (\bibinfo {collaboration} {CHIME/FRB}),\
  }\bibfield  {title} {\enquote {\bibinfo {title} {{A bright
  millisecond-duration radio burst from a Galactic magnetar}},}\ }\href
  {\doibase 10.1038/s41586-020-2863-y} {\bibfield  {journal} {\bibinfo
  {journal} {Nature}\ }\textbf {\bibinfo {volume} {587}},\ \bibinfo {pages}
  {54--58} (\bibinfo {year} {2020})},\ \Eprint
  {http://arxiv.org/abs/2005.10324} {arXiv:2005.10324 [astro-ph.HE]}
  \BibitemShut {NoStop}%
\bibitem [{\citenamefont {Bochenek}\ \emph {et~al.}(2020)\citenamefont
  {Bochenek}, \citenamefont {Ravi}, \citenamefont {Belov}, \citenamefont
  {Hallinan}, \citenamefont {Kocz}, \citenamefont {Kulkarni},\ and\
  \citenamefont {McKenna}}]{Bochenek:2020zxn}%
  \BibitemOpen
  \bibfield  {author} {\bibinfo {author} {\bibfnamefont {Christopher~D.}\
  \bibnamefont {Bochenek}}, \bibinfo {author} {\bibfnamefont {Vikram}\
  \bibnamefont {Ravi}}, \bibinfo {author} {\bibfnamefont {Konstantin~V.}\
  \bibnamefont {Belov}}, \bibinfo {author} {\bibfnamefont {Gregg}\ \bibnamefont
  {Hallinan}}, \bibinfo {author} {\bibfnamefont {Jonathon}\ \bibnamefont
  {Kocz}}, \bibinfo {author} {\bibfnamefont {Shri~R.}\ \bibnamefont
  {Kulkarni}}, \ and\ \bibinfo {author} {\bibfnamefont {Dan~L.}\ \bibnamefont
  {McKenna}},\ }\bibfield  {title} {\enquote {\bibinfo {title} {{A fast radio
  burst associated with a Galactic magnetar}},}\ }\href {\doibase
  10.1038/s41586-020-2872-x} {\bibfield  {journal} {\bibinfo  {journal}
  {Nature}\ }\textbf {\bibinfo {volume} {587}},\ \bibinfo {pages} {59--62}
  (\bibinfo {year} {2020})},\ \Eprint {http://arxiv.org/abs/2005.10828}
  {arXiv:2005.10828 [astro-ph.HE]} \BibitemShut {NoStop}%
\bibitem [{\citenamefont {Schutz}(1986)}]{Schutz:1986gp}%
  \BibitemOpen
  \bibfield  {author} {\bibinfo {author} {\bibfnamefont {Bernard~F.}\
  \bibnamefont {Schutz}},\ }\bibfield  {title} {\enquote {\bibinfo {title}
  {{Determining the Hubble Constant from Gravitational Wave Observations}},}\
  }\href {\doibase 10.1038/323310a0} {\bibfield  {journal} {\bibinfo  {journal}
  {Nature}\ }\textbf {\bibinfo {volume} {323}},\ \bibinfo {pages} {310--311}
  (\bibinfo {year} {1986})}\BibitemShut {NoStop}%
\bibitem [{\citenamefont {Holz}\ and\ \citenamefont
  {Hughes}(2005)}]{Holz:2005df}%
  \BibitemOpen
  \bibfield  {author} {\bibinfo {author} {\bibfnamefont {Daniel~E.}\
  \bibnamefont {Holz}}\ and\ \bibinfo {author} {\bibfnamefont {Scott~A.}\
  \bibnamefont {Hughes}},\ }\bibfield  {title} {\enquote {\bibinfo {title}
  {{Using gravitational-wave standard sirens}},}\ }\href {\doibase
  10.1086/431341} {\bibfield  {journal} {\bibinfo  {journal} {Astrophys. J.}\
  }\textbf {\bibinfo {volume} {629}},\ \bibinfo {pages} {15--22} (\bibinfo
  {year} {2005})},\ \Eprint {http://arxiv.org/abs/astro-ph/0504616}
  {arXiv:astro-ph/0504616} \BibitemShut {NoStop}%
\bibitem [{\citenamefont {Treu}\ and\ \citenamefont
  {Marshall}(2016)}]{Treu:2016ljm}%
  \BibitemOpen
  \bibfield  {author} {\bibinfo {author} {\bibfnamefont {Tommaso}\ \bibnamefont
  {Treu}}\ and\ \bibinfo {author} {\bibfnamefont {Philip~J.}\ \bibnamefont
  {Marshall}},\ }\bibfield  {title} {\enquote {\bibinfo {title} {{Time Delay
  Cosmography}},}\ }\href {\doibase 10.1007/s00159-016-0096-8} {\bibfield
  {journal} {\bibinfo  {journal} {Astron. Astrophys. Rev.}\ }\textbf {\bibinfo
  {volume} {24}},\ \bibinfo {pages} {11} (\bibinfo {year} {2016})},\ \Eprint
  {http://arxiv.org/abs/1605.05333} {arXiv:1605.05333 [astro-ph.CO]}
  \BibitemShut {NoStop}%
\bibitem [{\citenamefont {Bull}\ \emph {et~al.}(2015)\citenamefont {Bull},
  \citenamefont {Ferreira}, \citenamefont {Patel},\ and\ \citenamefont
  {Santos}}]{Bull:2014rha}%
  \BibitemOpen
  \bibfield  {author} {\bibinfo {author} {\bibfnamefont {Philip}\ \bibnamefont
  {Bull}}, \bibinfo {author} {\bibfnamefont {Pedro~G.}\ \bibnamefont
  {Ferreira}}, \bibinfo {author} {\bibfnamefont {Prina}\ \bibnamefont {Patel}},
  \ and\ \bibinfo {author} {\bibfnamefont {Mario~G.}\ \bibnamefont {Santos}},\
  }\bibfield  {title} {\enquote {\bibinfo {title} {{Late-time cosmology with
  21cm intensity mapping experiments}},}\ }\href {\doibase
  10.1088/0004-637X/803/1/21} {\bibfield  {journal} {\bibinfo  {journal}
  {Astrophys. J.}\ }\textbf {\bibinfo {volume} {803}},\ \bibinfo {pages} {21}
  (\bibinfo {year} {2015})},\ \Eprint {http://arxiv.org/abs/1405.1452}
  {arXiv:1405.1452 [astro-ph.CO]} \BibitemShut {NoStop}%
\bibitem [{\citenamefont {Chen}\ \emph {et~al.}(2019)\citenamefont {Chen},
  \citenamefont {Huang},\ and\ \citenamefont {Wang}}]{Chen:2018dbv}%
  \BibitemOpen
  \bibfield  {author} {\bibinfo {author} {\bibfnamefont {Lu}~\bibnamefont
  {Chen}}, \bibinfo {author} {\bibfnamefont {Qing-Guo}\ \bibnamefont {Huang}},
  \ and\ \bibinfo {author} {\bibfnamefont {Ke}~\bibnamefont {Wang}},\
  }\bibfield  {title} {\enquote {\bibinfo {title} {{Distance Priors from Planck
  Final Release}},}\ }\href {\doibase 10.1088/1475-7516/2019/02/028} {\bibfield
   {journal} {\bibinfo  {journal} {JCAP}\ }\textbf {\bibinfo {volume} {02}},\
  \bibinfo {pages} {028} (\bibinfo {year} {2019})},\ \Eprint
  {http://arxiv.org/abs/1808.05724} {arXiv:1808.05724 [astro-ph.CO]}
  \BibitemShut {NoStop}%
\bibitem [{\citenamefont {Meiksin}(2009)}]{Meiksin:2007rz}%
  \BibitemOpen
  \bibfield  {author} {\bibinfo {author} {\bibfnamefont {Avery~A.}\
  \bibnamefont {Meiksin}},\ }\bibfield  {title} {\enquote {\bibinfo {title}
  {{The Physics of the Intergalactic Medium}},}\ }\href {\doibase
  10.1103/RevModPhys.81.1405} {\bibfield  {journal} {\bibinfo  {journal} {Rev.
  Mod. Phys.}\ }\textbf {\bibinfo {volume} {81}},\ \bibinfo {pages}
  {1405--1469} (\bibinfo {year} {2009})},\ \Eprint
  {http://arxiv.org/abs/0711.3358} {arXiv:0711.3358 [astro-ph]} \BibitemShut
  {NoStop}%
\bibitem [{\citenamefont {Amiri}\ \emph {et~al.}(2021)\citenamefont {Amiri}
  \emph {et~al.}}]{CHIMEFRB:2021srp}%
  \BibitemOpen
  \bibfield  {author} {\bibinfo {author} {\bibfnamefont {Mandana}\ \bibnamefont
  {Amiri}} \emph {et~al.} (\bibinfo {collaboration} {CHIME/FRB}),\ }\bibfield
  {title} {\enquote {\bibinfo {title} {{The First CHIME/FRB Fast Radio Burst
  Catalog}},}\ }\href {\doibase 10.3847/1538-4365/ac33ab} {\bibfield  {journal}
  {\bibinfo  {journal} {Astrophys. J. Supp.}\ }\textbf {\bibinfo {volume}
  {257}},\ \bibinfo {pages} {59} (\bibinfo {year} {2021})},\ \Eprint
  {http://arxiv.org/abs/2106.04352} {arXiv:2106.04352 [astro-ph.HE]}
  \BibitemShut {NoStop}%
\bibitem [{\citenamefont {Jaroszynski}(2020)}]{Jaroszynski:2020kqy}%
  \BibitemOpen
  \bibfield  {author} {\bibinfo {author} {\bibfnamefont {M.}~\bibnamefont
  {Jaroszynski}},\ }\bibfield  {title} {\enquote {\bibinfo {title} {{FRBs: the
  Dispersion Measure of Host Galaxies}},}\ }\href {\doibase
  10.32023/0001-5237/70.2.1} {\bibfield  {journal} {\bibinfo  {journal} {Acta
  Astron.}\ }\textbf {\bibinfo {volume} {70}},\ \bibinfo {pages} {87--100}
  (\bibinfo {year} {2020})},\ \Eprint {http://arxiv.org/abs/2008.04634}
  {arXiv:2008.04634 [astro-ph.GA]} \BibitemShut {NoStop}%
\bibitem [{\citenamefont {Madau}\ and\ \citenamefont
  {Fragos}(2017)}]{Madau:2016jbv}%
  \BibitemOpen
  \bibfield  {author} {\bibinfo {author} {\bibfnamefont {Piero}\ \bibnamefont
  {Madau}}\ and\ \bibinfo {author} {\bibfnamefont {Tassos}\ \bibnamefont
  {Fragos}},\ }\bibfield  {title} {\enquote {\bibinfo {title} {{Radiation
  Backgrounds at Cosmic Dawn: X-Rays from Compact Binaries}},}\ }\href
  {\doibase 10.3847/1538-4357/aa6af9} {\bibfield  {journal} {\bibinfo
  {journal} {Astrophys. J.}\ }\textbf {\bibinfo {volume} {840}},\ \bibinfo
  {pages} {39} (\bibinfo {year} {2017})},\ \Eprint
  {http://arxiv.org/abs/1606.07887} {arXiv:1606.07887 [astro-ph.GA]}
  \BibitemShut {NoStop}%
\bibitem [{\citenamefont {Qiang}\ \emph {et~al.}(2022)\citenamefont {Qiang},
  \citenamefont {Li},\ and\ \citenamefont {Wei}}]{Qiang:2021ljr}%
  \BibitemOpen
  \bibfield  {author} {\bibinfo {author} {\bibfnamefont {Da-Chun}\ \bibnamefont
  {Qiang}}, \bibinfo {author} {\bibfnamefont {Shu-Ling}\ \bibnamefont {Li}}, \
  and\ \bibinfo {author} {\bibfnamefont {Hao}\ \bibnamefont {Wei}},\ }\bibfield
   {title} {\enquote {\bibinfo {title} {{Fast radio burst distributions
  consistent with the first CHIME/FRB catalog}},}\ }\href {\doibase
  10.1088/1475-7516/2022/01/040} {\bibfield  {journal} {\bibinfo  {journal}
  {JCAP}\ }\textbf {\bibinfo {volume} {01}},\ \bibinfo {pages} {040} (\bibinfo
  {year} {2022})},\ \Eprint {http://arxiv.org/abs/2111.07476} {arXiv:2111.07476
  [astro-ph.HE]} \BibitemShut {NoStop}%
\bibitem [{\citenamefont {Zhao}\ \emph {et~al.}(2011)\citenamefont {Zhao},
  \citenamefont {Van Den~Broeck}, \citenamefont {Baskaran},\ and\ \citenamefont
  {Li}}]{Zhao:2010sz}%
  \BibitemOpen
  \bibfield  {author} {\bibinfo {author} {\bibfnamefont {W.}~\bibnamefont
  {Zhao}}, \bibinfo {author} {\bibfnamefont {C.}~\bibnamefont {Van
  Den~Broeck}}, \bibinfo {author} {\bibfnamefont {D.}~\bibnamefont {Baskaran}},
  \ and\ \bibinfo {author} {\bibfnamefont {T.~G.~F.}\ \bibnamefont {Li}},\
  }\bibfield  {title} {\enquote {\bibinfo {title} {{Determination of Dark
  Energy by the Einstein Telescope: Comparing with CMB, BAO and SNIa
  Observations}},}\ }\href {\doibase 10.1103/PhysRevD.83.023005} {\bibfield
  {journal} {\bibinfo  {journal} {Phys. Rev. D}\ }\textbf {\bibinfo {volume}
  {83}},\ \bibinfo {pages} {023005} (\bibinfo {year} {2011})},\ \Eprint
  {http://arxiv.org/abs/1009.0206} {arXiv:1009.0206 [astro-ph.CO]} \BibitemShut
  {NoStop}%
\bibitem [{\citenamefont {Schneider}\ \emph {et~al.}(2001)\citenamefont
  {Schneider}, \citenamefont {Ferrari}, \citenamefont {Matarrese},\ and\
  \citenamefont {Portegies~Zwart}}]{Schneider:2000sg}%
  \BibitemOpen
  \bibfield  {author} {\bibinfo {author} {\bibfnamefont {Raffaella}\
  \bibnamefont {Schneider}}, \bibinfo {author} {\bibfnamefont {Valeria}\
  \bibnamefont {Ferrari}}, \bibinfo {author} {\bibfnamefont {Sabino}\
  \bibnamefont {Matarrese}}, \ and\ \bibinfo {author} {\bibfnamefont
  {Simon~F.}\ \bibnamefont {Portegies~Zwart}},\ }\bibfield  {title} {\enquote
  {\bibinfo {title} {{Gravitational waves from cosmological compact
  binaries}},}\ }\href {\doibase 10.1046/j.1365-8711.2001.04217.x} {\bibfield
  {journal} {\bibinfo  {journal} {Mon. Not. Roy. Astron. Soc.}\ }\textbf
  {\bibinfo {volume} {324}},\ \bibinfo {pages} {797} (\bibinfo {year}
  {2001})},\ \Eprint {http://arxiv.org/abs/astro-ph/0002055}
  {arXiv:astro-ph/0002055} \BibitemShut {NoStop}%
\bibitem [{\citenamefont {Cutler}\ and\ \citenamefont
  {Holz}(2009)}]{Cutler:2009qv}%
  \BibitemOpen
  \bibfield  {author} {\bibinfo {author} {\bibfnamefont {Curt}\ \bibnamefont
  {Cutler}}\ and\ \bibinfo {author} {\bibfnamefont {Daniel~E.}\ \bibnamefont
  {Holz}},\ }\bibfield  {title} {\enquote {\bibinfo {title} {{Ultra-high
  precision cosmology from gravitational waves}},}\ }\href {\doibase
  10.1103/PhysRevD.80.104009} {\bibfield  {journal} {\bibinfo  {journal} {Phys.
  Rev. D}\ }\textbf {\bibinfo {volume} {80}},\ \bibinfo {pages} {104009}
  (\bibinfo {year} {2009})},\ \Eprint {http://arxiv.org/abs/0906.3752}
  {arXiv:0906.3752 [astro-ph.CO]} \BibitemShut {NoStop}%
\bibitem [{\citenamefont {Zhang}\ \emph {et~al.}(2019)\citenamefont {Zhang},
  \citenamefont {Zhang}, \citenamefont {Jin}, \citenamefont {Qi},\ and\
  \citenamefont {Zhang}}]{Zhang:2019loq}%
  \BibitemOpen
  \bibfield  {author} {\bibinfo {author} {\bibfnamefont {Jing-Fei}\
  \bibnamefont {Zhang}}, \bibinfo {author} {\bibfnamefont {Ming}\ \bibnamefont
  {Zhang}}, \bibinfo {author} {\bibfnamefont {Shang-Jie}\ \bibnamefont {Jin}},
  \bibinfo {author} {\bibfnamefont {Jing-Zhao}\ \bibnamefont {Qi}}, \ and\
  \bibinfo {author} {\bibfnamefont {Xin}\ \bibnamefont {Zhang}},\ }\bibfield
  {title} {\enquote {\bibinfo {title} {{Cosmological parameter estimation with
  future gravitational wave standard siren observation from the Einstein
  Telescope}},}\ }\href {\doibase 10.1088/1475-7516/2019/09/068} {\bibfield
  {journal} {\bibinfo  {journal} {JCAP}\ }\textbf {\bibinfo {volume} {09}},\
  \bibinfo {pages} {068} (\bibinfo {year} {2019})},\ \Eprint
  {http://arxiv.org/abs/1907.03238} {arXiv:1907.03238 [astro-ph.CO]}
  \BibitemShut {NoStop}%
\bibitem [{\citenamefont {Jin}\ \emph {et~al.}(2020)\citenamefont {Jin},
  \citenamefont {He}, \citenamefont {Xu}, \citenamefont {Zhang},\ and\
  \citenamefont {Zhang}}]{Jin:2020hmc}%
  \BibitemOpen
  \bibfield  {author} {\bibinfo {author} {\bibfnamefont {Shang-Jie}\
  \bibnamefont {Jin}}, \bibinfo {author} {\bibfnamefont {Dong-Ze}\ \bibnamefont
  {He}}, \bibinfo {author} {\bibfnamefont {Yidong}\ \bibnamefont {Xu}},
  \bibinfo {author} {\bibfnamefont {Jing-Fei}\ \bibnamefont {Zhang}}, \ and\
  \bibinfo {author} {\bibfnamefont {Xin}\ \bibnamefont {Zhang}},\ }\bibfield
  {title} {\enquote {\bibinfo {title} {{Forecast for cosmological parameter
  estimation with gravitational-wave standard siren observation from the Cosmic
  Explorer}},}\ }\href {\doibase 10.1088/1475-7516/2020/03/051} {\bibfield
  {journal} {\bibinfo  {journal} {JCAP}\ }\textbf {\bibinfo {volume} {03}},\
  \bibinfo {pages} {051} (\bibinfo {year} {2020})},\ \Eprint
  {http://arxiv.org/abs/2001.05393} {arXiv:2001.05393 [astro-ph.CO]}
  \BibitemShut {NoStop}%
\bibitem [{\citenamefont {Wu}\ \emph {et~al.}(2023{\natexlab{a}})\citenamefont
  {Wu}, \citenamefont {Shao}, \citenamefont {Jin},\ and\ \citenamefont
  {Zhang}}]{Wu:2022dgy}%
  \BibitemOpen
  \bibfield  {author} {\bibinfo {author} {\bibfnamefont {Peng-Ju}\ \bibnamefont
  {Wu}}, \bibinfo {author} {\bibfnamefont {Yue}\ \bibnamefont {Shao}}, \bibinfo
  {author} {\bibfnamefont {Shang-Jie}\ \bibnamefont {Jin}}, \ and\ \bibinfo
  {author} {\bibfnamefont {Xin}\ \bibnamefont {Zhang}},\ }\bibfield  {title}
  {\enquote {\bibinfo {title} {{A path to precision cosmology: synergy between
  four promising late-universe cosmological probes}},}\ }\href {\doibase
  10.1088/1475-7516/2023/06/052} {\bibfield  {journal} {\bibinfo  {journal}
  {JCAP}\ }\textbf {\bibinfo {volume} {06}},\ \bibinfo {pages} {052} (\bibinfo
  {year} {2023}{\natexlab{a}})},\ \Eprint {http://arxiv.org/abs/2202.09726}
  {arXiv:2202.09726 [astro-ph.CO]} \BibitemShut {NoStop}%
\bibitem [{\citenamefont {Jin}\ \emph {et~al.}(2024)\citenamefont {Jin},
  \citenamefont {Zhang}, \citenamefont {Song}, \citenamefont {Zhang},\ and\
  \citenamefont {Zhang}}]{Jin:2023sfc}%
  \BibitemOpen
  \bibfield  {author} {\bibinfo {author} {\bibfnamefont {Shang-Jie}\
  \bibnamefont {Jin}}, \bibinfo {author} {\bibfnamefont {Ye-Zhu}\ \bibnamefont
  {Zhang}}, \bibinfo {author} {\bibfnamefont {Ji-Yu}\ \bibnamefont {Song}},
  \bibinfo {author} {\bibfnamefont {Jing-Fei}\ \bibnamefont {Zhang}}, \ and\
  \bibinfo {author} {\bibfnamefont {Xin}\ \bibnamefont {Zhang}},\ }\bibfield
  {title} {\enquote {\bibinfo {title} {{Taiji-TianQin-LISA network: Precisely
  measuring the Hubble constant using both bright and dark sirens}},}\ }\href
  {\doibase 10.1007/s11433-023-2276-1} {\bibfield  {journal} {\bibinfo
  {journal} {Sci. China Phys. Mech. Astron.}\ }\textbf {\bibinfo {volume}
  {67}},\ \bibinfo {pages} {220412} (\bibinfo {year} {2024})},\ \Eprint
  {http://arxiv.org/abs/2305.19714} {arXiv:2305.19714 [astro-ph.CO]}
  \BibitemShut {NoStop}%
\bibitem [{\citenamefont {Wang}\ \emph {et~al.}(2022)\citenamefont {Wang},
  \citenamefont {Jin}, \citenamefont {Zhang},\ and\ \citenamefont
  {Zhang}}]{Wang:2021srv}%
  \BibitemOpen
  \bibfield  {author} {\bibinfo {author} {\bibfnamefont {Ling-Feng}\
  \bibnamefont {Wang}}, \bibinfo {author} {\bibfnamefont {Shang-Jie}\
  \bibnamefont {Jin}}, \bibinfo {author} {\bibfnamefont {Jing-Fei}\
  \bibnamefont {Zhang}}, \ and\ \bibinfo {author} {\bibfnamefont {Xin}\
  \bibnamefont {Zhang}},\ }\bibfield  {title} {\enquote {\bibinfo {title}
  {{Forecast for cosmological parameter estimation with gravitational-wave
  standard sirens from the LISA-Taiji network}},}\ }\href {\doibase
  10.1007/s11433-021-1736-6} {\bibfield  {journal} {\bibinfo  {journal} {Sci.
  China Phys. Mech. Astron.}\ }\textbf {\bibinfo {volume} {65}},\ \bibinfo
  {pages} {210411} (\bibinfo {year} {2022})},\ \Eprint
  {http://arxiv.org/abs/2101.11882} {arXiv:2101.11882 [gr-qc]} \BibitemShut
  {NoStop}%
\bibitem [{\citenamefont {Xiao}\ \emph {et~al.}(2025)\citenamefont {Xiao},
  \citenamefont {Song}, \citenamefont {Shao}, \citenamefont {Wang},
  \citenamefont {Zhang},\ and\ \citenamefont {Zhang}}]{Xiao:2025mcg}%
  \BibitemOpen
  \bibfield  {author} {\bibinfo {author} {\bibfnamefont {Si-Ren}\ \bibnamefont
  {Xiao}}, \bibinfo {author} {\bibfnamefont {Ji-Yu}\ \bibnamefont {Song}},
  \bibinfo {author} {\bibfnamefont {Yue}\ \bibnamefont {Shao}}, \bibinfo
  {author} {\bibfnamefont {Ling-Feng}\ \bibnamefont {Wang}}, \bibinfo {author}
  {\bibfnamefont {Jing-Fei}\ \bibnamefont {Zhang}}, \ and\ \bibinfo {author}
  {\bibfnamefont {Xin}\ \bibnamefont {Zhang}},\ }\bibfield  {title} {\enquote
  {\bibinfo {title} {{Efficient pulsar distance measurement with multiple
  nanohertz gravitational-wave sources}},}\ }\href@noop {} {\  (\bibinfo {year}
  {2025})},\ \Eprint {http://arxiv.org/abs/2512.10729} {arXiv:2512.10729
  [gr-qc]} \BibitemShut {NoStop}%
\bibitem [{\citenamefont {Du}\ \emph {et~al.}(2025)\citenamefont {Du},
  \citenamefont {Song}, \citenamefont {Li}, \citenamefont {Jin}, \citenamefont
  {Wang}, \citenamefont {Zhang},\ and\ \citenamefont {Zhang}}]{Du:2025odq}%
  \BibitemOpen
  \bibfield  {author} {\bibinfo {author} {\bibfnamefont {Ya-Nan}\ \bibnamefont
  {Du}}, \bibinfo {author} {\bibfnamefont {Ji-Yu}\ \bibnamefont {Song}},
  \bibinfo {author} {\bibfnamefont {Yichao}\ \bibnamefont {Li}}, \bibinfo
  {author} {\bibfnamefont {Shang-Jie}\ \bibnamefont {Jin}}, \bibinfo {author}
  {\bibfnamefont {Ling-Feng}\ \bibnamefont {Wang}}, \bibinfo {author}
  {\bibfnamefont {Jing-Fei}\ \bibnamefont {Zhang}}, \ and\ \bibinfo {author}
  {\bibfnamefont {Xin}\ \bibnamefont {Zhang}},\ }\bibfield  {title} {\enquote
  {\bibinfo {title} {{Synergy between CSST and third-generation
  gravitational-wave detectors: Inferring cosmological parameters using
  cross-correlation of dark sirens and galaxies}},}\ }\href@noop {} {\
  (\bibinfo {year} {2025})},\ \Eprint {http://arxiv.org/abs/2510.21521}
  {arXiv:2510.21521 [astro-ph.CO]} \BibitemShut {NoStop}%
\bibitem [{\citenamefont {Hirata}\ \emph {et~al.}(2010)\citenamefont {Hirata},
  \citenamefont {Holz},\ and\ \citenamefont {Cutler}}]{Hirata:2010ba}%
  \BibitemOpen
  \bibfield  {author} {\bibinfo {author} {\bibfnamefont {Christopher~M.}\
  \bibnamefont {Hirata}}, \bibinfo {author} {\bibfnamefont {Daniel~E.}\
  \bibnamefont {Holz}}, \ and\ \bibinfo {author} {\bibfnamefont {Curt}\
  \bibnamefont {Cutler}},\ }\bibfield  {title} {\enquote {\bibinfo {title}
  {{Reducing the weak lensing noise for the gravitational wave Hubble diagram
  using the non-Gaussianity of the magnification distribution}},}\ }\href
  {\doibase 10.1103/PhysRevD.81.124046} {\bibfield  {journal} {\bibinfo
  {journal} {Phys. Rev. D}\ }\textbf {\bibinfo {volume} {81}},\ \bibinfo
  {pages} {124046} (\bibinfo {year} {2010})},\ \Eprint
  {http://arxiv.org/abs/1004.3988} {arXiv:1004.3988 [astro-ph.CO]} \BibitemShut
  {NoStop}%
\bibitem [{\citenamefont {Gordon}\ \emph {et~al.}(2007)\citenamefont {Gordon},
  \citenamefont {Land},\ and\ \citenamefont {Slosar}}]{Gordon:2007zw}%
  \BibitemOpen
  \bibfield  {author} {\bibinfo {author} {\bibfnamefont {Christopher}\
  \bibnamefont {Gordon}}, \bibinfo {author} {\bibfnamefont {Kate}\ \bibnamefont
  {Land}}, \ and\ \bibinfo {author} {\bibfnamefont {Anze}\ \bibnamefont
  {Slosar}},\ }\bibfield  {title} {\enquote {\bibinfo {title} {{Cosmological
  Constraints from Type Ia Supernovae Peculiar Velocity Measurements}},}\
  }\href {\doibase 10.1103/PhysRevLett.99.081301} {\bibfield  {journal}
  {\bibinfo  {journal} {Phys. Rev. Lett.}\ }\textbf {\bibinfo {volume} {99}},\
  \bibinfo {pages} {081301} (\bibinfo {year} {2007})},\ \Eprint
  {http://arxiv.org/abs/0705.1718} {arXiv:0705.1718 [astro-ph]} \BibitemShut
  {NoStop}%
\bibitem [{\citenamefont {Cai}\ and\ \citenamefont {Yang}(2017)}]{Cai:2016sby}%
  \BibitemOpen
  \bibfield  {author} {\bibinfo {author} {\bibfnamefont {Rong-Gen}\
  \bibnamefont {Cai}}\ and\ \bibinfo {author} {\bibfnamefont {Tao}\
  \bibnamefont {Yang}},\ }\bibfield  {title} {\enquote {\bibinfo {title}
  {{Estimating cosmological parameters by the simulated data of gravitational
  waves from the Einstein Telescope}},}\ }\href {\doibase
  10.1103/PhysRevD.95.044024} {\bibfield  {journal} {\bibinfo  {journal} {Phys.
  Rev. D}\ }\textbf {\bibinfo {volume} {95}},\ \bibinfo {pages} {044024}
  (\bibinfo {year} {2017})},\ \Eprint {http://arxiv.org/abs/1608.08008}
  {arXiv:1608.08008 [astro-ph.CO]} \BibitemShut {NoStop}%
\bibitem [{\citenamefont {Ivezi{\'c}}\ \emph {et~al.}(2019)\citenamefont
  {Ivezi{\'c}} \emph {et~al.}}]{LSST:2008ijt}%
  \BibitemOpen
  \bibfield  {author} {\bibinfo {author} {\bibfnamefont {{\v{Z}}eljko}\
  \bibnamefont {Ivezi{\'c}}} \emph {et~al.} (\bibinfo {collaboration} {LSST}),\
  }\bibfield  {title} {\enquote {\bibinfo {title} {{LSST: from Science Drivers
  to Reference Design and Anticipated Data Products}},}\ }\href {\doibase
  10.3847/1538-4357/ab042c} {\bibfield  {journal} {\bibinfo  {journal}
  {Astrophys. J.}\ }\textbf {\bibinfo {volume} {873}},\ \bibinfo {pages} {111}
  (\bibinfo {year} {2019})},\ \Eprint {http://arxiv.org/abs/0805.2366}
  {arXiv:0805.2366 [astro-ph]} \BibitemShut {NoStop}%
\bibitem [{\citenamefont {Collett}(2015)}]{Collett:2015roa}%
  \BibitemOpen
  \bibfield  {author} {\bibinfo {author} {\bibfnamefont {Thomas~E}\
  \bibnamefont {Collett}},\ }\bibfield  {title} {\enquote {\bibinfo {title}
  {{The population of galaxy-galaxy strong lenses in forthcoming optical
  imaging surveys}},}\ }\href {\doibase 10.1088/0004-637X/811/1/20} {\bibfield
  {journal} {\bibinfo  {journal} {Astrophys. J.}\ }\textbf {\bibinfo {volume}
  {811}},\ \bibinfo {pages} {20} (\bibinfo {year} {2015})},\ \Eprint
  {http://arxiv.org/abs/1507.02657} {arXiv:1507.02657 [astro-ph.CO]}
  \BibitemShut {NoStop}%
\bibitem [{\citenamefont {Qi}\ \emph {et~al.}(2019)\citenamefont {Qi},
  \citenamefont {Cao}, \citenamefont {Biesiada}, \citenamefont {Ding},
  \citenamefont {Zhu},\ and\ \citenamefont {Zheng}}]{Qi:2018atg}%
  \BibitemOpen
  \bibfield  {author} {\bibinfo {author} {\bibfnamefont {Jingzhao}\
  \bibnamefont {Qi}}, \bibinfo {author} {\bibfnamefont {Shuo}\ \bibnamefont
  {Cao}}, \bibinfo {author} {\bibfnamefont {Marek}\ \bibnamefont {Biesiada}},
  \bibinfo {author} {\bibfnamefont {Xuheng}\ \bibnamefont {Ding}}, \bibinfo
  {author} {\bibfnamefont {Zong-Hong}\ \bibnamefont {Zhu}}, \ and\ \bibinfo
  {author} {\bibfnamefont {Xiaogang}\ \bibnamefont {Zheng}},\ }\bibfield
  {title} {\enquote {\bibinfo {title} {{Strongly gravitationally lensed type Ia
  supernovae: Direct test of the Friedman-Lema{\^\i}tre-Robertson-Walker
  metric}},}\ }\href {\doibase 10.1103/PhysRevD.100.023530} {\bibfield
  {journal} {\bibinfo  {journal} {Phys. Rev. D}\ }\textbf {\bibinfo {volume}
  {100}},\ \bibinfo {pages} {023530} (\bibinfo {year} {2019})},\ \Eprint
  {http://arxiv.org/abs/1802.05532} {arXiv:1802.05532 [astro-ph.CO]}
  \BibitemShut {NoStop}%
\bibitem [{\citenamefont {Newburgh}\ \emph {et~al.}(2016)\citenamefont
  {Newburgh} \emph {et~al.}}]{Newburgh:2016mwi}%
  \BibitemOpen
  \bibfield  {author} {\bibinfo {author} {\bibfnamefont {L.~B.}\ \bibnamefont
  {Newburgh}} \emph {et~al.},\ }\bibfield  {title} {\enquote {\bibinfo {title}
  {{HIRAX: A Probe of Dark Energy and Radio Transients}},}\ }\href {\doibase
  10.1117/12.2234286} {\bibfield  {journal} {\bibinfo  {journal} {Proc. SPIE
  Int. Soc. Opt. Eng.}\ }\textbf {\bibinfo {volume} {9906}},\ \bibinfo {pages}
  {99065X} (\bibinfo {year} {2016})},\ \Eprint
  {http://arxiv.org/abs/1607.02059} {arXiv:1607.02059 [astro-ph.IM]}
  \BibitemShut {NoStop}%
\bibitem [{\citenamefont {Lewis}\ \emph {et~al.}(2000)\citenamefont {Lewis},
  \citenamefont {Challinor},\ and\ \citenamefont {Lasenby}}]{Lewis:1999bs}%
  \BibitemOpen
  \bibfield  {author} {\bibinfo {author} {\bibfnamefont {Antony}\ \bibnamefont
  {Lewis}}, \bibinfo {author} {\bibfnamefont {Anthony}\ \bibnamefont
  {Challinor}}, \ and\ \bibinfo {author} {\bibfnamefont {Anthony}\ \bibnamefont
  {Lasenby}},\ }\bibfield  {title} {\enquote {\bibinfo {title} {{Efficient
  computation of CMB anisotropies in closed FRW models}},}\ }\href {\doibase
  10.1086/309179} {\bibfield  {journal} {\bibinfo  {journal} {Astrophys. J.}\
  }\textbf {\bibinfo {volume} {538}},\ \bibinfo {pages} {473--476} (\bibinfo
  {year} {2000})},\ \Eprint {http://arxiv.org/abs/astro-ph/9911177}
  {arXiv:astro-ph/9911177} \BibitemShut {NoStop}%
\bibitem [{\citenamefont {Santos}\ \emph {et~al.}(2005)\citenamefont {Santos},
  \citenamefont {Cooray},\ and\ \citenamefont {Knox}}]{Santos:2004ju}%
  \BibitemOpen
  \bibfield  {author} {\bibinfo {author} {\bibfnamefont {Mario~G.}\
  \bibnamefont {Santos}}, \bibinfo {author} {\bibfnamefont {Asantha}\
  \bibnamefont {Cooray}}, \ and\ \bibinfo {author} {\bibfnamefont {Lloyd}\
  \bibnamefont {Knox}},\ }\bibfield  {title} {\enquote {\bibinfo {title}
  {{Multifrequency analysis of 21 cm fluctuations from the era of
  reionization}},}\ }\href {\doibase 10.1086/429857} {\bibfield  {journal}
  {\bibinfo  {journal} {Astrophys. J.}\ }\textbf {\bibinfo {volume} {625}},\
  \bibinfo {pages} {575--587} (\bibinfo {year} {2005})},\ \Eprint
  {http://arxiv.org/abs/astro-ph/0408515} {arXiv:astro-ph/0408515} \BibitemShut
  {NoStop}%
\bibitem [{\citenamefont {Masui}\ and\ \citenamefont
  {Sigurdson}(2015)}]{Masui:2015ola}%
  \BibitemOpen
  \bibfield  {author} {\bibinfo {author} {\bibfnamefont {Kiyoshi~Wesley}\
  \bibnamefont {Masui}}\ and\ \bibinfo {author} {\bibfnamefont {Kris}\
  \bibnamefont {Sigurdson}},\ }\bibfield  {title} {\enquote {\bibinfo {title}
  {{Dispersion Distance and the Matter Distribution of the Universe in
  Dispersion Space}},}\ }\href {\doibase 10.1103/PhysRevLett.115.121301}
  {\bibfield  {journal} {\bibinfo  {journal} {Phys. Rev. Lett.}\ }\textbf
  {\bibinfo {volume} {115}},\ \bibinfo {pages} {121301} (\bibinfo {year}
  {2015})},\ \Eprint {http://arxiv.org/abs/1506.01704} {arXiv:1506.01704
  [astro-ph.CO]} \BibitemShut {NoStop}%
\bibitem [{\citenamefont {Maity}(2024)}]{Maity:2024zwa}%
  \BibitemOpen
  \bibfield  {author} {\bibinfo {author} {\bibfnamefont {Barun}\ \bibnamefont
  {Maity}},\ }\bibfield  {title} {\enquote {\bibinfo {title} {{Prospects of
  fast radio bursts and large-scale 21 cm power spectra in constraining the
  Epoch of Reionization}},}\ }\href {\doibase 10.1051/0004-6361/202451160}
  {\bibfield  {journal} {\bibinfo  {journal} {Astron. Astrophys.}\ }\textbf
  {\bibinfo {volume} {689}},\ \bibinfo {pages} {A340} (\bibinfo {year}
  {2024})},\ \Eprint {http://arxiv.org/abs/2408.05722} {arXiv:2408.05722
  [astro-ph.CO]} \BibitemShut {NoStop}%
\bibitem [{\citenamefont {Mingarelli}\ \emph {et~al.}(2015)\citenamefont
  {Mingarelli}, \citenamefont {Levin},\ and\ \citenamefont
  {Lazio}}]{Mingarelli:2015bpo}%
  \BibitemOpen
  \bibfield  {author} {\bibinfo {author} {\bibfnamefont {Chiara M.~F.}\
  \bibnamefont {Mingarelli}}, \bibinfo {author} {\bibfnamefont {Janna}\
  \bibnamefont {Levin}}, \ and\ \bibinfo {author} {\bibfnamefont
  {T.~Joseph~W.}\ \bibnamefont {Lazio}},\ }\bibfield  {title} {\enquote
  {\bibinfo {title} {{Fast Radio Bursts and Radio Transients from Black Hole
  Batteries}},}\ }\href {\doibase 10.1088/2041-8205/814/2/L20} {\bibfield
  {journal} {\bibinfo  {journal} {Astrophys. J. Lett.}\ }\textbf {\bibinfo
  {volume} {814}},\ \bibinfo {pages} {L20} (\bibinfo {year} {2015})},\ \Eprint
  {http://arxiv.org/abs/1511.02870} {arXiv:1511.02870 [astro-ph.HE]}
  \BibitemShut {NoStop}%
\bibitem [{\citenamefont {Wang}\ \emph {et~al.}(2016)\citenamefont {Wang},
  \citenamefont {Yang}, \citenamefont {Wu}, \citenamefont {Dai},\ and\
  \citenamefont {Wang}}]{Wang:2016dgs}%
  \BibitemOpen
  \bibfield  {author} {\bibinfo {author} {\bibfnamefont {Jie-Shuang}\
  \bibnamefont {Wang}}, \bibinfo {author} {\bibfnamefont {Yuan-Pei}\
  \bibnamefont {Yang}}, \bibinfo {author} {\bibfnamefont {Xue-Feng}\
  \bibnamefont {Wu}}, \bibinfo {author} {\bibfnamefont {Zi-Gao}\ \bibnamefont
  {Dai}}, \ and\ \bibinfo {author} {\bibfnamefont {Fa-Yin}\ \bibnamefont
  {Wang}},\ }\bibfield  {title} {\enquote {\bibinfo {title} {{Fast Radio Bursts
  from the Inspiral of Double Neutron Stars}},}\ }\href {\doibase
  10.3847/2041-8205/822/1/L7} {\bibfield  {journal} {\bibinfo  {journal}
  {Astrophys. J. Lett.}\ }\textbf {\bibinfo {volume} {822}},\ \bibinfo {pages}
  {L7} (\bibinfo {year} {2016})},\ \Eprint {http://arxiv.org/abs/1603.02014}
  {arXiv:1603.02014 [astro-ph.HE]} \BibitemShut {NoStop}%
\bibitem [{\citenamefont {Margalit}\ \emph {et~al.}(2019)\citenamefont
  {Margalit}, \citenamefont {Berger},\ and\ \citenamefont
  {Metzger}}]{Margalit:2019hke}%
  \BibitemOpen
  \bibfield  {author} {\bibinfo {author} {\bibfnamefont {Ben}\ \bibnamefont
  {Margalit}}, \bibinfo {author} {\bibfnamefont {Edo}\ \bibnamefont {Berger}},
  \ and\ \bibinfo {author} {\bibfnamefont {Brian~D.}\ \bibnamefont {Metzger}},\
  }\bibfield  {title} {\enquote {\bibinfo {title} {{Fast Radio Bursts from
  Magnetars Born in Binary Neutron Star Mergers and Accretion Induced
  Collapse}},}\ }\href {\doibase 10.3847/1538-4357/ab4c31} {\  (\bibinfo {year}
  {2019}),\ 10.3847/1538-4357/ab4c31},\ \Eprint
  {http://arxiv.org/abs/1907.00016} {arXiv:1907.00016 [astro-ph.HE]}
  \BibitemShut {NoStop}%
\bibitem [{\citenamefont {Most}\ and\ \citenamefont
  {Philippov}(2023)}]{Most:2023unc}%
  \BibitemOpen
  \bibfield  {author} {\bibinfo {author} {\bibfnamefont {Elias~R.}\
  \bibnamefont {Most}}\ and\ \bibinfo {author} {\bibfnamefont {Alexander~A.}\
  \bibnamefont {Philippov}},\ }\bibfield  {title} {\enquote {\bibinfo {title}
  {{Electromagnetic Precursors to Black Hole{\textendash}Neutron Star
  Gravitational Wave Events: Flares and Reconnection-powered Fast Radio
  Transients from the Late Inspiral}},}\ }\href {\doibase
  10.3847/2041-8213/acfdae} {\bibfield  {journal} {\bibinfo  {journal}
  {Astrophys. J. Lett.}\ }\textbf {\bibinfo {volume} {956}},\ \bibinfo {pages}
  {L33} (\bibinfo {year} {2023})},\ \Eprint {http://arxiv.org/abs/2309.04271}
  {arXiv:2309.04271 [astro-ph.HE]} \BibitemShut {NoStop}%
\bibitem [{\citenamefont {Zhang}(2024)}]{Zhang:2024rqu}%
  \BibitemOpen
  \bibfield  {author} {\bibinfo {author} {\bibfnamefont {Bing}\ \bibnamefont
  {Zhang}},\ }\bibfield  {title} {\enquote {\bibinfo {title} {{Multiwavelength
  and Multimessenger Counterparts of Fast Radio Bursts}},}\ }\href {\doibase
  10.1146/annurev-nucl-102020-124444} {\bibfield  {journal} {\bibinfo
  {journal} {Ann. Rev. Nucl. Part. Sci.}\ }\textbf {\bibinfo {volume} {74}},\
  \bibinfo {pages} {89--112} (\bibinfo {year} {2024})},\ \Eprint
  {http://arxiv.org/abs/2410.02216} {arXiv:2410.02216 [astro-ph.HE]}
  \BibitemShut {NoStop}%
\bibitem [{\citenamefont {Li}\ \emph {et~al.}(2018)\citenamefont {Li},
  \citenamefont {Gao}, \citenamefont {Ding}, \citenamefont {Wang},\ and\
  \citenamefont {Zhang}}]{Li:2017mek}%
  \BibitemOpen
  \bibfield  {author} {\bibinfo {author} {\bibfnamefont {Zheng-Xiang}\
  \bibnamefont {Li}}, \bibinfo {author} {\bibfnamefont {He}~\bibnamefont
  {Gao}}, \bibinfo {author} {\bibfnamefont {Xu-Heng}\ \bibnamefont {Ding}},
  \bibinfo {author} {\bibfnamefont {Guo-Jian}\ \bibnamefont {Wang}}, \ and\
  \bibinfo {author} {\bibfnamefont {Bing}\ \bibnamefont {Zhang}},\ }\bibfield
  {title} {\enquote {\bibinfo {title} {{Strongly lensed repeating fast radio
  bursts as precision probes of the universe}},}\ }\href {\doibase
  10.1038/s41467-018-06303-0} {\bibfield  {journal} {\bibinfo  {journal}
  {Nature Commun.}\ }\textbf {\bibinfo {volume} {9}},\ \bibinfo {pages} {3833}
  (\bibinfo {year} {2018})},\ \Eprint {http://arxiv.org/abs/1708.06357}
  {arXiv:1708.06357 [astro-ph.CO]} \BibitemShut {NoStop}%
\bibitem [{\citenamefont {Liu}\ \emph {et~al.}(2019)\citenamefont {Liu},
  \citenamefont {Li}, \citenamefont {Gao},\ and\ \citenamefont
  {Zhu}}]{Liu:2019jka}%
  \BibitemOpen
  \bibfield  {author} {\bibinfo {author} {\bibfnamefont {Bin}\ \bibnamefont
  {Liu}}, \bibinfo {author} {\bibfnamefont {Zhengxiang}\ \bibnamefont {Li}},
  \bibinfo {author} {\bibfnamefont {He}~\bibnamefont {Gao}}, \ and\ \bibinfo
  {author} {\bibfnamefont {Zong-Hong}\ \bibnamefont {Zhu}},\ }\bibfield
  {title} {\enquote {\bibinfo {title} {{Prospects of strongly lensed repeating
  fast radio bursts: Complementary constraints on dark energy evolution}},}\
  }\href {\doibase 10.1103/PhysRevD.99.123517} {\bibfield  {journal} {\bibinfo
  {journal} {Phys. Rev. D}\ }\textbf {\bibinfo {volume} {99}},\ \bibinfo
  {pages} {123517} (\bibinfo {year} {2019})},\ \Eprint
  {http://arxiv.org/abs/1907.10488} {arXiv:1907.10488 [astro-ph.CO]}
  \BibitemShut {NoStop}%
\bibitem [{\citenamefont {Wucknitz}\ \emph {et~al.}(2021)\citenamefont
  {Wucknitz}, \citenamefont {Spitler},\ and\ \citenamefont
  {Pen}}]{Wucknitz:2020spz}%
  \BibitemOpen
  \bibfield  {author} {\bibinfo {author} {\bibfnamefont {O.}~\bibnamefont
  {Wucknitz}}, \bibinfo {author} {\bibfnamefont {L.~G.}\ \bibnamefont
  {Spitler}}, \ and\ \bibinfo {author} {\bibfnamefont {U.~L.}\ \bibnamefont
  {Pen}},\ }\bibfield  {title} {\enquote {\bibinfo {title} {{Cosmology with
  gravitationally lensed repeating Fast Radio Bursts}},}\ }\href {\doibase
  10.1051/0004-6361/202038248} {\bibfield  {journal} {\bibinfo  {journal}
  {Astron. Astrophys.}\ }\textbf {\bibinfo {volume} {645}},\ \bibinfo {pages}
  {A44} (\bibinfo {year} {2021})},\ \Eprint {http://arxiv.org/abs/2004.11643}
  {arXiv:2004.11643 [astro-ph.CO]} \BibitemShut {NoStop}%
\bibitem [{\citenamefont {Zhao}\ \emph {et~al.}(2021)\citenamefont {Zhao},
  \citenamefont {Liu}, \citenamefont {Li},\ and\ \citenamefont
  {Gao}}]{Zhao:2021jeb}%
  \BibitemOpen
  \bibfield  {author} {\bibinfo {author} {\bibfnamefont {Shaoxin}\ \bibnamefont
  {Zhao}}, \bibinfo {author} {\bibfnamefont {Bin}\ \bibnamefont {Liu}},
  \bibinfo {author} {\bibfnamefont {Zhengxiang}\ \bibnamefont {Li}}, \ and\
  \bibinfo {author} {\bibfnamefont {He}~\bibnamefont {Gao}},\ }\bibfield
  {title} {\enquote {\bibinfo {title} {{Model-independent Estimation of H 0 and
  {\ensuremath{\Omega}} K from Strongly Lensed Fast Radio Bursts}},}\ }\href
  {\doibase 10.3847/1538-4357/abfa91} {\bibfield  {journal} {\bibinfo
  {journal} {Astrophys. J.}\ }\textbf {\bibinfo {volume} {916}},\ \bibinfo
  {pages} {70} (\bibinfo {year} {2021})}\BibitemShut {NoStop}%
\bibitem [{\citenamefont {Gao}\ \emph {et~al.}(2022)\citenamefont {Gao},
  \citenamefont {Li},\ and\ \citenamefont {Gao}}]{Gao:2022ifq}%
  \BibitemOpen
  \bibfield  {author} {\bibinfo {author} {\bibfnamefont {Ran}\ \bibnamefont
  {Gao}}, \bibinfo {author} {\bibfnamefont {Zhengxiang}\ \bibnamefont {Li}}, \
  and\ \bibinfo {author} {\bibfnamefont {He}~\bibnamefont {Gao}},\ }\bibfield
  {title} {\enquote {\bibinfo {title} {{Prospects of strongly lensed fast radio
  bursts: simultaneous measurement of post-Newtonian parameter and Hubble
  constant}},}\ }\href {\doibase 10.1093/mnras/stac2270} {\bibfield  {journal}
  {\bibinfo  {journal} {Mon. Not. Roy. Astron. Soc.}\ }\textbf {\bibinfo
  {volume} {516}},\ \bibinfo {pages} {1977--1982} (\bibinfo {year} {2022})},\
  \Eprint {http://arxiv.org/abs/2208.10175} {arXiv:2208.10175 [astro-ph.CO]}
  \BibitemShut {NoStop}%
\bibitem [{\citenamefont {Wang}\ \emph {et~al.}(1996)\citenamefont {Wang},
  \citenamefont {Stebbins},\ and\ \citenamefont {Turner}}]{Wang:1996as}%
  \BibitemOpen
  \bibfield  {author} {\bibinfo {author} {\bibfnamefont {Yun}\ \bibnamefont
  {Wang}}, \bibinfo {author} {\bibfnamefont {Albert}\ \bibnamefont {Stebbins}},
  \ and\ \bibinfo {author} {\bibfnamefont {Edwin~L.}\ \bibnamefont {Turner}},\
  }\bibfield  {title} {\enquote {\bibinfo {title} {{Gravitational lensing of
  gravitational waves from merging neutron star binaries}},}\ }\href {\doibase
  10.1103/PhysRevLett.77.2875} {\bibfield  {journal} {\bibinfo  {journal}
  {Phys. Rev. Lett.}\ }\textbf {\bibinfo {volume} {77}},\ \bibinfo {pages}
  {2875--2878} (\bibinfo {year} {1996})},\ \Eprint
  {http://arxiv.org/abs/astro-ph/9605140} {arXiv:astro-ph/9605140} \BibitemShut
  {NoStop}%
\bibitem [{\citenamefont {Nakamura}(1998)}]{Nakamura:1997sw}%
  \BibitemOpen
  \bibfield  {author} {\bibinfo {author} {\bibfnamefont {Takahiro~T.}\
  \bibnamefont {Nakamura}},\ }\bibfield  {title} {\enquote {\bibinfo {title}
  {{Gravitational lensing of gravitational waves from inspiraling binaries by a
  point mass lens}},}\ }\href {\doibase 10.1103/PhysRevLett.80.1138} {\bibfield
   {journal} {\bibinfo  {journal} {Phys. Rev. Lett.}\ }\textbf {\bibinfo
  {volume} {80}},\ \bibinfo {pages} {1138--1141} (\bibinfo {year}
  {1998})}\BibitemShut {NoStop}%
\bibitem [{\citenamefont {Liao}\ \emph {et~al.}(2017)\citenamefont {Liao},
  \citenamefont {Fan}, \citenamefont {Ding}, \citenamefont {Biesiada},\ and\
  \citenamefont {Zhu}}]{Liao:2017ioi}%
  \BibitemOpen
  \bibfield  {author} {\bibinfo {author} {\bibfnamefont {Kai}\ \bibnamefont
  {Liao}}, \bibinfo {author} {\bibfnamefont {Xi-Long}\ \bibnamefont {Fan}},
  \bibinfo {author} {\bibfnamefont {Xu-Heng}\ \bibnamefont {Ding}}, \bibinfo
  {author} {\bibfnamefont {Marek}\ \bibnamefont {Biesiada}}, \ and\ \bibinfo
  {author} {\bibfnamefont {Zong-Hong}\ \bibnamefont {Zhu}},\ }\bibfield
  {title} {\enquote {\bibinfo {title} {{Precision cosmology from future lensed
  gravitational wave and electromagnetic signals}},}\ }\href {\doibase
  10.1038/s41467-017-01152-9} {\bibfield  {journal} {\bibinfo  {journal}
  {Nature Commun.}\ }\textbf {\bibinfo {volume} {8}},\ \bibinfo {pages} {1148}
  (\bibinfo {year} {2017})},\ \bibinfo {note} {[Erratum: Nature Commun. 8, 2136
  (2017)]},\ \Eprint {http://arxiv.org/abs/1703.04151} {arXiv:1703.04151
  [astro-ph.CO]} \BibitemShut {NoStop}%
\bibitem [{\citenamefont {Diego}\ \emph {et~al.}(2021)\citenamefont {Diego},
  \citenamefont {Broadhurst},\ and\ \citenamefont {Smoot}}]{Diego:2021fyd}%
  \BibitemOpen
  \bibfield  {author} {\bibinfo {author} {\bibfnamefont {Jose~M.}\ \bibnamefont
  {Diego}}, \bibinfo {author} {\bibfnamefont {Tom}\ \bibnamefont {Broadhurst}},
  \ and\ \bibinfo {author} {\bibfnamefont {George}\ \bibnamefont {Smoot}},\
  }\bibfield  {title} {\enquote {\bibinfo {title} {{Evidence for lensing of
  gravitational waves from LIGO-Virgo data}},}\ }\href {\doibase
  10.1103/PhysRevD.104.103529} {\bibfield  {journal} {\bibinfo  {journal}
  {Phys. Rev. D}\ }\textbf {\bibinfo {volume} {104}},\ \bibinfo {pages}
  {103529} (\bibinfo {year} {2021})},\ \Eprint
  {http://arxiv.org/abs/2106.06545} {arXiv:2106.06545 [gr-qc]} \BibitemShut
  {NoStop}%
\bibitem [{\citenamefont {Wei}\ and\ \citenamefont {Wu}(2017)}]{Wei:2017emo}%
  \BibitemOpen
  \bibfield  {author} {\bibinfo {author} {\bibfnamefont {Jun-Jie}\ \bibnamefont
  {Wei}}\ and\ \bibinfo {author} {\bibfnamefont {Xue-Feng}\ \bibnamefont
  {Wu}},\ }\bibfield  {title} {\enquote {\bibinfo {title} {{Strongly Lensed
  Gravitational Waves and Electromagnetic Signals as Powerful Cosmic
  Rulers}},}\ }\href {\doibase 10.1093/mnras/stx2210} {\bibfield  {journal}
  {\bibinfo  {journal} {Mon. Not. Roy. Astron. Soc.}\ }\textbf {\bibinfo
  {volume} {472}},\ \bibinfo {pages} {2906--2912} (\bibinfo {year} {2017})},\
  \Eprint {http://arxiv.org/abs/1707.04152} {arXiv:1707.04152 [astro-ph.CO]}
  \BibitemShut {NoStop}%
\bibitem [{\citenamefont {Huang}\ \emph {et~al.}(2023)\citenamefont {Huang},
  \citenamefont {Hu}, \citenamefont {Chen}, \citenamefont {Zhang},
  \citenamefont {Li}, \citenamefont {Gao},\ and\ \citenamefont
  {Lin}}]{Huang:2023prq}%
  \BibitemOpen
  \bibfield  {author} {\bibinfo {author} {\bibfnamefont {Shun-Jia}\
  \bibnamefont {Huang}}, \bibinfo {author} {\bibfnamefont {Yi-Ming}\
  \bibnamefont {Hu}}, \bibinfo {author} {\bibfnamefont {Xian}\ \bibnamefont
  {Chen}}, \bibinfo {author} {\bibfnamefont {Jian-dong}\ \bibnamefont {Zhang}},
  \bibinfo {author} {\bibfnamefont {En-Kun}\ \bibnamefont {Li}}, \bibinfo
  {author} {\bibfnamefont {Zucheng}\ \bibnamefont {Gao}}, \ and\ \bibinfo
  {author} {\bibfnamefont {Xin-Yi}\ \bibnamefont {Lin}},\ }\bibfield  {title}
  {\enquote {\bibinfo {title} {{Measuring the Hubble constant using strongly
  lensed gravitational wave signals}},}\ }\href {\doibase
  10.1088/1475-7516/2023/08/003} {\bibfield  {journal} {\bibinfo  {journal}
  {JCAP}\ }\textbf {\bibinfo {volume} {08}},\ \bibinfo {pages} {003} (\bibinfo
  {year} {2023})},\ \Eprint {http://arxiv.org/abs/2304.10435} {arXiv:2304.10435
  [astro-ph.CO]} \BibitemShut {NoStop}%
\bibitem [{\citenamefont {Sereno}\ \emph {et~al.}(2011)\citenamefont {Sereno},
  \citenamefont {Jetzer}, \citenamefont {Sesana},\ and\ \citenamefont
  {Volonteri}}]{Sereno:2011ty}%
  \BibitemOpen
  \bibfield  {author} {\bibinfo {author} {\bibfnamefont {M.}~\bibnamefont
  {Sereno}}, \bibinfo {author} {\bibfnamefont {Ph.}\ \bibnamefont {Jetzer}},
  \bibinfo {author} {\bibfnamefont {A.}~\bibnamefont {Sesana}}, \ and\ \bibinfo
  {author} {\bibfnamefont {M.}~\bibnamefont {Volonteri}},\ }\bibfield  {title}
  {\enquote {\bibinfo {title} {{Cosmography with strong lensing of LISA
  gravitational wave sources}},}\ }\href {\doibase
  10.1111/j.1365-2966.2011.18895.x} {\bibfield  {journal} {\bibinfo  {journal}
  {Mon. Not. Roy. Astron. Soc.}\ }\textbf {\bibinfo {volume} {415}},\ \bibinfo
  {pages} {2773} (\bibinfo {year} {2011})},\ \Eprint
  {http://arxiv.org/abs/1104.1977} {arXiv:1104.1977 [astro-ph.CO]} \BibitemShut
  {NoStop}%
\bibitem [{\citenamefont {Zazzera}\ \emph {et~al.}(2025)\citenamefont
  {Zazzera}, \citenamefont {Fonseca}, \citenamefont {Baker},\ and\
  \citenamefont {Clarkson}}]{Zazzera:2025ord}%
  \BibitemOpen
  \bibfield  {author} {\bibinfo {author} {\bibfnamefont {Stefano}\ \bibnamefont
  {Zazzera}}, \bibinfo {author} {\bibfnamefont {Jos{\'e}}\ \bibnamefont
  {Fonseca}}, \bibinfo {author} {\bibfnamefont {Tessa}\ \bibnamefont {Baker}},
  \ and\ \bibinfo {author} {\bibfnamefont {Chris}\ \bibnamefont {Clarkson}},\
  }\bibfield  {title} {\enquote {\bibinfo {title} {{Exploring future synergies
  for large-scale structure between gravitational waves and radio sources}},}\
  }\href {\doibase 10.1093/mnras/stag307} {\  (\bibinfo {year} {2025}),\
  10.1093/mnras/stag307},\ \Eprint {http://arxiv.org/abs/2505.15645}
  {arXiv:2505.15645 [astro-ph.CO]} \BibitemShut {NoStop}%
\bibitem [{\citenamefont {Scelfo}\ \emph {et~al.}(2023)\citenamefont {Scelfo},
  \citenamefont {Berti}, \citenamefont {Silvestri},\ and\ \citenamefont
  {Viel}}]{Scelfo:2022lsx}%
  \BibitemOpen
  \bibfield  {author} {\bibinfo {author} {\bibfnamefont {Giulio}\ \bibnamefont
  {Scelfo}}, \bibinfo {author} {\bibfnamefont {Maria}\ \bibnamefont {Berti}},
  \bibinfo {author} {\bibfnamefont {Alessandra}\ \bibnamefont {Silvestri}}, \
  and\ \bibinfo {author} {\bibfnamefont {Matteo}\ \bibnamefont {Viel}},\
  }\bibfield  {title} {\enquote {\bibinfo {title} {{Testing gravity with
  gravitational waves $\times$ electromagnetic probes cross-correlations}},}\
  }\href {\doibase 10.1088/1475-7516/2023/02/010} {\bibfield  {journal}
  {\bibinfo  {journal} {JCAP}\ }\textbf {\bibinfo {volume} {02}},\ \bibinfo
  {pages} {010} (\bibinfo {year} {2023})},\ \Eprint
  {http://arxiv.org/abs/2210.02460} {arXiv:2210.02460 [astro-ph.CO]}
  \BibitemShut {NoStop}%
\bibitem [{\citenamefont {Scelfo}\ \emph {et~al.}(2022)\citenamefont {Scelfo},
  \citenamefont {Spinelli}, \citenamefont {Raccanelli}, \citenamefont {Boco},
  \citenamefont {Lapi},\ and\ \citenamefont {Viel}}]{Scelfo:2021fqe}%
  \BibitemOpen
  \bibfield  {author} {\bibinfo {author} {\bibfnamefont {Giulio}\ \bibnamefont
  {Scelfo}}, \bibinfo {author} {\bibfnamefont {Marta}\ \bibnamefont
  {Spinelli}}, \bibinfo {author} {\bibfnamefont {Alvise}\ \bibnamefont
  {Raccanelli}}, \bibinfo {author} {\bibfnamefont {Lumen}\ \bibnamefont
  {Boco}}, \bibinfo {author} {\bibfnamefont {Andrea}\ \bibnamefont {Lapi}}, \
  and\ \bibinfo {author} {\bibfnamefont {Matteo}\ \bibnamefont {Viel}},\
  }\bibfield  {title} {\enquote {\bibinfo {title} {{Gravitational waves
  $\times$ HI intensity mapping: cosmological and astrophysical
  applications}},}\ }\href {\doibase 10.1088/1475-7516/2022/01/004} {\bibfield
  {journal} {\bibinfo  {journal} {JCAP}\ }\textbf {\bibinfo {volume} {01}},\
  \bibinfo {pages} {004} (\bibinfo {year} {2022})},\ \Eprint
  {http://arxiv.org/abs/2106.09786} {arXiv:2106.09786 [astro-ph.CO]}
  \BibitemShut {NoStop}%
\bibitem [{\citenamefont {Sangka}\ and\ \citenamefont
  {Bacon}(2024)}]{Sangka:2024vfg}%
  \BibitemOpen
  \bibfield  {author} {\bibinfo {author} {\bibfnamefont {Anut}\ \bibnamefont
  {Sangka}}\ and\ \bibinfo {author} {\bibfnamefont {David}\ \bibnamefont
  {Bacon}},\ }\bibfield  {title} {\enquote {\bibinfo {title} {{The feasibility
  of weak lensing and 21cm intensity mapping cross-correlation
  measurements}},}\ }\href {\doibase 10.1093/mnras/stae1542} {\bibfield
  {journal} {\bibinfo  {journal} {Mon. Not. Roy. Astron. Soc.}\ }\textbf
  {\bibinfo {volume} {532}},\ \bibinfo {pages} {996--1009} (\bibinfo {year}
  {2024})},\ \Eprint {http://arxiv.org/abs/2406.14475} {arXiv:2406.14475
  [astro-ph.CO]} \BibitemShut {NoStop}%
\bibitem [{\citenamefont {Cao}\ \emph {et~al.}(2026)\citenamefont {Cao} \emph
  {et~al.}}]{Euclid:2026ulk}%
  \BibitemOpen
  \bibfield  {author} {\bibinfo {author} {\bibfnamefont {Y.}~\bibnamefont
  {Cao}} \emph {et~al.} (\bibinfo {collaboration} {Euclid}),\ }\bibfield
  {title} {\enquote {\bibinfo {title} {{Euclid preparation. Decomposing
  components of the extragalactic background light using multi-band intensity
  mapping cross-correlations}},}\ }\href@noop {} {\  (\bibinfo {year}
  {2026})},\ \Eprint {http://arxiv.org/abs/2601.21111} {arXiv:2601.21111
  [astro-ph.CO]} \BibitemShut {NoStop}%
\bibitem [{\citenamefont {Chevallier}\ and\ \citenamefont
  {Polarski}(2001)}]{Chevallier:2000qy}%
  \BibitemOpen
  \bibfield  {author} {\bibinfo {author} {\bibfnamefont {Michel}\ \bibnamefont
  {Chevallier}}\ and\ \bibinfo {author} {\bibfnamefont {David}\ \bibnamefont
  {Polarski}},\ }\bibfield  {title} {\enquote {\bibinfo {title} {{Accelerating
  universes with scaling dark matter}},}\ }\href {\doibase
  10.1142/S0218271801000822} {\bibfield  {journal} {\bibinfo  {journal} {Int.
  J. Mod. Phys. D}\ }\textbf {\bibinfo {volume} {10}},\ \bibinfo {pages}
  {213--224} (\bibinfo {year} {2001})},\ \Eprint
  {http://arxiv.org/abs/gr-qc/0009008} {arXiv:gr-qc/0009008} \BibitemShut
  {NoStop}%
\bibitem [{\citenamefont {Linder}(2003)}]{Linder:2002et}%
  \BibitemOpen
  \bibfield  {author} {\bibinfo {author} {\bibfnamefont {Eric~V.}\ \bibnamefont
  {Linder}},\ }\bibfield  {title} {\enquote {\bibinfo {title} {{Exploring the
  expansion history of the universe}},}\ }\href {\doibase
  10.1103/PhysRevLett.90.091301} {\bibfield  {journal} {\bibinfo  {journal}
  {Phys. Rev. Lett.}\ }\textbf {\bibinfo {volume} {90}},\ \bibinfo {pages}
  {091301} (\bibinfo {year} {2003})},\ \Eprint
  {http://arxiv.org/abs/astro-ph/0208512} {arXiv:astro-ph/0208512} \BibitemShut
  {NoStop}%
\bibitem [{\citenamefont {Qi}\ \emph {et~al.}(2022)\citenamefont {Qi},
  \citenamefont {Hu}, \citenamefont {Cui}, \citenamefont {Zhang},\ and\
  \citenamefont {Zhang}}]{Qi:2022kfg}%
  \BibitemOpen
  \bibfield  {author} {\bibinfo {author} {\bibfnamefont {Jing-Zhao}\
  \bibnamefont {Qi}}, \bibinfo {author} {\bibfnamefont {Wei-Hong}\ \bibnamefont
  {Hu}}, \bibinfo {author} {\bibfnamefont {Yu}~\bibnamefont {Cui}}, \bibinfo
  {author} {\bibfnamefont {Jing-Fei}\ \bibnamefont {Zhang}}, \ and\ \bibinfo
  {author} {\bibfnamefont {Xin}\ \bibnamefont {Zhang}},\ }\bibfield  {title}
  {\enquote {\bibinfo {title} {{Cosmological Parameter Estimation Using Current
  and Future Observations of Strong Gravitational Lensing}},}\ }\href {\doibase
  10.3390/universe8050254} {\bibfield  {journal} {\bibinfo  {journal}
  {Universe}\ }\textbf {\bibinfo {volume} {8}},\ \bibinfo {pages} {254}
  (\bibinfo {year} {2022})},\ \Eprint {http://arxiv.org/abs/2203.10862}
  {arXiv:2203.10862 [astro-ph.CO]} \BibitemShut {NoStop}%
\bibitem [{\citenamefont {Gruber}\ and\ \citenamefont
  {Luongo}(2014)}]{Gruber:2013wua}%
  \BibitemOpen
  \bibfield  {author} {\bibinfo {author} {\bibfnamefont {Christine}\
  \bibnamefont {Gruber}}\ and\ \bibinfo {author} {\bibfnamefont {Orlando}\
  \bibnamefont {Luongo}},\ }\bibfield  {title} {\enquote {\bibinfo {title}
  {{Cosmographic analysis of the equation of state of the universe through
  Pad{\'e} approximations}},}\ }\href {\doibase 10.1103/PhysRevD.89.103506}
  {\bibfield  {journal} {\bibinfo  {journal} {Phys. Rev. D}\ }\textbf {\bibinfo
  {volume} {89}},\ \bibinfo {pages} {103506} (\bibinfo {year} {2014})},\
  \Eprint {http://arxiv.org/abs/1309.3215} {arXiv:1309.3215 [gr-qc]}
  \BibitemShut {NoStop}%
\bibitem [{\citenamefont {Wei}\ \emph {et~al.}(2014)\citenamefont {Wei},
  \citenamefont {Yan},\ and\ \citenamefont {Zhou}}]{Wei:2013jya}%
  \BibitemOpen
  \bibfield  {author} {\bibinfo {author} {\bibfnamefont {Hao}\ \bibnamefont
  {Wei}}, \bibinfo {author} {\bibfnamefont {Xiao-Peng}\ \bibnamefont {Yan}}, \
  and\ \bibinfo {author} {\bibfnamefont {Ya-Nan}\ \bibnamefont {Zhou}},\
  }\bibfield  {title} {\enquote {\bibinfo {title} {{Cosmological Applications
  of Pad{\'e} Approximant}},}\ }\href {\doibase 10.1088/1475-7516/2014/01/045}
  {\bibfield  {journal} {\bibinfo  {journal} {JCAP}\ }\textbf {\bibinfo
  {volume} {01}},\ \bibinfo {pages} {045} (\bibinfo {year} {2014})},\ \Eprint
  {http://arxiv.org/abs/1312.1117} {arXiv:1312.1117 [astro-ph.CO]} \BibitemShut
  {NoStop}%
\bibitem [{\citenamefont {Capozziello}\ \emph {et~al.}(2020)\citenamefont
  {Capozziello}, \citenamefont {D'Agostino},\ and\ \citenamefont
  {Luongo}}]{Capozziello:2020ctn}%
  \BibitemOpen
  \bibfield  {author} {\bibinfo {author} {\bibfnamefont {S.}~\bibnamefont
  {Capozziello}}, \bibinfo {author} {\bibfnamefont {R.}~\bibnamefont
  {D'Agostino}}, \ and\ \bibinfo {author} {\bibfnamefont {O.}~\bibnamefont
  {Luongo}},\ }\bibfield  {title} {\enquote {\bibinfo {title} {{High-redshift
  cosmography: auxiliary variables versus Pad{\'e} polynomials}},}\ }\href
  {\doibase 10.1093/mnras/staa871} {\bibfield  {journal} {\bibinfo  {journal}
  {Mon. Not. Roy. Astron. Soc.}\ }\textbf {\bibinfo {volume} {494}},\ \bibinfo
  {pages} {2576--2590} (\bibinfo {year} {2020})},\ \Eprint
  {http://arxiv.org/abs/2003.09341} {arXiv:2003.09341 [astro-ph.CO]}
  \BibitemShut {NoStop}%
\bibitem [{\citenamefont {D'Agostino}\ and\ \citenamefont
  {Nunes}(2023)}]{DAgostino:2023cgx}%
  \BibitemOpen
  \bibfield  {author} {\bibinfo {author} {\bibfnamefont {Rocco}\ \bibnamefont
  {D'Agostino}}\ and\ \bibinfo {author} {\bibfnamefont {Rafael~C.}\
  \bibnamefont {Nunes}},\ }\bibfield  {title} {\enquote {\bibinfo {title}
  {{Cosmographic view on the H0 and {\ensuremath{\sigma}}8 tensions}},}\ }\href
  {\doibase 10.1103/PhysRevD.108.023523} {\bibfield  {journal} {\bibinfo
  {journal} {Phys. Rev. D}\ }\textbf {\bibinfo {volume} {108}},\ \bibinfo
  {pages} {023523} (\bibinfo {year} {2023})},\ \Eprint
  {http://arxiv.org/abs/2307.13464} {arXiv:2307.13464 [astro-ph.CO]}
  \BibitemShut {NoStop}%
\bibitem [{\citenamefont {Seikel}\ \emph {et~al.}(2012)\citenamefont {Seikel},
  \citenamefont {Clarkson},\ and\ \citenamefont {Smith}}]{Seikel:2012uu}%
  \BibitemOpen
  \bibfield  {author} {\bibinfo {author} {\bibfnamefont {Marina}\ \bibnamefont
  {Seikel}}, \bibinfo {author} {\bibfnamefont {Chris}\ \bibnamefont
  {Clarkson}}, \ and\ \bibinfo {author} {\bibfnamefont {Mathew}\ \bibnamefont
  {Smith}},\ }\bibfield  {title} {\enquote {\bibinfo {title} {{Reconstruction
  of dark energy and expansion dynamics using Gaussian processes}},}\ }\href
  {\doibase 10.1088/1475-7516/2012/06/036} {\bibfield  {journal} {\bibinfo
  {journal} {JCAP}\ }\textbf {\bibinfo {volume} {06}},\ \bibinfo {pages} {036}
  (\bibinfo {year} {2012})},\ \Eprint {http://arxiv.org/abs/1204.2832}
  {arXiv:1204.2832 [astro-ph.CO]} \BibitemShut {NoStop}%
\bibitem [{\citenamefont {Wang}\ \emph {et~al.}(2020)\citenamefont {Wang},
  \citenamefont {Ma}, \citenamefont {Li},\ and\ \citenamefont
  {Xia}}]{Wang:2019vxv}%
  \BibitemOpen
  \bibfield  {author} {\bibinfo {author} {\bibfnamefont {Guo-Jian}\
  \bibnamefont {Wang}}, \bibinfo {author} {\bibfnamefont {Xiao-Jiao}\
  \bibnamefont {Ma}}, \bibinfo {author} {\bibfnamefont {Si-Yao}\ \bibnamefont
  {Li}}, \ and\ \bibinfo {author} {\bibfnamefont {Jun-Qing}\ \bibnamefont
  {Xia}},\ }\bibfield  {title} {\enquote {\bibinfo {title} {{Reconstructing
  Functions and Estimating Parameters with Artificial Neural Networks: A Test
  with a Hubble Parameter and SNe Ia}},}\ }\href {\doibase
  10.3847/1538-4365/ab620b} {\bibfield  {journal} {\bibinfo  {journal}
  {Astrophys. J. Suppl.}\ }\textbf {\bibinfo {volume} {246}},\ \bibinfo {pages}
  {13} (\bibinfo {year} {2020})},\ \Eprint {http://arxiv.org/abs/1910.03636}
  {arXiv:1910.03636 [astro-ph.CO]} \BibitemShut {NoStop}%
\bibitem [{\citenamefont {Nesseris}\ and\ \citenamefont
  {Garcia-Bellido}(2012)}]{Nesseris:2012tt}%
  \BibitemOpen
  \bibfield  {author} {\bibinfo {author} {\bibfnamefont {Savvas}\ \bibnamefont
  {Nesseris}}\ and\ \bibinfo {author} {\bibfnamefont {Juan}\ \bibnamefont
  {Garcia-Bellido}},\ }\bibfield  {title} {\enquote {\bibinfo {title} {{A new
  perspective on Dark Energy modeling via Genetic Algorithms}},}\ }\href
  {\doibase 10.1088/1475-7516/2012/11/033} {\bibfield  {journal} {\bibinfo
  {journal} {JCAP}\ }\textbf {\bibinfo {volume} {11}},\ \bibinfo {pages} {033}
  (\bibinfo {year} {2012})},\ \Eprint {http://arxiv.org/abs/1205.0364}
  {arXiv:1205.0364 [astro-ph.CO]} \BibitemShut {NoStop}%
\bibitem [{\citenamefont {Zhu}\ and\ \citenamefont {Feng}(2021)}]{Zhu:2020prt}%
  \BibitemOpen
  \bibfield  {author} {\bibinfo {author} {\bibfnamefont {Weishan}\ \bibnamefont
  {Zhu}}\ and\ \bibinfo {author} {\bibfnamefont {Long-Long}\ \bibnamefont
  {Feng}},\ }\bibfield  {title} {\enquote {\bibinfo {title} {{The Dispersion
  Measure and Scattering of Fast Radio Bursts: Contributions from the
  Intergalactic Medium, Foreground Halos, and Hosts}},}\ }\href {\doibase
  10.3847/1538-4357/abcb90} {\bibfield  {journal} {\bibinfo  {journal}
  {Astrophys. J.}\ }\textbf {\bibinfo {volume} {906}},\ \bibinfo {pages} {95}
  (\bibinfo {year} {2021})},\ \Eprint {http://arxiv.org/abs/2011.08519}
  {arXiv:2011.08519 [astro-ph.HE]} \BibitemShut {NoStop}%
\bibitem [{\citenamefont {Cordes}\ and\ \citenamefont
  {Lazio}(2002)}]{Cordes:2002wz}%
  \BibitemOpen
  \bibfield  {author} {\bibinfo {author} {\bibfnamefont {James~M.}\
  \bibnamefont {Cordes}}\ and\ \bibinfo {author} {\bibfnamefont {T.~J.~W.}\
  \bibnamefont {Lazio}},\ }\bibfield  {title} {\enquote {\bibinfo {title}
  {{NE2001. 1. A New model for the galactic distribution of free electrons and
  its fluctuations}},}\ }\href@noop {} {\  (\bibinfo {year} {2002})},\ \Eprint
  {http://arxiv.org/abs/astro-ph/0207156} {arXiv:astro-ph/0207156} \BibitemShut
  {NoStop}%
\bibitem [{\citenamefont {Yao}\ \emph {et~al.}(2017)\citenamefont {Yao},
  \citenamefont {Manchester},\ and\ \citenamefont {Wang}}]{Yao:2017kcp}%
  \BibitemOpen
  \bibfield  {author} {\bibinfo {author} {\bibfnamefont {J.~M.}\ \bibnamefont
  {Yao}}, \bibinfo {author} {\bibfnamefont {R.~N.}\ \bibnamefont {Manchester}},
  \ and\ \bibinfo {author} {\bibfnamefont {N.}~\bibnamefont {Wang}},\
  }\bibfield  {title} {\enquote {\bibinfo {title} {{A new Electron-density
  Model for Estimation of Pulsar and frb Distances}},}\ }\href {\doibase
  10.3847/1538-4357/835/1/29} {\bibfield  {journal} {\bibinfo  {journal}
  {Astrophys. J.}\ }\textbf {\bibinfo {volume} {835}},\ \bibinfo {pages} {29}
  (\bibinfo {year} {2017})}\BibitemShut {NoStop}%
\bibitem [{\citenamefont {Xue}\ \emph {et~al.}(2023)\citenamefont {Xue},
  \citenamefont {Zhu}, \citenamefont {Wu}, \citenamefont {Xu},\ and\
  \citenamefont {Wang}}]{Xue:2023hix}%
  \BibitemOpen
  \bibfield  {author} {\bibinfo {author} {\bibfnamefont {Mengyao}\ \bibnamefont
  {Xue}}, \bibinfo {author} {\bibfnamefont {Weiwei}\ \bibnamefont {Zhu}},
  \bibinfo {author} {\bibfnamefont {Xiangping}\ \bibnamefont {Wu}}, \bibinfo
  {author} {\bibfnamefont {Renxin}\ \bibnamefont {Xu}}, \ and\ \bibinfo
  {author} {\bibfnamefont {Hongguang}\ \bibnamefont {Wang}},\ }\bibfield
  {title} {\enquote {\bibinfo {title} {{Pulsar Discovery Prospect of FASTA}},}\
  }\href {\doibase 10.1088/1674-4527/acdbbd} {\bibfield  {journal} {\bibinfo
  {journal} {Res. Astron. Astrophys.}\ }\textbf {\bibinfo {volume} {23}},\
  \bibinfo {pages} {095005} (\bibinfo {year} {2023})},\ \Eprint
  {http://arxiv.org/abs/2307.03422} {arXiv:2307.03422 [astro-ph.HE]}
  \BibitemShut {NoStop}%
\bibitem [{\citenamefont {Hallinan}\ \emph {et~al.}(2019)\citenamefont
  {Hallinan} \emph {et~al.}}]{Hallinan:2019qyo}%
  \BibitemOpen
  \bibfield  {author} {\bibinfo {author} {\bibfnamefont {G.}~\bibnamefont
  {Hallinan}} \emph {et~al.},\ }\bibfield  {title} {\enquote {\bibinfo {title}
  {{The DSA-2000 -- A Radio Survey Camera}},}\ }\href@noop {} {\  (\bibinfo
  {year} {2019})},\ \Eprint {http://arxiv.org/abs/1907.07648} {arXiv:1907.07648
  [astro-ph.IM]} \BibitemShut {NoStop}%
\bibitem [{\citenamefont {Vanderlinde}\ \emph {et~al.}(2019)\citenamefont
  {Vanderlinde} \emph {et~al.}}]{Vanderlinde:2019tjt}%
  \BibitemOpen
  \bibfield  {author} {\bibinfo {author} {\bibfnamefont {K.}~\bibnamefont
  {Vanderlinde}} \emph {et~al.},\ }\bibfield  {title} {\enquote {\bibinfo
  {title} {{LRP 2020 Whitepaper: The Canadian Hydrogen Observatory and
  Radio-transient Detector (CHORD)}},}\ }\href {\doibase
  10.5281/zenodo.3765414} {\  (\bibinfo {year} {2019}),\
  10.5281/zenodo.3765414},\ \Eprint {http://arxiv.org/abs/1911.01777}
  {arXiv:1911.01777 [astro-ph.IM]} \BibitemShut {NoStop}%
\bibitem [{\citenamefont {Lin}\ \emph {et~al.}(2022)\citenamefont {Lin} \emph
  {et~al.}}]{Lin:2022wgp}%
  \BibitemOpen
  \bibfield  {author} {\bibinfo {author} {\bibfnamefont {Hsiu-Hsien}\
  \bibnamefont {Lin}} \emph {et~al.},\ }\bibfield  {title} {\enquote {\bibinfo
  {title} {{BURSTT: Bustling Universe Radio Survey Telescope in Taiwan}},}\
  }\href {\doibase 10.1088/1538-3873/ac8f71} {\bibfield  {journal} {\bibinfo
  {journal} {Publ. Astron. Soc. Pac.}\ }\textbf {\bibinfo {volume} {134}},\
  \bibinfo {pages} {094106} (\bibinfo {year} {2022})},\ \Eprint
  {http://arxiv.org/abs/2206.08983} {arXiv:2206.08983 [astro-ph.IM]}
  \BibitemShut {NoStop}%
\bibitem [{\citenamefont {Reitze}\ \emph {et~al.}(2019)\citenamefont {Reitze}
  \emph {et~al.}}]{Reitze:2019iox}%
  \BibitemOpen
  \bibfield  {author} {\bibinfo {author} {\bibfnamefont {David}\ \bibnamefont
  {Reitze}} \emph {et~al.},\ }\bibfield  {title} {\enquote {\bibinfo {title}
  {{Cosmic Explorer: The U.S. Contribution to Gravitational-Wave Astronomy
  beyond LIGO}},}\ }\href@noop {} {\bibfield  {journal} {\bibinfo  {journal}
  {Bull. Am. Astron. Soc.}\ }\textbf {\bibinfo {volume} {51}},\ \bibinfo
  {pages} {035} (\bibinfo {year} {2019})},\ \Eprint
  {http://arxiv.org/abs/1907.04833} {arXiv:1907.04833 [astro-ph.IM]}
  \BibitemShut {NoStop}%
\bibitem [{\citenamefont {Amaro-Seoane}\ \emph {et~al.}(2017)\citenamefont
  {Amaro-Seoane} \emph {et~al.}}]{LISA:2017pwj}%
  \BibitemOpen
  \bibfield  {author} {\bibinfo {author} {\bibfnamefont {Pau}\ \bibnamefont
  {Amaro-Seoane}} \emph {et~al.} (\bibinfo {collaboration} {LISA}),\ }\bibfield
   {title} {\enquote {\bibinfo {title} {{Laser Interferometer Space
  Antenna}},}\ }\href@noop {} {\  (\bibinfo {year} {2017})},\ \Eprint
  {http://arxiv.org/abs/1702.00786} {arXiv:1702.00786 [astro-ph.IM]}
  \BibitemShut {NoStop}%
\bibitem [{\citenamefont {Robson}\ \emph {et~al.}(2019)\citenamefont {Robson},
  \citenamefont {Cornish},\ and\ \citenamefont {Liu}}]{Robson:2018ifk}%
  \BibitemOpen
  \bibfield  {author} {\bibinfo {author} {\bibfnamefont {Travis}\ \bibnamefont
  {Robson}}, \bibinfo {author} {\bibfnamefont {Neil~J.}\ \bibnamefont
  {Cornish}}, \ and\ \bibinfo {author} {\bibfnamefont {Chang}\ \bibnamefont
  {Liu}},\ }\bibfield  {title} {\enquote {\bibinfo {title} {{The construction
  and use of LISA sensitivity curves}},}\ }\href {\doibase
  10.1088/1361-6382/ab1101} {\bibfield  {journal} {\bibinfo  {journal} {Class.
  Quant. Grav.}\ }\textbf {\bibinfo {volume} {36}},\ \bibinfo {pages} {105011}
  (\bibinfo {year} {2019})},\ \Eprint {http://arxiv.org/abs/1803.01944}
  {arXiv:1803.01944 [astro-ph.HE]} \BibitemShut {NoStop}%
\bibitem [{\citenamefont {Auclair}\ \emph {et~al.}(2023)\citenamefont {Auclair}
  \emph {et~al.}}]{LISACosmologyWorkingGroup:2022jok}%
  \BibitemOpen
  \bibfield  {author} {\bibinfo {author} {\bibfnamefont {Pierre}\ \bibnamefont
  {Auclair}} \emph {et~al.} (\bibinfo {collaboration} {LISA Cosmology Working
  Group}),\ }\bibfield  {title} {\enquote {\bibinfo {title} {{Cosmology with
  the Laser Interferometer Space Antenna}},}\ }\href {\doibase
  10.1007/s41114-023-00045-2} {\bibfield  {journal} {\bibinfo  {journal}
  {Living Rev. Rel.}\ }\textbf {\bibinfo {volume} {26}},\ \bibinfo {pages} {5}
  (\bibinfo {year} {2023})},\ \Eprint {http://arxiv.org/abs/2204.05434}
  {arXiv:2204.05434 [astro-ph.CO]} \BibitemShut {NoStop}%
\bibitem [{\citenamefont {Ruan}\ \emph {et~al.}(2020)\citenamefont {Ruan},
  \citenamefont {Guo}, \citenamefont {Cai},\ and\ \citenamefont
  {Zhang}}]{Ruan:2018tsw}%
  \BibitemOpen
  \bibfield  {author} {\bibinfo {author} {\bibfnamefont {Wen-Hong}\
  \bibnamefont {Ruan}}, \bibinfo {author} {\bibfnamefont {Zong-Kuan}\
  \bibnamefont {Guo}}, \bibinfo {author} {\bibfnamefont {Rong-Gen}\
  \bibnamefont {Cai}}, \ and\ \bibinfo {author} {\bibfnamefont {Yuan-Zhong}\
  \bibnamefont {Zhang}},\ }\bibfield  {title} {\enquote {\bibinfo {title}
  {{Taiji program: Gravitational-wave sources}},}\ }\href {\doibase
  10.1142/S0217751X2050075X} {\bibfield  {journal} {\bibinfo  {journal} {Int.
  J. Mod. Phys. A}\ }\textbf {\bibinfo {volume} {35}},\ \bibinfo {pages}
  {2050075} (\bibinfo {year} {2020})},\ \Eprint
  {http://arxiv.org/abs/1807.09495} {arXiv:1807.09495 [gr-qc]} \BibitemShut
  {NoStop}%
\bibitem [{\citenamefont {Luo}\ \emph {et~al.}(2016)\citenamefont {Luo} \emph
  {et~al.}}]{TianQin:2015yph}%
  \BibitemOpen
  \bibfield  {author} {\bibinfo {author} {\bibfnamefont {Jun}\ \bibnamefont
  {Luo}} \emph {et~al.} (\bibinfo {collaboration} {TianQin}),\ }\bibfield
  {title} {\enquote {\bibinfo {title} {{TianQin: a space-borne gravitational
  wave detector}},}\ }\href {\doibase 10.1088/0264-9381/33/3/035010} {\bibfield
   {journal} {\bibinfo  {journal} {Class. Quant. Grav.}\ }\textbf {\bibinfo
  {volume} {33}},\ \bibinfo {pages} {035010} (\bibinfo {year} {2016})},\
  \Eprint {http://arxiv.org/abs/1512.02076} {arXiv:1512.02076 [astro-ph.IM]}
  \BibitemShut {NoStop}%
\bibitem [{\citenamefont {Liu}\ \emph {et~al.}(2020)\citenamefont {Liu},
  \citenamefont {Hu}, \citenamefont {Zhang},\ and\ \citenamefont
  {Mei}}]{Liu:2020eko}%
  \BibitemOpen
  \bibfield  {author} {\bibinfo {author} {\bibfnamefont {Shuai}\ \bibnamefont
  {Liu}}, \bibinfo {author} {\bibfnamefont {Yi-Ming}\ \bibnamefont {Hu}},
  \bibinfo {author} {\bibfnamefont {Jian-dong}\ \bibnamefont {Zhang}}, \ and\
  \bibinfo {author} {\bibfnamefont {Jianwei}\ \bibnamefont {Mei}},\ }\bibfield
  {title} {\enquote {\bibinfo {title} {{Science with the TianQin observatory:
  Preliminary results on stellar-mass binary black holes}},}\ }\href {\doibase
  10.1103/PhysRevD.101.103027} {\bibfield  {journal} {\bibinfo  {journal}
  {Phys. Rev. D}\ }\textbf {\bibinfo {volume} {101}},\ \bibinfo {pages}
  {103027} (\bibinfo {year} {2020})},\ \Eprint
  {http://arxiv.org/abs/2004.14242} {arXiv:2004.14242 [astro-ph.HE]}
  \BibitemShut {NoStop}%
\bibitem [{\citenamefont {Wang}\ \emph {et~al.}(2019)\citenamefont {Wang} \emph
  {et~al.}}]{Wang:2019ryf}%
  \BibitemOpen
  \bibfield  {author} {\bibinfo {author} {\bibfnamefont {Hai-Tian}\
  \bibnamefont {Wang}} \emph {et~al.},\ }\bibfield  {title} {\enquote {\bibinfo
  {title} {{Science with the TianQin observatory: Preliminary results on
  massive black hole binaries}},}\ }\href {\doibase
  10.1103/PhysRevD.100.043003} {\bibfield  {journal} {\bibinfo  {journal}
  {Phys. Rev. D}\ }\textbf {\bibinfo {volume} {100}},\ \bibinfo {pages}
  {043003} (\bibinfo {year} {2019})},\ \Eprint
  {http://arxiv.org/abs/1902.04423} {arXiv:1902.04423 [astro-ph.HE]}
  \BibitemShut {NoStop}%
\bibitem [{\citenamefont {Mei}\ \emph {et~al.}(2021)\citenamefont {Mei} \emph
  {et~al.}}]{TianQin:2020hid}%
  \BibitemOpen
  \bibfield  {author} {\bibinfo {author} {\bibfnamefont {Jianwei}\ \bibnamefont
  {Mei}} \emph {et~al.} (\bibinfo {collaboration} {TianQin}),\ }\bibfield
  {title} {\enquote {\bibinfo {title} {{The TianQin project: current progress
  on science and technology}},}\ }\href {\doibase 10.1093/ptep/ptaa114}
  {\bibfield  {journal} {\bibinfo  {journal} {PTEP}\ }\textbf {\bibinfo
  {volume} {2021}},\ \bibinfo {pages} {05A107} (\bibinfo {year} {2021})},\
  \Eprint {http://arxiv.org/abs/2008.10332} {arXiv:2008.10332 [gr-qc]}
  \BibitemShut {NoStop}%
\bibitem [{\citenamefont {Kawamura}\ \emph {et~al.}(2011)\citenamefont
  {Kawamura} \emph {et~al.}}]{Kawamura:2011zz}%
  \BibitemOpen
  \bibfield  {author} {\bibinfo {author} {\bibfnamefont {Seiji}\ \bibnamefont
  {Kawamura}} \emph {et~al.},\ }\bibfield  {title} {\enquote {\bibinfo {title}
  {{The Japanese space gravitational wave antenna: DECIGO}},}\ }\href {\doibase
  10.1088/0264-9381/28/9/094011} {\bibfield  {journal} {\bibinfo  {journal}
  {Class. Quant. Grav.}\ }\textbf {\bibinfo {volume} {28}},\ \bibinfo {pages}
  {094011} (\bibinfo {year} {2011})}\BibitemShut {NoStop}%
\bibitem [{\citenamefont {Jin}\ \emph {et~al.}(2023)\citenamefont {Jin},
  \citenamefont {Xing}, \citenamefont {Shao}, \citenamefont {Zhang},\ and\
  \citenamefont {Zhang}}]{Jin:2023zhi}%
  \BibitemOpen
  \bibfield  {author} {\bibinfo {author} {\bibfnamefont {Shang-Jie}\
  \bibnamefont {Jin}}, \bibinfo {author} {\bibfnamefont {Shuang-Shuang}\
  \bibnamefont {Xing}}, \bibinfo {author} {\bibfnamefont {Yue}\ \bibnamefont
  {Shao}}, \bibinfo {author} {\bibfnamefont {Jing-Fei}\ \bibnamefont {Zhang}},
  \ and\ \bibinfo {author} {\bibfnamefont {Xin}\ \bibnamefont {Zhang}},\
  }\bibfield  {title} {\enquote {\bibinfo {title} {{Joint constraints on
  cosmological parameters using future multi-band gravitational wave standard
  siren observations*}},}\ }\href {\doibase 10.1088/1674-1137/acc8be}
  {\bibfield  {journal} {\bibinfo  {journal} {Chin. Phys. C}\ }\textbf
  {\bibinfo {volume} {47}},\ \bibinfo {pages} {065104} (\bibinfo {year}
  {2023})},\ \Eprint {http://arxiv.org/abs/2301.06722} {arXiv:2301.06722
  [astro-ph.CO]} \BibitemShut {NoStop}%
\bibitem [{\citenamefont {Jin}\ \emph {et~al.}(2025)\citenamefont {Jin},
  \citenamefont {Song}, \citenamefont {Sun}, \citenamefont {Xiao},
  \citenamefont {Wang}, \citenamefont {Wang}, \citenamefont {Zhang},\ and\
  \citenamefont {Zhang}}]{Jin:2025dvf}%
  \BibitemOpen
  \bibfield  {author} {\bibinfo {author} {\bibfnamefont {Shang-Jie}\
  \bibnamefont {Jin}}, \bibinfo {author} {\bibfnamefont {Ji-Yu}\ \bibnamefont
  {Song}}, \bibinfo {author} {\bibfnamefont {Tian-Yang}\ \bibnamefont {Sun}},
  \bibinfo {author} {\bibfnamefont {Si-Ren}\ \bibnamefont {Xiao}}, \bibinfo
  {author} {\bibfnamefont {He}~\bibnamefont {Wang}}, \bibinfo {author}
  {\bibfnamefont {Ling-Feng}\ \bibnamefont {Wang}}, \bibinfo {author}
  {\bibfnamefont {Jing-Fei}\ \bibnamefont {Zhang}}, \ and\ \bibinfo {author}
  {\bibfnamefont {Xin}\ \bibnamefont {Zhang}},\ }\bibfield  {title} {\enquote
  {\bibinfo {title} {{Gravitational wave standard sirens: A brief review of
  cosmological parameter estimation}},}\ }\href@noop {} {\  (\bibinfo {year}
  {2025})},\ \Eprint {http://arxiv.org/abs/2507.12965} {arXiv:2507.12965
  [astro-ph.CO]} \BibitemShut {NoStop}%
\bibitem [{\citenamefont {Cao}\ \emph {et~al.}(2024)\citenamefont {Cao} \emph
  {et~al.}}]{Cao:2023bnl}%
  \BibitemOpen
  \bibfield  {author} {\bibinfo {author} {\bibfnamefont {Xiaoyue}\ \bibnamefont
  {Cao}} \emph {et~al.},\ }\bibfield  {title} {\enquote {\bibinfo {title}
  {{CSST strong lensing preparation: forecasting the galaxy{\textendash}galaxy
  strong lensing population for the China space station telescope}},}\ }\href
  {\doibase 10.1093/mnras/stae1865} {\bibfield  {journal} {\bibinfo  {journal}
  {Mon. Not. Roy. Astron. Soc.}\ }\textbf {\bibinfo {volume} {533}},\ \bibinfo
  {pages} {1960--1975} (\bibinfo {year} {2024})},\ \Eprint
  {http://arxiv.org/abs/2312.06239} {arXiv:2312.06239 [astro-ph.GA]}
  \BibitemShut {NoStop}%
\bibitem [{\citenamefont {Santos}\ \emph {et~al.}(2015)\citenamefont {Santos}
  \emph {et~al.}}]{Santos:2015gra}%
  \BibitemOpen
  \bibfield  {author} {\bibinfo {author} {\bibfnamefont {Mario~G.}\
  \bibnamefont {Santos}} \emph {et~al.},\ }\bibfield  {title} {\enquote
  {\bibinfo {title} {{Cosmology from a SKA HI intensity mapping survey}},}\
  }\href {\doibase 10.22323/1.215.0019} {\bibfield  {journal} {\bibinfo
  {journal} {PoS}\ }\textbf {\bibinfo {volume} {AASKA14}},\ \bibinfo {pages}
  {019} (\bibinfo {year} {2015})},\ \Eprint {http://arxiv.org/abs/1501.03989}
  {arXiv:1501.03989 [astro-ph.CO]} \BibitemShut {NoStop}%
\bibitem [{\citenamefont {Bacon}\ \emph {et~al.}(2020)\citenamefont {Bacon}
  \emph {et~al.}}]{SKA:2018ckk}%
  \BibitemOpen
  \bibfield  {author} {\bibinfo {author} {\bibfnamefont {David~J.}\
  \bibnamefont {Bacon}} \emph {et~al.} (\bibinfo {collaboration} {SKA}),\
  }\bibfield  {title} {\enquote {\bibinfo {title} {{Cosmology with Phase 1 of
  the Square Kilometre Array: Red Book 2018: Technical specifications and
  performance forecasts}},}\ }\href {\doibase 10.1017/pasa.2019.51} {\bibfield
  {journal} {\bibinfo  {journal} {Publ. Astron. Soc. Austral.}\ }\textbf
  {\bibinfo {volume} {37}},\ \bibinfo {pages} {e007} (\bibinfo {year}
  {2020})},\ \Eprint {http://arxiv.org/abs/1811.02743} {arXiv:1811.02743
  [astro-ph.CO]} \BibitemShut {NoStop}%
\bibitem [{\citenamefont {Wu}\ and\ \citenamefont {Zhang}(2022)}]{Wu:2021vfz}%
  \BibitemOpen
  \bibfield  {author} {\bibinfo {author} {\bibfnamefont {Peng-Ju}\ \bibnamefont
  {Wu}}\ and\ \bibinfo {author} {\bibfnamefont {Xin}\ \bibnamefont {Zhang}},\
  }\bibfield  {title} {\enquote {\bibinfo {title} {{Prospects for measuring
  dark energy with 21 cm intensity mapping experiments}},}\ }\href {\doibase
  10.1088/1475-7516/2022/01/060} {\bibfield  {journal} {\bibinfo  {journal}
  {JCAP}\ }\textbf {\bibinfo {volume} {01}},\ \bibinfo {pages} {060} (\bibinfo
  {year} {2022})},\ \Eprint {http://arxiv.org/abs/2108.03552} {arXiv:2108.03552
  [astro-ph.CO]} \BibitemShut {NoStop}%
\bibitem [{\citenamefont {Amiri}\ \emph {et~al.}(2022)\citenamefont {Amiri}
  \emph {et~al.}}]{CHIME:2022dwe}%
  \BibitemOpen
  \bibfield  {author} {\bibinfo {author} {\bibfnamefont {Mandana}\ \bibnamefont
  {Amiri}} \emph {et~al.} (\bibinfo {collaboration} {CHIME}),\ }\bibfield
  {title} {\enquote {\bibinfo {title} {{An Overview of CHIME, the Canadian
  Hydrogen Intensity Mapping Experiment}},}\ }\href {\doibase
  10.3847/1538-4365/ac6fd9} {\bibfield  {journal} {\bibinfo  {journal}
  {Astrophys. J. Supp.}\ }\textbf {\bibinfo {volume} {261}},\ \bibinfo {pages}
  {29} (\bibinfo {year} {2022})},\ \Eprint {http://arxiv.org/abs/2201.07869}
  {arXiv:2201.07869 [astro-ph.IM]} \BibitemShut {NoStop}%
\bibitem [{\citenamefont {Pan}\ \emph {et~al.}(2025)\citenamefont {Pan},
  \citenamefont {Wu}, \citenamefont {Du}, \citenamefont {Li},\ and\
  \citenamefont {Zhang}}]{Pan:2024xoj}%
  \BibitemOpen
  \bibfield  {author} {\bibinfo {author} {\bibfnamefont {Jun-Da}\ \bibnamefont
  {Pan}}, \bibinfo {author} {\bibfnamefont {Peng-Ju}\ \bibnamefont {Wu}},
  \bibinfo {author} {\bibfnamefont {Guo-Hong}\ \bibnamefont {Du}}, \bibinfo
  {author} {\bibfnamefont {Yichao}\ \bibnamefont {Li}}, \ and\ \bibinfo
  {author} {\bibfnamefont {Xin}\ \bibnamefont {Zhang}},\ }\bibfield  {title}
  {\enquote {\bibinfo {title} {{Prospects for cosmological research with the
  FAST array: 21-cm intensity mapping survey observation strategies}},}\ }\href
  {\doibase 10.1088/1475-7516/2025/01/080} {\bibfield  {journal} {\bibinfo
  {journal} {JCAP}\ }\textbf {\bibinfo {volume} {01}},\ \bibinfo {pages} {080}
  (\bibinfo {year} {2025})},\ \Eprint {http://arxiv.org/abs/2408.00268}
  {arXiv:2408.00268 [astro-ph.CO]} \BibitemShut {NoStop}%
\bibitem [{\citenamefont {Wu}\ \emph {et~al.}(2023{\natexlab{b}})\citenamefont
  {Wu}, \citenamefont {Li}, \citenamefont {Zhang},\ and\ \citenamefont
  {Zhang}}]{Wu:2022jkf}%
  \BibitemOpen
  \bibfield  {author} {\bibinfo {author} {\bibfnamefont {Peng-Ju}\ \bibnamefont
  {Wu}}, \bibinfo {author} {\bibfnamefont {Yichao}\ \bibnamefont {Li}},
  \bibinfo {author} {\bibfnamefont {Jing-Fei}\ \bibnamefont {Zhang}}, \ and\
  \bibinfo {author} {\bibfnamefont {Xin}\ \bibnamefont {Zhang}},\ }\bibfield
  {title} {\enquote {\bibinfo {title} {{Prospects for measuring dark energy
  with 21 cm intensity mapping experiments: A joint survey strategy}},}\ }\href
  {\doibase 10.1007/s11433-022-2104-7} {\bibfield  {journal} {\bibinfo
  {journal} {Sci. China Phys. Mech. Astron.}\ }\textbf {\bibinfo {volume}
  {66}},\ \bibinfo {pages} {270413} (\bibinfo {year} {2023}{\natexlab{b}})},\
  \Eprint {http://arxiv.org/abs/2212.07681} {arXiv:2212.07681 [astro-ph.CO]}
  \BibitemShut {NoStop}%
\bibitem [{\citenamefont {Abdul~Karim}\ \emph {et~al.}(2025)\citenamefont
  {Abdul~Karim} \emph {et~al.}}]{DESI:2025zgx}%
  \BibitemOpen
  \bibfield  {author} {\bibinfo {author} {\bibfnamefont {M.}~\bibnamefont
  {Abdul~Karim}} \emph {et~al.} (\bibinfo {collaboration} {DESI}),\ }\bibfield
  {title} {\enquote {\bibinfo {title} {{DESI DR2 results. II. Measurements of
  baryon acoustic oscillations and cosmological constraints}},}\ }\href
  {\doibase 10.1103/tr6y-kpc6} {\bibfield  {journal} {\bibinfo  {journal}
  {Phys. Rev. D}\ }\textbf {\bibinfo {volume} {112}},\ \bibinfo {pages}
  {083515} (\bibinfo {year} {2025})},\ \Eprint
  {http://arxiv.org/abs/2503.14738} {arXiv:2503.14738 [astro-ph.CO]}
  \BibitemShut {NoStop}%
\bibitem [{\citenamefont {Amendola}\ \emph {et~al.}(2018)\citenamefont
  {Amendola} \emph {et~al.}}]{Amendola:2016saw}%
  \BibitemOpen
  \bibfield  {author} {\bibinfo {author} {\bibfnamefont {Luca}\ \bibnamefont
  {Amendola}} \emph {et~al.},\ }\bibfield  {title} {\enquote {\bibinfo {title}
  {{Cosmology and fundamental physics with the Euclid satellite}},}\ }\href
  {\doibase 10.1007/s41114-017-0010-3} {\bibfield  {journal} {\bibinfo
  {journal} {Living Rev. Rel.}\ }\textbf {\bibinfo {volume} {21}},\ \bibinfo
  {pages} {2} (\bibinfo {year} {2018})},\ \Eprint
  {http://arxiv.org/abs/1606.00180} {arXiv:1606.00180 [astro-ph.CO]}
  \BibitemShut {NoStop}%
\bibitem [{\citenamefont {Cao}\ and\ \citenamefont
  {Ratra}(2022)}]{Cao:2022ugh}%
  \BibitemOpen
  \bibfield  {author} {\bibinfo {author} {\bibfnamefont {Shulei}\ \bibnamefont
  {Cao}}\ and\ \bibinfo {author} {\bibfnamefont {Bharat}\ \bibnamefont
  {Ratra}},\ }\bibfield  {title} {\enquote {\bibinfo {title} {{Using lower
  redshift, non-CMB, data to constrain the Hubble constant and other
  cosmological parameters}},}\ }\href {\doibase 10.1093/mnras/stac1184}
  {\bibfield  {journal} {\bibinfo  {journal} {Mon. Not. Roy. Astron. Soc.}\
  }\textbf {\bibinfo {volume} {513}},\ \bibinfo {pages} {5686--5700} (\bibinfo
  {year} {2022})},\ \Eprint {http://arxiv.org/abs/2203.10825} {arXiv:2203.10825
  [astro-ph.CO]} \BibitemShut {NoStop}%
\bibitem [{\citenamefont {Cao}\ \emph {et~al.}(2022{\natexlab{a}})\citenamefont
  {Cao}, \citenamefont {Khadka},\ and\ \citenamefont {Ratra}}]{Cao:2021irf}%
  \BibitemOpen
  \bibfield  {author} {\bibinfo {author} {\bibfnamefont {Shulei}\ \bibnamefont
  {Cao}}, \bibinfo {author} {\bibfnamefont {Narayan}\ \bibnamefont {Khadka}}, \
  and\ \bibinfo {author} {\bibfnamefont {Bharat}\ \bibnamefont {Ratra}},\
  }\bibfield  {title} {\enquote {\bibinfo {title} {{Standardizing
  Dainotti-correlated gamma-ray bursts, and using them with standardized
  Amati-correlated gamma-ray bursts to constrain cosmological model
  parameters}},}\ }\href {\doibase 10.1093/mnras/stab3559} {\bibfield
  {journal} {\bibinfo  {journal} {Mon. Not. Roy. Astron. Soc.}\ }\textbf
  {\bibinfo {volume} {510}},\ \bibinfo {pages} {2928--2947} (\bibinfo {year}
  {2022}{\natexlab{a}})},\ \Eprint {http://arxiv.org/abs/2110.14840}
  {arXiv:2110.14840 [astro-ph.CO]} \BibitemShut {NoStop}%
\bibitem [{\citenamefont {Cao}\ \emph {et~al.}(2022{\natexlab{b}})\citenamefont
  {Cao}, \citenamefont {Dainotti},\ and\ \citenamefont {Ratra}}]{Cao:2022wlg}%
  \BibitemOpen
  \bibfield  {author} {\bibinfo {author} {\bibfnamefont {Shulei}\ \bibnamefont
  {Cao}}, \bibinfo {author} {\bibfnamefont {Maria}\ \bibnamefont {Dainotti}}, \
  and\ \bibinfo {author} {\bibfnamefont {Bharat}\ \bibnamefont {Ratra}},\
  }\bibfield  {title} {\enquote {\bibinfo {title} {{Standardizing Platinum
  Dainotti-correlated gamma-ray bursts, and using them with standardized
  Amati-correlated gamma-ray bursts to constrain cosmological model
  parameters}},}\ }\href {\doibase 10.1093/mnras/stac517} {\bibfield  {journal}
  {\bibinfo  {journal} {Mon. Not. Roy. Astron. Soc.}\ }\textbf {\bibinfo
  {volume} {512}},\ \bibinfo {pages} {439--454} (\bibinfo {year}
  {2022}{\natexlab{b}})},\ \Eprint {http://arxiv.org/abs/2201.05245}
  {arXiv:2201.05245 [astro-ph.CO]} \BibitemShut {NoStop}%
\bibitem [{\citenamefont {Dainotti}\ \emph
  {et~al.}(2023{\natexlab{a}})\citenamefont {Dainotti}, \citenamefont
  {Bargiacchi}, \citenamefont {Bogdan}, \citenamefont {Lenart}, \citenamefont
  {Iwasaki}, \citenamefont {Capozziello}, \citenamefont {Zhang},\ and\
  \citenamefont {Fraija}}]{Dainotti:2023bwq}%
  \BibitemOpen
  \bibfield  {author} {\bibinfo {author} {\bibfnamefont {Maria~Giovanna}\
  \bibnamefont {Dainotti}}, \bibinfo {author} {\bibfnamefont {Giada}\
  \bibnamefont {Bargiacchi}}, \bibinfo {author} {\bibfnamefont {Ma{\l}gorzata}\
  \bibnamefont {Bogdan}}, \bibinfo {author} {\bibfnamefont
  {Aleksander~{\L}ukasz}\ \bibnamefont {Lenart}}, \bibinfo {author}
  {\bibfnamefont {Kazunari}\ \bibnamefont {Iwasaki}}, \bibinfo {author}
  {\bibfnamefont {Salvatore}\ \bibnamefont {Capozziello}}, \bibinfo {author}
  {\bibfnamefont {Bing}\ \bibnamefont {Zhang}}, \ and\ \bibinfo {author}
  {\bibfnamefont {Nissim}\ \bibnamefont {Fraija}},\ }\bibfield  {title}
  {\enquote {\bibinfo {title} {{Reducing the Uncertainty on the Hubble Constant
  up to 35{\%} with an Improved Statistical Analysis: Different Best-fit
  Likelihoods for Type Ia Supernovae, Baryon Acoustic Oscillations, Quasars,
  and Gamma-Ray Bursts}},}\ }\href {\doibase 10.3847/1538-4357/acd63f}
  {\bibfield  {journal} {\bibinfo  {journal} {Astrophys. J.}\ }\textbf
  {\bibinfo {volume} {951}},\ \bibinfo {pages} {63} (\bibinfo {year}
  {2023}{\natexlab{a}})},\ \Eprint {http://arxiv.org/abs/2305.10030}
  {arXiv:2305.10030 [astro-ph.CO]} \BibitemShut {NoStop}%
\bibitem [{\citenamefont {Dainotti}\ \emph
  {et~al.}(2023{\natexlab{b}})\citenamefont {Dainotti}, \citenamefont {Lenart},
  \citenamefont {Chraya}, \citenamefont {Sarracino}, \citenamefont {Nagataki},
  \citenamefont {Fraija}, \citenamefont {Capozziello},\ and\ \citenamefont
  {Bogdan}}]{Dainotti:2022ked}%
  \BibitemOpen
  \bibfield  {author} {\bibinfo {author} {\bibfnamefont {M.~G.}\ \bibnamefont
  {Dainotti}}, \bibinfo {author} {\bibfnamefont {A.~{\L}.}\ \bibnamefont
  {Lenart}}, \bibinfo {author} {\bibfnamefont {A.}~\bibnamefont {Chraya}},
  \bibinfo {author} {\bibfnamefont {G.}~\bibnamefont {Sarracino}}, \bibinfo
  {author} {\bibfnamefont {S.}~\bibnamefont {Nagataki}}, \bibinfo {author}
  {\bibfnamefont {N.}~\bibnamefont {Fraija}}, \bibinfo {author} {\bibfnamefont
  {S.}~\bibnamefont {Capozziello}}, \ and\ \bibinfo {author} {\bibfnamefont
  {M.}~\bibnamefont {Bogdan}},\ }\bibfield  {title} {\enquote {\bibinfo {title}
  {{The gamma-ray bursts fundamental plane correlation as a cosmological
  tool}},}\ }\href {\doibase 10.1093/mnras/stac2752} {\bibfield  {journal}
  {\bibinfo  {journal} {Mon. Not. Roy. Astron. Soc.}\ }\textbf {\bibinfo
  {volume} {518}},\ \bibinfo {pages} {2201--2240} (\bibinfo {year}
  {2023}{\natexlab{b}})},\ \Eprint {http://arxiv.org/abs/2209.08675}
  {arXiv:2209.08675 [astro-ph.HE]} \BibitemShut {NoStop}%
\bibitem [{\citenamefont {Dainotti}\ \emph {et~al.}(2022)\citenamefont
  {Dainotti}, \citenamefont {Nielson}, \citenamefont {Sarracino}, \citenamefont
  {Rinaldi}, \citenamefont {Nagataki}, \citenamefont {Capozziello},
  \citenamefont {Gnedin},\ and\ \citenamefont {Bargiacchi}}]{Dainotti:2022wli}%
  \BibitemOpen
  \bibfield  {author} {\bibinfo {author} {\bibfnamefont {Maria~Giovanna}\
  \bibnamefont {Dainotti}}, \bibinfo {author} {\bibfnamefont {Via}\
  \bibnamefont {Nielson}}, \bibinfo {author} {\bibfnamefont {Giuseppe}\
  \bibnamefont {Sarracino}}, \bibinfo {author} {\bibfnamefont {Enrico}\
  \bibnamefont {Rinaldi}}, \bibinfo {author} {\bibfnamefont {Shigehiro}\
  \bibnamefont {Nagataki}}, \bibinfo {author} {\bibfnamefont {Salvatore}\
  \bibnamefont {Capozziello}}, \bibinfo {author} {\bibfnamefont {Oleg~Y.}\
  \bibnamefont {Gnedin}}, \ and\ \bibinfo {author} {\bibfnamefont {Giada}\
  \bibnamefont {Bargiacchi}},\ }\bibfield  {title} {\enquote {\bibinfo {title}
  {{Optical and X-ray GRB Fundamental Planes as cosmological distance
  indicators}},}\ }\href {\doibase 10.1093/mnras/stac1141} {\bibfield
  {journal} {\bibinfo  {journal} {Mon. Not. Roy. Astron. Soc.}\ }\textbf
  {\bibinfo {volume} {514}},\ \bibinfo {pages} {1828--1856} (\bibinfo {year}
  {2022})},\ \Eprint {http://arxiv.org/abs/2203.15538} {arXiv:2203.15538
  [astro-ph.CO]} \BibitemShut {NoStop}%
\bibitem [{\citenamefont {Bargiacchi}\ \emph {et~al.}(2023)\citenamefont
  {Bargiacchi}, \citenamefont {Dainotti},\ and\ \citenamefont
  {Capozziello}}]{Bargiacchi:2023rfd}%
  \BibitemOpen
  \bibfield  {author} {\bibinfo {author} {\bibfnamefont {G.}~\bibnamefont
  {Bargiacchi}}, \bibinfo {author} {\bibfnamefont {M.~G.}\ \bibnamefont
  {Dainotti}}, \ and\ \bibinfo {author} {\bibfnamefont {S.}~\bibnamefont
  {Capozziello}},\ }\bibfield  {title} {\enquote {\bibinfo {title} {{Tensions
  with the flat $\boldsymbol{\Lambda}$CDM model from high-redshift
  cosmography}},}\ }\href {\doibase 10.1093/mnras/stad2326} {\bibfield
  {journal} {\bibinfo  {journal} {Mon. Not. Roy. Astron. Soc.}\ }\textbf
  {\bibinfo {volume} {525}},\ \bibinfo {pages} {3104--3116} (\bibinfo {year}
  {2023})},\ \Eprint {http://arxiv.org/abs/2307.15359} {arXiv:2307.15359
  [astro-ph.CO]} \BibitemShut {NoStop}%
\bibitem [{\citenamefont {Troja}\ \emph {et~al.}(2017)\citenamefont {Troja}
  \emph {et~al.}}]{Troja:2017nqp}%
  \BibitemOpen
  \bibfield  {author} {\bibinfo {author} {\bibfnamefont {E.}~\bibnamefont
  {Troja}} \emph {et~al.},\ }\bibfield  {title} {\enquote {\bibinfo {title}
  {{The X-ray counterpart to the gravitational wave event GW 170817}},}\ }\href
  {\doibase 10.1038/nature24290} {\bibfield  {journal} {\bibinfo  {journal}
  {Nature}\ }\textbf {\bibinfo {volume} {551}},\ \bibinfo {pages} {71--74}
  (\bibinfo {year} {2017})},\ \Eprint {http://arxiv.org/abs/1710.05433}
  {arXiv:1710.05433 [astro-ph.HE]} \BibitemShut {NoStop}%
\end{thebibliography}%
\end{document}